\documentclass{aastex6}

\usepackage{graphics,graphicx}
\usepackage{natbib}
\usepackage{amsmath}

\usepackage{color}
\definecolor{red}{rgb}{1.,0.0,0.}

\begin{document}

\title{The Orbit and Dynamical Mass of Polaris: Observations with the CHARA Array
} 


\author{Nancy Remage Evans}
\affil{Smithsonian Astrophysical Observatory,
MS 4, 60 Garden St., Cambridge, MA 02138; nevans@cfa.harvard.edu}

\author{Gail H.\ Schaefer}
\affil{CHARA Array of Georgia State University, Mount Wilson, CA 91023}  

\author{Alexandre Gallenne}
\affiliation{Instituto de Astrof\'isica, Departamento de Ciencias F\'/sicas, Facultad de Ciencias Exactas, Universidad Andr\'es Bello, Fern\'andez Concha 700, Las Condes, Santiago, Chile and French-Chilean Laboratory for Astronomy, IRL 3386, CNRS, Casilla 36-D, Santiago, Chile}

\author{Guillermo Torres}
\affil{Smithsonian Astrophysical Observatory,
 60 Garden St., Cambridge, MA 02138}

\author{Elliott P. Horch}
\affil{ Department of Physics, Southern Connecticut State Univ,
501 Crescent Street, New Haven, CT 06515 }

\author{Richard I Anderson}
 \affil{Institute of Physics, \'Ecole Polytechnique F\'ed\'erale de Lausanne (EPFL), Observatoire de Sauverny, 1290 Versoix, Switzerland}

\author{John D.\ Monnier} 
\affiliation{University of Michigan, Department of Astronomy,
1085 S. University, Ann Arbor, MI 48109}

\author{Rachael M. Roettenbacher} 
\affiliation{University of Michigan, Department of Astronomy,
1085 S. University, Ann Arbor, MI 48109}

\author{Fabien Baron} 
\affiliation{Center for High Angular Resolution Astronomy and Department of Physics and Astronomy, Georgia State University, P.O. Box 5060, Atlanta, GA 30302-5060}

\author{Narsireddy Anugu}
\affil{CHARA Array of Georgia State University, Mount Wilson, CA 91023}

\author{James W. Davidson, Jr.}
\affil{ Department of Astronomy, Univ. of Virginia, 530 McCormick Rd.,
  Charlottesville, VA 22904}

\author{Pierre Kervella}
\affil{LESIA, Observatoire de Paris, Universit\'e PSL, CNRS, Sorbonne Universit\'e, Universit\'e Paris Cit\'e, 5 place Jules Janssen, 92195 Meudon, France}

\author{Garance Bras}
\affil{LESIA, Observatoire de Paris, Universit\'e PSL, CNRS, Sorbonne Universit\'e, Universit\'e Paris Cit\'e, 5 place Jules Janssen, 92195 Meudon, France}

\author{Charles Proffitt}
\affiliation{Space Telescope Science Institute, 3700 San Martin Drive, Baltimore, MD 21218}

\author{Antoine M\'erand}
 \affil{European Southern Observatory, Karl-Schwarzschild-Str. 2, 85748 Garching, Germany}

\author{Margarita Karovska}
\affil{Smithsonian Astrophysical Observatory,
MS 4, 60 Garden St., Cambridge, MA 02138}

\author{Jeremy Jones}
\affil{Center for High Angular Resolution Astronomy and Department of Physics and Astronomy, Georgia State University, P.O. Box 5060, Atlanta, GA 30302-5060}

\author{Cyprien Lanthermann}
\affil{CHARA Array of Georgia State University, Mount Wilson, CA 91023}

\author{Stefan Kraus}
\affil{University of Exeter, School of Physics and Astronomy, Astrophysics Group, Stocker Road, Exeter, EX4 4QL, UK }

\author{Isabelle Codron}
\affil{University of Exeter, School of Physics and Astronomy, Astrophysics Group, Stocker Road, Exeter, EX4 4QL, UK }

\author{Howard E. Bond}
\affil{ Department of Astronomy \& Astrophysics,  Pennsylvania State University, University Park, PA, 16802, USA
  and Space Telescope Science Institute, 3700 San Martin Dr., Baltimore, MD 21218, USA }

\author{ Giordano Viviani}
 \affil{Institute of Physics, \'Ecole Polytechnique F\'ed\'erale de Lausanne (EPFL), Observatoire de Sauverny, 1290 Versoix, Switzerland}





\begin{abstract} 
  The 30 year orbit of the Cepheid Polaris has been followed with observations by the
  CHARA Array (Center for High Angular Resolution Astronomy) from 2016 through
  2021.  An additional
  measurement has been made with speckle interferometry at the Apache Point Observatory.
 Detection of the companion is complicated 
  by  its comparative faintness--an extreme flux ratio.  Angular diameter
  measurements appear to show some variation with pulsation phase.  
  Astrometric positions of the companion were measured with a custom 
grid-based model-fitting procedure   and confirmed with the
  CANDID software.  These positions were combined with the extensive radial velocities
  discussed by Torres (2023) to fit an orbit.  Because of the imbalance of the sizes
  of the astrometry and radial velocity datasets, several methods of weighting
  are discussed.  The resulting mass of the Cepheid 
  is 5.13$\pm$ 0.28 $M_\odot$.
  Because of the comparatively large eccentricity of the orbit (0.63), the mass derived
  is sensitive to the value found for the eccentricity. 
  The mass combined with the distance shows that the Cepheid
   is more luminous than predicted for this mass from evolutionary tracks.
   The identification
  of surface spots is discussed.  This would give credence to the identification of
  photometric variation with a period of approximately 120 days as a rotation period.
  Polaris has some unusual properties (rapid period change, a phase jump,
  variable amplitude, unusual polarization). However, a
  pulsation scenario involving pulsation mode,
  orbital periastron passage (Torres 2023), and low pulsation amplitude can explain
   these characteristics within the framework of pulsation seen in Cepheids. 
  
\end{abstract}


\keywords{stars: masses Cepheids: Polaris: binaries; stars:massive; stars: variable}


\section{Introduction}

The importance of Cepheid variable stars as the first step in the cosmological distance scale has
been emphasized recently in the discussion of the tension between the Hubble constant {\it H$_0$}
from Cepheids and Type Ia supernovae
and that from the early universe based on {\it Planck\/} CMB  observations 
(Riess et al.\ 2021).

Another property of Cepheids,  their masses, 
provides pivotal tests of 
stellar evolution and predictions of the production of  end-stage 
objects such as neutron stars. 
 In stellar evolution calculations,  the luminosity at the Cepheid stage
depends on main-sequence core convective overshooting and 
rotation in Cepheid progenitors (B stars; Anderson et al.\ 2014)
and also mass loss (see eg. Neilson et al. 2013).  
Asteroseismology from space-based observations is revolutionizing stellar evolution 
calculations 
by providing information about the internal structure of stars (Kurtz 2022).  
Comparing the Mass-Luminosity (ML) relation from calculations with observed quantities
is an important test.   The persistent
``Cepheid mass problem'' (the disagreement between masses inferred from evolutionary tracks and 
from pulsation calculations) has been reduced to about 10\% (Bono et al.\ 2001, Neilson et al.
2011) but still remains. 
While calibration of Cepheid luminosities will
 be significantly improved in the final
 {\it Gaia} release, their masses can only be  measured directly
 in binary systems, and hence only a
limited number are available, one of which is Polaris.

 A further motivation for determining the masses of Milky Way
(MW) Cepheids as accurately as possible is that six Cepheids have been found to be members of eclipsing
binaries in the Large Magellanic Cloud (LMC; Pilecki et al.\ 2018, 2021), so comparison between
the  ML relations for the metallicities of the MW and the LMC can now
be made.

 Cepheids have typical masses of about 4--$5\, M_\odot$, 
but the longest-period Cepheids may be as massive as $\sim$$11\, M_\odot$ 
(Anderson et al.\ 2014). Most Cepheids are fated to become white dwarfs; however, the most
massive may become supernovae, and end their lives as 
neutron stars.
As an example of the importance of the calibration of the Cepheid 
ML relation, V1334 Cyg is the  Cepheid in the MW which has
the most accurate 
dynamical mass (accurate to 3\%; Gallenne et al. 2018).
This mass is smaller than 
predicted by evolutionary tracks. This means that  the fraction of 
Cepheids which are massive enough to become neutron stars 
 could be smaller than expected, leading to implications for the frequency of 
neutron stars.

Mass determination starts with a radial velocity orbit and
pulsation curve for a binary containing a Cepheid.
At the present time several new capabilities are available
which  make it possible to directly measure   model-independent,
dynamical  masses for  Cepheids using this orbit.  
Interferometry has resolved a number of systems, providing the semi-major axis, $a=a_1 +
a_2$, and the inclination, $i$.
CHARA(Center for High Angular Resolution Astronomy) Array
and the Very Large Telescope Interferometer (VLTI)
make it possible to reach stars in both hemispheres.
In addition, high-resolution spectra in
the  ultraviolet (UV) from {\it Hubble Space Telescope (HST)}
allow the orbital radial-velocity amplitude of hot companions
of Cepheids to be measured.
The {\it Gaia} spacecraft can provide both distances
and proper motions.  Astrometry is modeled by a sum of constant proper motion and an astrometric orbit, including the inclination.
Final versions of proper
motions and distances will not be available {\bf for Cepheids generally}
  until the {\it Gaia} DR4 data release
which includes orbital solutions in the astrometry {\bf  routinely}. 

The accuracy of inputs from any of these measurements depends on many
characteristics of the star: brightness, orbital period, inclination, 
and separation, distance, and mass ratio
of the components. This means that each Cepheid system is unique and has
to be analyzed independently.  The system under discussion here (Polaris),
for instance, has a comparatively low mass companion.  It is not possible
to measure a velocity of the companion with {\it HST} UV spectra.  It is also
too bright to be accurately measured by  {\it Gaia}, so the distance
determination depends on a third star in the system, as discussed below.
However, because it is the nearest Cepheid, and has a long period orbit,
the system has been resolved by  {\it HST}, CHARA and speckle interferometry
at APO.  The long orbital
period, however, results in a low velocity amplitude and
the need to make use of a combination
of radial velocities from many instruments.  Both the low orbital velocity amplitude and
the use of data from many sources  
are challenges for the
basic orbit determination, which have recently been addressed in the
spectroscopic study of Torres (2023).

This study is part of a series to incorporate current capabilities into
studies of Cepheid masses. 

\subsection{Polaris:}\label{intro.pol}

Polaris ($\alpha$ UMi, HR 424,  HD 8890, 
V = 2.02 mag) is the nearest and brightest classical Cepheid. It is a
member of a triple system with a resolved 8$^{th}$ magnitude physical companion
at a separation of 18$\arcsec$.  The Cepheid has been known for many years to be
a single-lined spectroscopic binary with a period of about 30 years (Roemer 1965;
Kamper 1996), with components designated Aa (the Cepheid) and Ab.
A thorough compilation of radial velocity data has recently
been produced by Torres (2023), resulting in a definitive orbit.

While Polaris Aa is a Cepheid, albeit with a small amplitude, it has
several characteristics which are unusual.

$\bullet$ It pulsates in the first overtone (Feast and Catchpole 1997)

$\bullet$ It has an unusually rapid period change (Neilson et al. 2012).  For fundamental mode pulsators,
the period change agrees with expectations of evolution through the instability strip.  For
overtone pulsators, some additional factor seems to be involved (Evans et al. 2018).  

$\bullet$ The pulsation amplitude can vary.  Specifically, it  decreased from about 1960 to 1990
(Arellano Ferro 1983). However, it did not die out completely (cease
pulsation), but in about 2000 began to increase (e.g. Bruntt et al. 2008;
Anderson 2019).  Anderson, indeed, found the amplitude to be very stable over
the course of 7 years. 

$\bullet$ In addition, there was a ``glitch'' (phase jump) in the O-C period residuals about 1960 (Arellano Ferro 1983; Turner et al. 2005).

$\bullet$  The instability strip crossing in which Polaris is located has been extensively
discussed.  Based on the most recent distance and the sign of the period
change, Evans et al. (2018) concluded it is on the third crossing, although it has
been argued that it is on the first crossing (Anderson 2018).  

$\bullet$ The possibility that the Cepheid itself may be the product of a merger is
discussed by Bond et al. (2018),  Evans et al. (2018) and Anderson (2018), based on
the fact that the distant component Polaris B is too cool to match the
isochrone of 100 Myr which fits the Cepheid, Polaris Aa.

$\bullet$ The single observation of Polaris to detect polarization (Barron et al. 2022)
found much more complicated Stokes V profile than for other Cepheids implying
a complex magnetic field.  However, the
profile is similar to the non-variable supergiant $\alpha$ Per (Grunhut et al. 2010), which perhaps
reflects the very low pulsation amplitude.

$\bullet$ Several studies have searched for additional periodicities 
 in the radial velocities (Lee et al. (2008)
Anderson (2019), and other studies discussed therein). 
A periodicity of approximately 120 days was found by Lee et al., for instance,  which
they identify as due to rotation in a star with spots. Two periodicities (40.2 and 60 d)
found by Anderson; 120d is an integral multiple of these.

These topics are fully discussed and references to all the velocity data sources are
provided by Torres (2023).  In this discussion these apparently exceptional properties
are placed in the context of Cepheids.   The times  of periastron passage of Polaris
 coincide with the amplitude changes and also the phase
jump at about 1960.
In the extensive study of period changes in Cepheids, Csornyei et al. (2022) find
a number with phase jumps, as was found by Szabados previously (1991, 1989, and
references therein).  They are often found in short period overtone
pulsators, and it was suggested by Szabados (1992) that they are likely to
be found in binaries.  
This would link a number of the peculiarities related to Polaris to phenomena
seen in other Cepheids, and provide a physical explanation.  On one hand, this
is surprising because the Polaris Aa and Ab system is comparatively wide.
Torres estimates that the periastron separation between the two stars
is 6.2 au, or approximately 29 times the Cepheid radius.    On the other
hand, the orbit is eccentric (e = 0.635), providing a variation of conditions
around the orbit.  The wide orbit would make it surprising for a characteristic
such as a phase jump to be produced by simple tidal interaction.  However,
overtone pulsators are more erratic in their periods than fundamental mode
pulsators, presumably for reasons we do not completely understand.  
Of course, these peculiarities may exist in other stars but have not been
identified because they do not have such  long and well covered datasets.

An important parameter in determining properties of Polaris is its distance.
Specifically, the determination of the total mass in the system depends on the distance
to convert the angular separation to au.  The distance 
has been controversial, 
partly because of difficulties because Polaris is very bright.
Recent distance determinations are summarized
by Engle et al. (2018).  Polaris B is the wide companion at a distance
of 18$\arcsec$ from the Polaris A system.  It can be used for distance determination.
Bond et al. (2018) argued that Polaris A and B form a gravitationally bound
system, which allows us to use Polaris B for distance determination. 
The distance to Polaris B using {\it Gaia} DR3 ({\it Gaia} Collaboration et al. 2023)
including the Lindegren parallax offset is 136.90 $\pm$ 0.34 pc
\citep{gaia23,Lindegren21}.  The distance will be reviewed when the {\it Gaia} DR4
is available, including corrections such as discussed in Khan et al. (2023).


Several previous steps have led up to the  measurement of the mass of the
Cepheid.  The binary sytem discussed here is  Polaris A made up of the Cepheid
Polaris Aa and the companion Polaris Ab.
Kamper (1996)   published an orbit, (with additional velocities by Kamper and Fernie [1998])
In addition, Wielen et al.
(2000) determined the inclination and the position angle of the line of nodes by
comparing the proper motion from the {\it Hipparcos} satellite with the average
long term ground-based proper motion.  The final parameter to determine the mass
was the separation of Aa and Ab.  This system was  resolved  with  
 {\it HST} (Evans et al. 2008) using the ACS camera.  This provided
the mass of the Cepheid of 4.5$^{+2.2}_{-1.4}M_\odot$.  This was followed up with
three observations  between 2007 and 2014 using the WFPC2 and WFC3
since the ACS HRC (high resolution channel) was no longer available (Evans et al. 2018).
The astrometry
from these observations was less accurate. However since they cover a quarter of the
orbit, the inclination could be determined from them as well as the separation
between the components. The present paper discusses  observations around the orbit
 continued with
  CHARA and a speckle observation from Apache Point Observatory (APO).

 The sections below contain the CHARA interferometric observations
 including data reduction
 and analysis, the diameter and surface imaging of Polaris,
 APO speckle observations, the orbit fitting, and discussion of
the results.


\section{Interferometric observations and data reduction} \label{section_data}

We collected long-baseline optical interferometric data at Georgia State University's Center for High Angular Resolution Astronomy (CHARA) Array \citep{tenbrummelaar05} located at Mount Wilson Observatory. The CHARA Array consists of six 1\,m aperture telescopes in an Y-shaped configuration with two telescopes along each arm, oriented to the east (E1, E2), west (W1,W2) and south (S1, S2), offering good coverage of the $(u, v)$ plane. The baselines range from 34\,m to 331\,m, providing an angular resolution down to 0.5 mas at $\lambda = 1.6\,\mu$m. The data on Polaris were collected with the Michigan InfraRed Combiner \citep[MIRC;][]{monnier04} before 2017.5 and the upgraded Michigan InfraRed Combiner-eXeter \citep[MIRC-X,][]{anugu20} after 2017.5. MIRC and MIRC-X combine the light from all six telescopes simultaneously in the $H$-band, providing up to 15 fringe visibilities and 20 closure phase measurements across multiple spectral channels.

At the location of Polaris in the sky, we can combine only four telescopes at a time because of limitations on the length of the CHARA delay lines. Initially we used the low spectral resolution mode (prism $R=\lambda/\Delta\lambda=50$)
with MIRC and MIRC-X. Bandwidth smearing limits the effective field of view to $\sim$ 50 mas (given by $\lambda^2/\Delta \lambda$) for the low spectral resolution mode. This resolution was sufficient to resolve the Ab companion during closest approach, however, detecting the companion became more challenging as the relative separation increased from orbital motion. In September 2019, we switched to higher spectral resolution modes (prism $R$ = 102, grism $R$ = 190) to extend the interferometric field of view to 100 mas and 190 mas, respectively. A log of our observations is available in Table~\ref{table_log}.

\begin{deluxetable*}{lcccccccc}
		\tablecaption{\label{table_log} Log of the MIRC and MIRC-X observations at the CHARA Array.}
		\tablewidth{0pt}
		\tablehead{
			\colhead{UT Date} & \colhead{Combiner} & \colhead{Configuration} & \colhead{Mode} & \colhead{$N_{\rm sets}$} & \colhead{$N_\mathrm{spec}$} & \colhead{$N_\mathrm{vis}$} & \colhead{$N_\mathrm{CP}$} & \colhead{Calibrators}  }
		\startdata
\hline
2016Sep12$^*$ & MIRC   & W1E1E2W2 &  R=50  & 1 &  8 & 1472 &  976  & 1, 2    \\
2016Nov18     & MIRC   & W1E1E2W2 &  R=50  & 1 &  8 & 1488 &  960  & 1, 2    \\
2018Aug27     & MIRC-X & W1E1E2W2 &  R=50  & 3 & 10 & 5560 & 3660  & 1, 3    \\
              &        & W1S2E1E2 &  R=50  & 3 & 10 & 5740 & 3780  & 1, 3    \\
2019Apr09     & MIRC-X & E1W2W1E2 &  R=50  & 4 &  6 & 3216 & 2016  & 1, 3    \\
              &        & E1W1S2E2 &  R=50  & 3 &  6 & 2310 & 1212  & 1, 3    \\
2019Sep02$^*$ & MIRC-X & E1W2W1E2 &  R=102 & 1 & 13 & 1287 &  816  & 1, 2    \\
              &        & E1W2W1E2 &  R=190 & 2 & 31 & 6076 & 3870  & 1, 2    \\
2021Apr02$^*$ & MIRC-X & E1W2W1E2 &  R=190 & 3 & 32 & 8256 & 5115  & 1, 3, 4  \\
              &        & E1W1S2E2 &  R=190 & 2 & 32 & 9664 & 6169  & 1, 3, 4  \\
2021Apr03$^*$ & MIRC-X & E1W2W1E2 &  R=190 & 5 & 33 & 11187 & 5920  & 1, 3, 4  \\
2021Apr04$^*$ & MIRC-X & E1W2W1E2 &  R=190 & 4 & 33 &  9174 & 4576  & 1, 3, 4  \\
              &        & E1W1S2E2 &  R=190 & 4 & 33 & 11649 & 7520  & 1, 3, 4  \\
		\enddata
		\tablecomments{The nights of 2018--2021 include two different telescope configurations and/or different spectral modes. The table gives the number of observing sets ($N_{\rm sets}$) and the number of spectral channels ($N_\mathrm{spec}$). It also lists the number of visibility ($N_\mathrm{vis}$) and closure phase ($N_\mathrm{CP}$) measurements for the reduction using the 30 second integration time. We adopted uniform disk diameters in the $H$-band for the calibrators from the JMMC angular diameter catalog \citep{bourges14}: 1: HD 6319, $\theta\mathrm{_{UDH}} = 0.740\pm0.078$\,mas, 2: HD 12918, $\theta\mathrm{_{UDH}} = 0.622\pm0.058$\,mas, 3: HD 42855, $\theta\mathrm{_{UDH}} = 0.758\pm0.078$\,mas, and 4: HD 204149 $\theta\mathrm{_{UDH}} = 0.671\pm0.065$\,mas. An asterisk next to the UT date indicates nights where we detected the Ab companion. \\}
	\end{deluxetable*}

Each observation consisted of recording 10 minutes of fringe data on Polaris followed by a shutter sequence to measure backgrounds, foregrounds, and the ratio of light between the fringe data and the photometric channels for each telescope. We monitored the interferometric transfer function by observing unresolved calibrator stars before and after each observation of Polaris. The calibrators were selected using the SearchCal software\footnote{\url{http://www.jmmc.fr/searchcal}} \citep{bonneau11} provided by the Jean-Marie Mariotti Center (JMMC).
The calibrators and their adopted angular diameters \citep{bourges14} are listed in Table~\ref{table_log}.
	
\begin{figure*}[ht]
  \centering
  \resizebox{\hsize}{!}{\includegraphics[width=.7\textwidth]{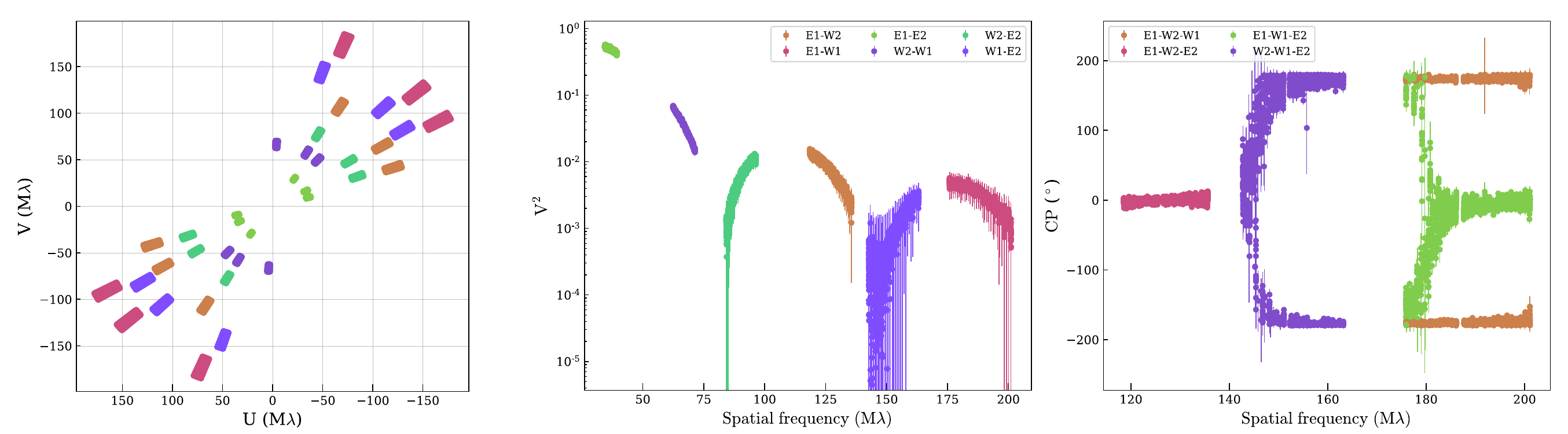}}
 \caption{$(u,v)$ coverage (left), squared visibilities (middle), and closure phase data (right)
   for the CHARA MIRC-X observations of Polaris on UT 2019Sep02.  The closure phase data is plotted against the spatial frequency of the longest baseline in the closed triangle.
 }
  \label{figure_data}
\end{figure*}

The data were reduced using the standard pipelines for MIRC \citep{monnier07} and MIRC-X\footnote{ \url{https://gitlab.chara.gsu.edu/lebouquj/mircx_pipeline.git}} \citep[version 1.3.5;][]{anugu20}. The pipelines produce squared visibilities and triple products for each baseline and spectral channel.  We used an integration time of 30 sec for improved detection of the faint companion located at separations that ranged from 30--85 mas. We also produced OIFITS files (see Section~\ref{surf.img})
that were averaged over each 10 min observing block for measuring the angular diameter and imaging the surface of Polaris Aa.  We used a calibration script written in IDL by J.D. Monnier with the ``deep cleaning'' option to remove outliers.
The reduced and calibrated OIFITS files will be available through the JMMC Optical Interferometry Database\footnote{\url{https://www.jmmc.fr/english/tools/data-bases/oidb/}} (OIDB)
An example of data collected on UT 2019Sep02 are displayed in Figure~\ref{figure_data}.


\begin{figure*}[ht]
  \centering
  \scalebox{0.5}{\includegraphics{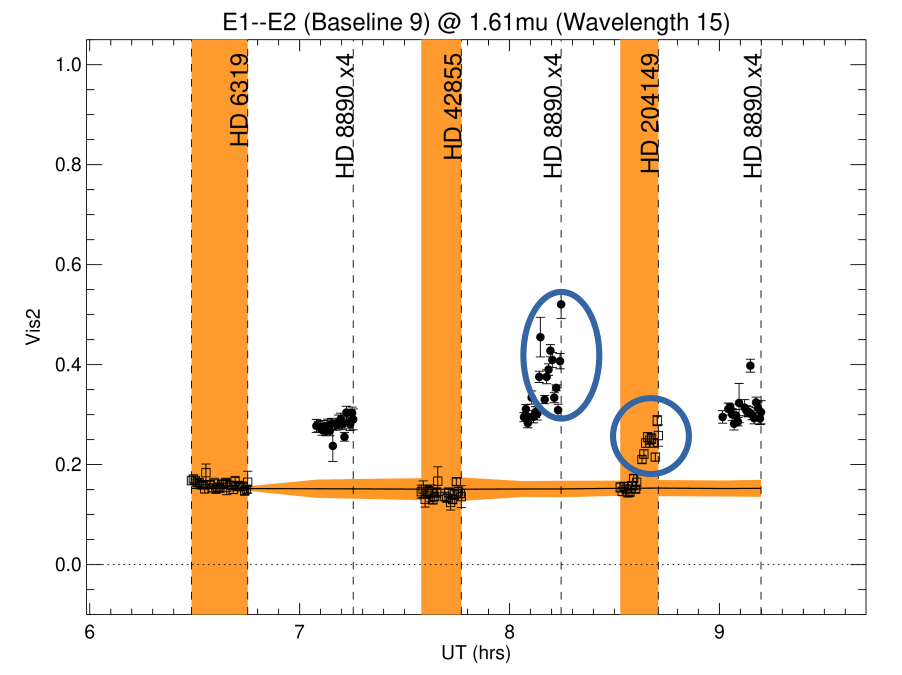}}
  \caption{Visibility transfer function for the E1-E2 baseline on UT 2021 Apr 02 for the E1E2W1W2 configuration. The calibrators are marked by the orange shaded regions. The visibilities of Polaris have been multiplied by a factor of four to make them more visible on the plot. Circled in blue are two regions showing discrepant visibilities on Polaris and calibrator HD 204149.}
  \label{figure_xfer}
\end{figure*}

There were occasional offsets and more scatter than expected in some of the uncalibrated visibilities, particularly on the baselines including the E2 telescope in April 2021. These discrepancies showed up between different calibrators and sometimes within a data set on the same target. Figure~\ref{figure_xfer} shows an example where the visibilities on E1-E2 baseline scatter upwards during the observation of calibrator HD 204149 and during the second set on Polaris. In both of these cases, the discrepant visibilities appeared after a brief pause in the data collection to realign the starlight into the MIRC-X fibers. We could not find a definitive cause for the discrepant observations; it could be related to vibrations in the E2 delay line cart while tracking at very slow speeds in the north or telescope oscillations while pointing at the pole. We inspected the visibility transfer functions from each night to reject measurements that were clearly discrepant. 
	
The angular diameters and binary separations computed in Section~\ref{section_analysis} were multiplied by factors of 1.004 $\pm$ 0.0025 for MIRC and divided by 1.0054 $\pm$ 0.0006 for MIRC-X (J.D. Monnier, private communication). This is equivalent to adjusting the respective wavelengths reported in the OIFITS files by the same factors.

\section{Data Analysis}  \label{section_analysis}

\subsection{Diameter Measurements of Polaris Aa}

We began our analysis of the interferometric data by fitting the squared visibility amplitudes for the angular diameter of Polaris Aa. The limb-darkening was parameterized with the linear function $I_\lambda (\mu) = I_\lambda (1) [ 1-u_\lambda (1-\mu) ]$ \citep{hanbury-brown74}, with $\mu = \cos{\zeta}$, where $\zeta$ is the angle between the line of sight and a surface element of the star. We solved for the limb-darkened diameter ($\theta_{\rm LD}$), limb-darkening coefficient ($u_\lambda$), and visibility scaling factor ($V_0$) to account for visibility miscalibrations or the presence of extended flux. The squared visibilities of the limb-darkened model are multiplied by $V_0$ before comparing with the measured values. We first performed a grid search through the three parameters to find the best fit and then followed a Monte Carlo bootstrap approach with 1000 iterations to determine uncertainties. For each bootstrap iteration, we randomly selected measurements with repetition to construct a new set of visibilities with some measurements repeated and some left out. We then applied Gaussian uncertainties to the resampled measurements and fit a limb-darkened model using the IDL MPFIT package \citep{markwardt09} to optimize the fit for each iteration. We adopted the standard deviation of the bootstrap distributions as the uncertainties on each parameter.  The angular diameters had been corrected  using the wavelength calibration factors discussed in Section~\ref{section_data}. To minimize scatter in the observations, the angular diameters were fit to the squared visibilities averaged over the 10-minute observing sets. The results from the diameter fits are presented in Table~\ref{table_diam}.

We measured a mean diameter of $\theta_{\rm LD} = 3.143 \pm 0.027$ mas and a mean limb-darkening coefficient of $u_\lambda = 0.120 \pm 0.043$. At the {\it Gaia} distance of $136.90 \pm 0.34$ pc  this corresponds to a mean radius of $46.27 \pm 0.42$ R$_\odot$. On most nights, the scaling factor $V_0$ was greater than 1.0, indicating that the $V^2$ measurements are higher than the limb-darkened model. This implies a ``negative'' background flux, indicating that the scaling term likely arises because of a miscalibration of the visibilities, rather than a physical cause like extended flux outside the field of view. The calibration issues discussed in Section~\ref{section_data}, as well as uncertainties in the partially resolved calibrator diameters, could have contributed to the miscalibration. The mean diameter is consistent within the uncertainties with the interferometric diameter measured with the FLUOR instrument at the CHARA Array \citep{merand06}. However, the calibration of the MIRC/MIRC-X visibilities was not accurate enough to confirm the presence of the extended circumstellar envelop reported by \citet{merand06}.

\begin{figure*}[ht]
  \centering
 \scalebox{0.16}{\includegraphics{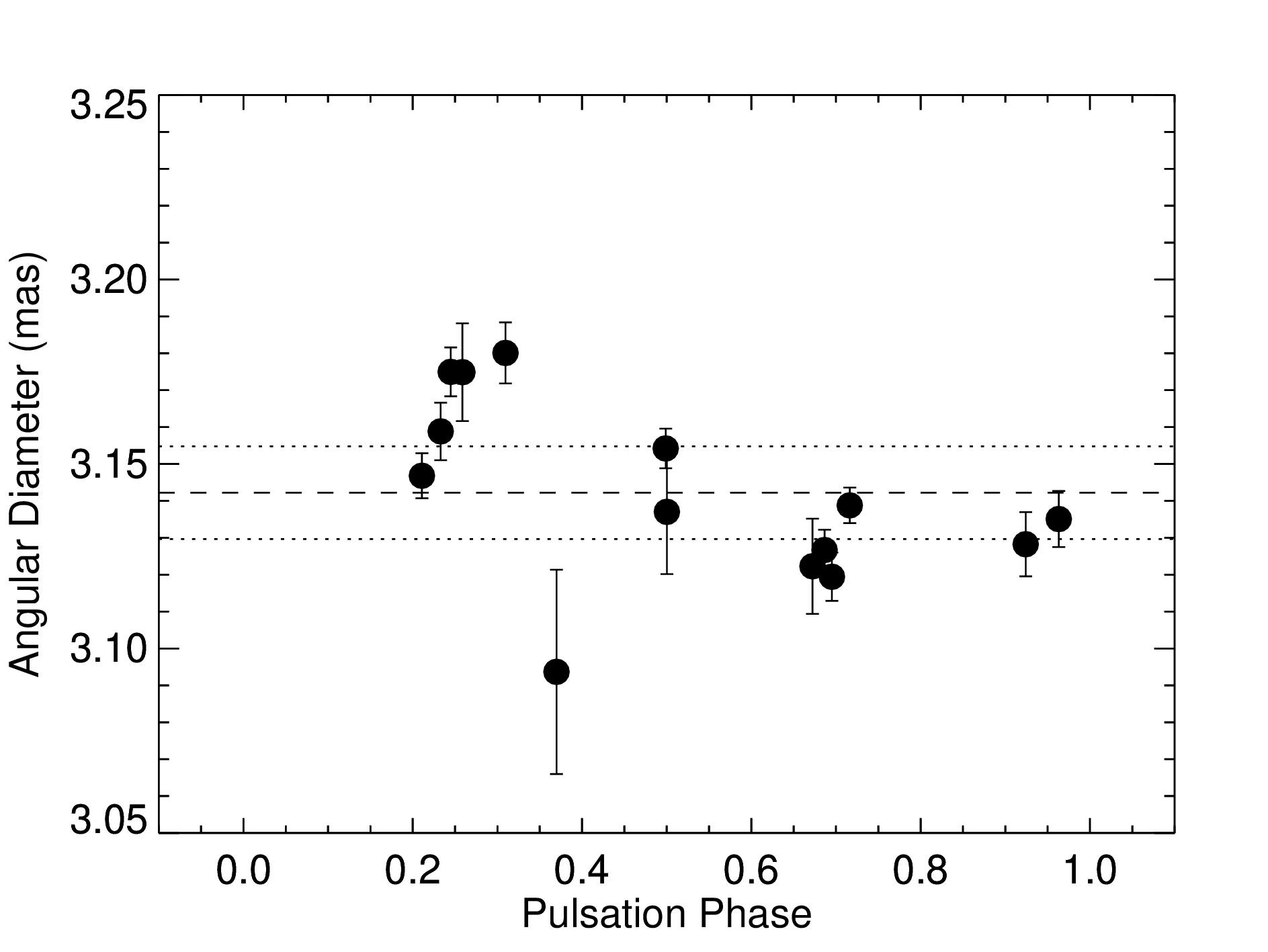}}  
  \caption{Limb-darkened diameters of Polaris Aa plotted against the pulsation phase. The dashed line represents the mean angular diameter and the dotted lines show 0.4\% variation in diameter expected from pulsation \citep{moskalik05}. }
  \label{figure_diam}
\end{figure*}

To investigate whether we see changes in the diameter over the 4-day pulsation cycle, we fit the diameters separately for each configuration observed on each night as shown in Table~\ref{table_diam}. We adopted the ephemeris of \citet{berdnikov95} for the times of maximum light:
\begin{equation}
\mathrm{HJD_{max}} = 2,428,260.727 + 3.969251 E
\end{equation}
 To account for the changing pulsation period, we used the O-C diagram in Figure~6 of \cite{torres23} to calculate the times of maximum light nearest to each angular diameter measurement. The columns in Table 2 are the date of the observation, the HJD -2,400,000, the time of maximum light used to compute the phase, the angular diameter and the limb darkening coefficient, the scaling factor, the angular diameter from a fixed limb darkening coefficient $u_\lambda$ = 0.120, and the number of visibility measurements. We plot the limb-darkened diameters with the fixed limb darkening coefficient against the pulsation phase in Figure~\ref{figure_diam}. The dashed line represents the mean angular diameter and the dotted lines show 0.4\% variation in diameter expected from pulsation \citep{moskalik05}. The standard deviation of the mean diameter is 2.1 times larger than the variation expected from pulsation. Cepheids are expected to reach a minimum {\bf angular diameter}  around phase 0.8, which is roughly consistent with our measurements. However, the variations in the measured diameters could also be impacted by the visibility miscalibrations discussed in Section~\ref{section_data}.

\begin{deluxetable*}{cccccccc}
  \tablecaption{\label{table_diam} Limb-darkened diameter fits for Polaris Aa.}
  \tablewidth{0pt}
  \tablehead{
    \colhead{UT Date} & \colhead{HJD} & \colhead{T$_{max}$} 
    & \colhead{$\theta_{\rm LD}$ (mas)} & \colhead{$u_\lambda$} & \colhead{$V_0$}
    & \colhead{$\theta_{fx u}$ (mas)} 
    & \colhead{$N_\mathrm{vis}$}  }
\startdata
2016Sep12 & 57643.736 & 57641.7504 & 3.1257 $\pm$ 0.0169 & 0.083 $\pm$ 0.044 & 1.2089 $\pm$ 0.0298 &  3.1370 $\pm$ 0.0169  & 96  \\
2016Nov18 & 57710.744 & 57709.2759 &  3.1042 $\pm$ 0.0277 & 0.156 $\pm$ 0.064 & 0.8628 $\pm$ 0.0501 & 3.0937 $\pm$ 0.0277  & 96  \\
2018Aug27 & 58357.744 & 58356.7171 & 3.1769 $\pm$ 0.0133 & 0.127 $\pm$ 0.037 & 1.2848 $\pm$ 0.0350 &  3.1749 $\pm$ 0.0133  & 360 \\
2018Aug27 & 58357.946 & 58356.7171 & 3.1811 $\pm$ 0.0083 & 0.125 $\pm$ 0.033 & 1.2695 $\pm$ 0.0296 &  3.1801 $\pm$ 0.0083  & 360 \\
2019Apr09 & 58582.815 & 58579.1471 & 3.1278 $\pm$ 0.0087 & 0.119 $\pm$ 0.024 & 1.2352 $\pm$ 0.0145 &  3.1283 $\pm$ 0.0087  & 216 \\
2019Apr09 & 58582.970 & 58579.1471 & 3.1386 $\pm$ 0.0076 & 0.136 $\pm$ 0.023 & 1.3896 $\pm$ 0.0184 &  3.1351 $\pm$ 0.0076  & 180 \\
2019Sep02 & 58728.777 & 58726.1087 & 3.1095 $\pm$ 0.0129 & 0.074 $\pm$ 0.038 & 1.0904 $\pm$ 0.0182 &  3.1223 $\pm$ 0.0129  & 78  \\
2019Sep02 & 58728.868 & 58726.1087 & 3.1392 $\pm$ 0.0065 & 0.186 $\pm$ 0.018 & 1.1135 $\pm$ 0.0079 &  3.1194 $\pm$ 0.0065  & 372 \\
2021Apr02 & 59306.839 & 59306.0020 & 3.1377 $\pm$ 0.0061 & 0.089 $\pm$ 0.018 & 0.9053 $\pm$ 0.0066 &  3.1468 $\pm$ 0.0061  & 576 \\
2021Apr02 & 59306.927 & 59306.0020 & 3.1484 $\pm$ 0.0078 & 0.065 $\pm$ 0.025 & 1.0072 $\pm$ 0.0108 &  3.1588 $\pm$ 0.0078  & 192 \\
2021Apr02 & 59306.974 & 59306.0020 & 3.1848 $\pm$ 0.0066 & 0.167 $\pm$ 0.018 & 0.8198 $\pm$ 0.0064 &  3.1750 $\pm$ 0.0066  & 384 \\
2021Apr03 & 59307.982 & 59306.0020 & 3.1747 $\pm$ 0.0054 & 0.189 $\pm$ 0.014 & 1.1104 $\pm$ 0.0090 &  3.1542 $\pm$ 0.0054  & 660 \\
2021Apr04 & 59308.727 & 59306.0020 & 3.1179 $\pm$ 0.0055 & 0.089 $\pm$ 0.015 & 1.1564 $\pm$ 0.0111 &  3.1267 $\pm$ 0.0055  & 594 \\
2021Apr04 & 59308.845 & 59306.0020 & 3.1289 $\pm$ 0.0048 & 0.069 $\pm$ 0.016 & 1.0364 $\pm$ 0.0088 &  3.1388 $\pm$ 0.0048  & 792 \\
\enddata
\end{deluxetable*}

\subsection{Surface Imaging of Polaris Aa}\label{surf.img}

    The limb-darkened diameter model does a reasonable job fitting the squared visibility amplitudes. However, the strong non-zero and non-180$^\circ$ closure phases (see Figure~\ref{figure_data}) suggest the presence of asymmetries on the surface of Polaris Aa. Initial modeling of the data indicated that these strong deviations in the closure phases are not produced by the faint binary companion Ab. Modeling with PMOIRED\footnote{\url{https://github.com/amerand/PMOIRED}} \citep{merand22} and a custom-made routine in IDL, indicates that a few starspots on the surface of Polaris Aa can account for most of the closure phase signal. However, the size, location, and contrast of the spots are not well-constrained due to the limited ($u,v$) coverage during the observations.

    We also reconstructed images of the stellar surface of Polaris Aa using SQUEEZE\footnote{\url{https://github.com/fabienbaron/squeeze}} \citep{baron10,baron12}, SURFING \citep[e.g.,][]{roettenbacher16}, and ROTIR\footnote{\url{https://github.com/fabienbaron/ROTIR.jl}} \citep{baron20}. Both SURFING and ROTIR perform the image reconstructions on the surface of a spheroid, constraining the size and shape of the star which is important given the limited ($u,v$) coverage of the observations. ROTIR is written in Julia and uses the OITOOLS libraries\footnote{\url{https://github.com/fabienbaron/OITOOLS.jl}}, a package for visualizing, modeling, and imaging interferometric data. We used the total variation regularizer with a weight of 0.01-0.05 for ROTIR.  To prepare the OIFITS files for imaging, we corrected the visibilities for each individual data set using $V_0$ scaling factors computed from limb-darkened diameter fits; we used the data averaged over the 10-minute observing sets. We merged together the datasets from all configurations on a given night to improve the ($u,v$) coverage for the image reconstructions. For the data from April 2021, we merged the data together from the three consecutive nights.

\begin{figure*}[ht]
  \begin{center}
    \scalebox{0.45}{\includegraphics{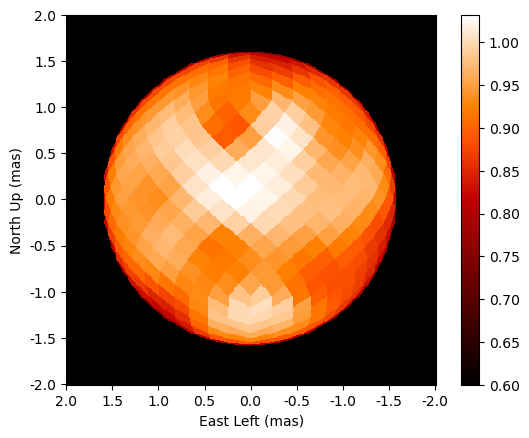}}
     \scalebox{0.3}{\includegraphics{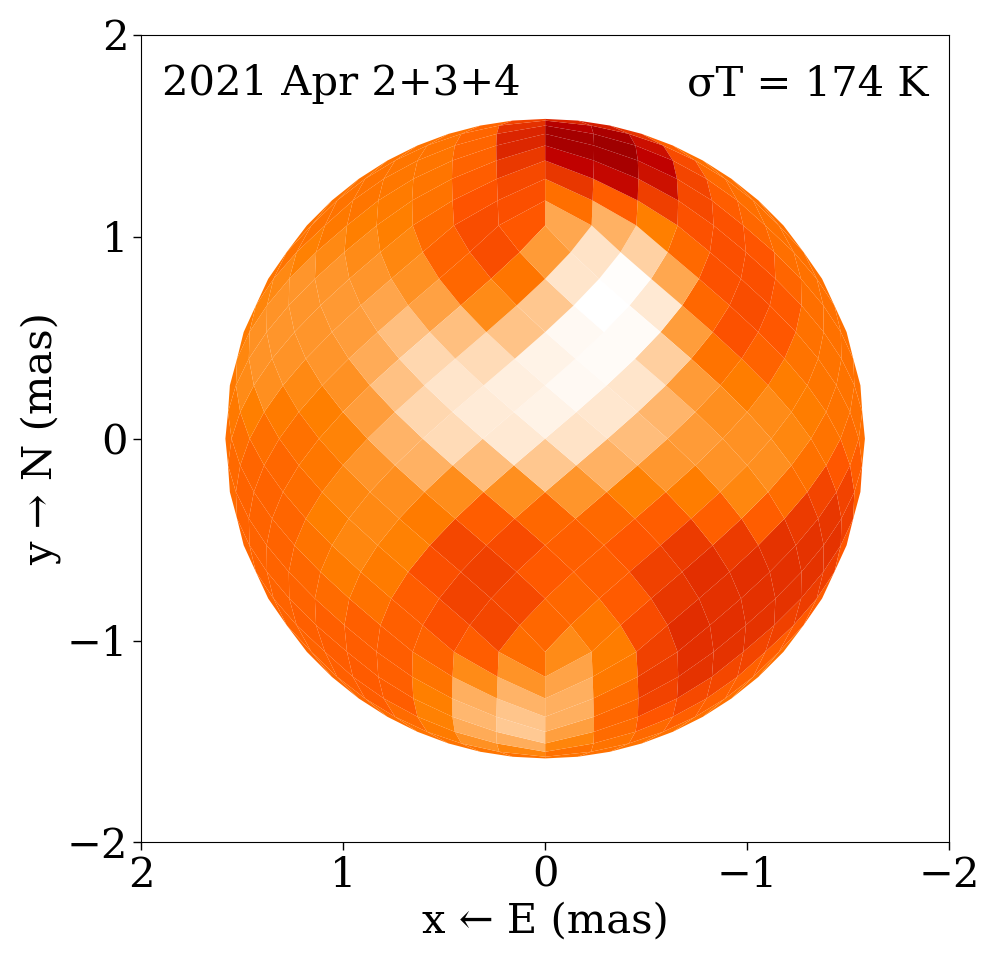}}   
  \end{center}
  \caption{Surface images of Polaris Aa reconstructed using SURFING (left) and ROTIR (right) on the merged data set from UT 2021Apr02, 2021Apr03, and 2021Apr04.}
  \label{figure_bestim}
\end{figure*}
\begin{figure*}[ht]
  \begin{center}
  \scalebox{0.2}{\includegraphics{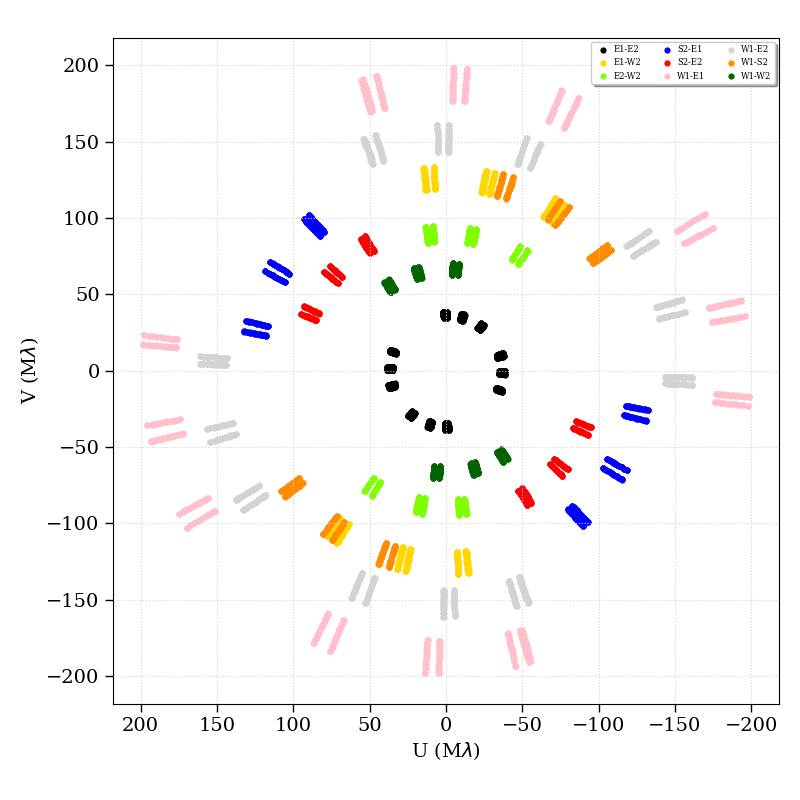}}
  \scalebox{0.2}{\includegraphics{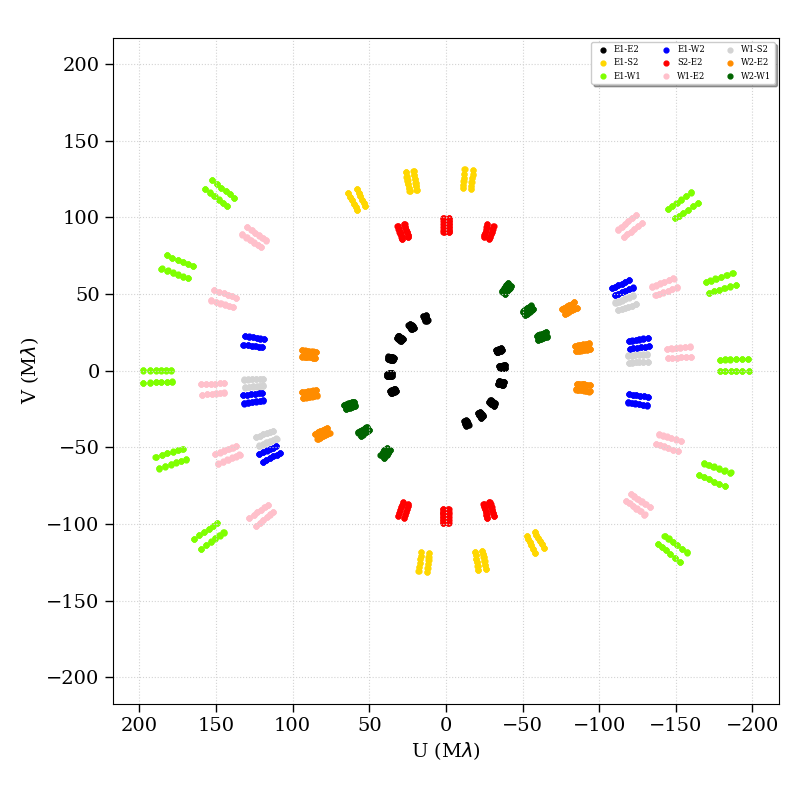}}
  \scalebox{0.2}{\includegraphics{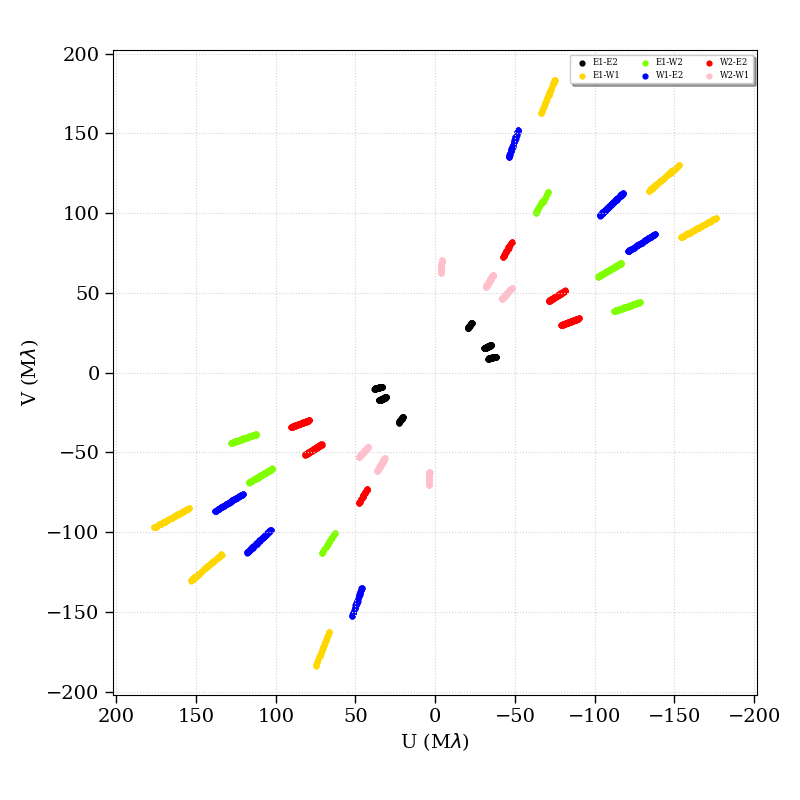}}
  \scalebox{0.2}{\includegraphics{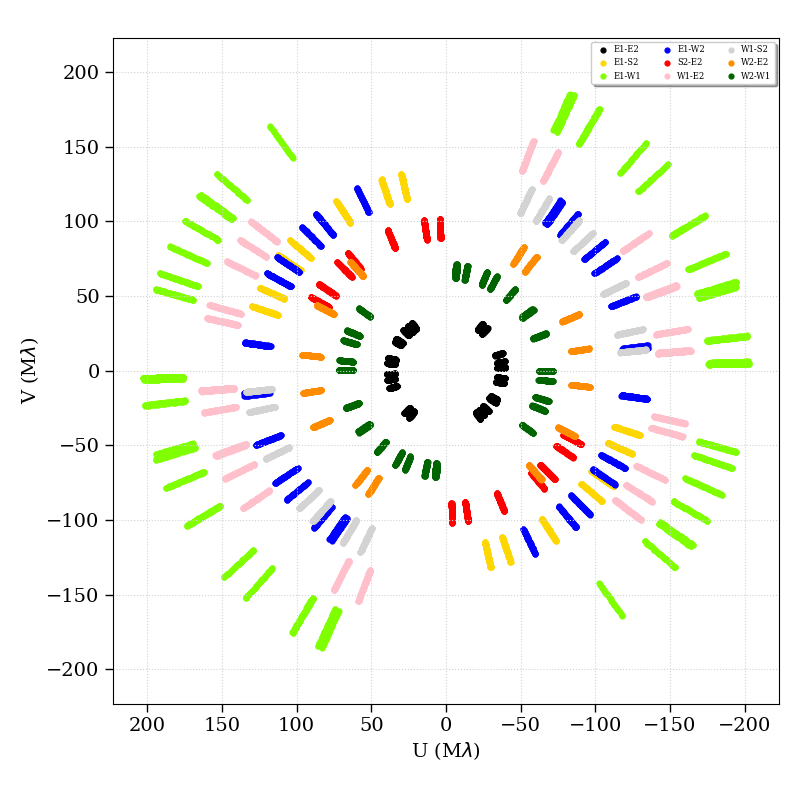}} \\
  \scalebox{0.22}{\includegraphics{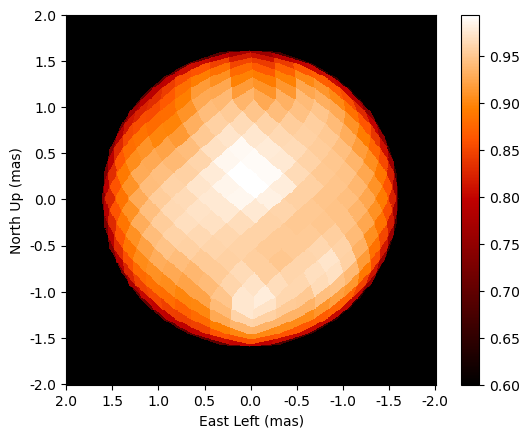}}
  \scalebox{0.22}{\includegraphics{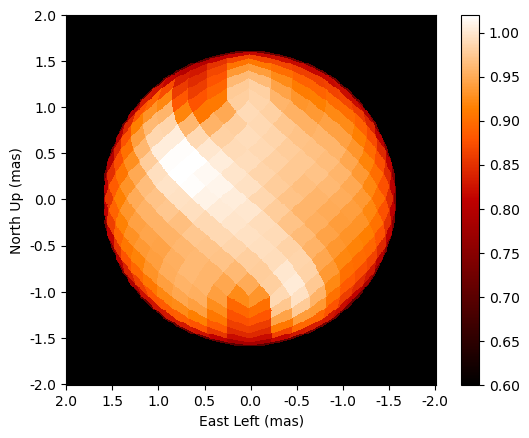}}
  \scalebox{0.22}{\includegraphics{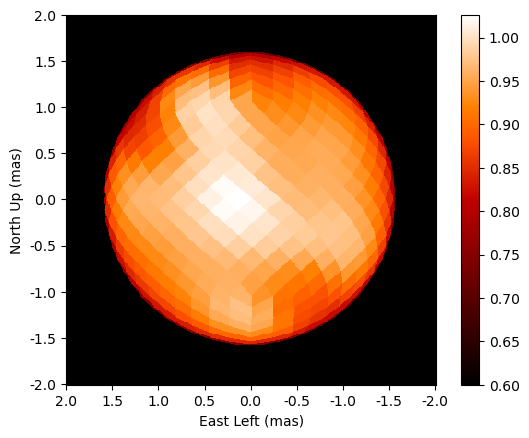}}
  \scalebox{0.22}{\includegraphics{SURFING_Polaris_2021Apr_20.png}} \\
  \scalebox{0.158}{\includegraphics{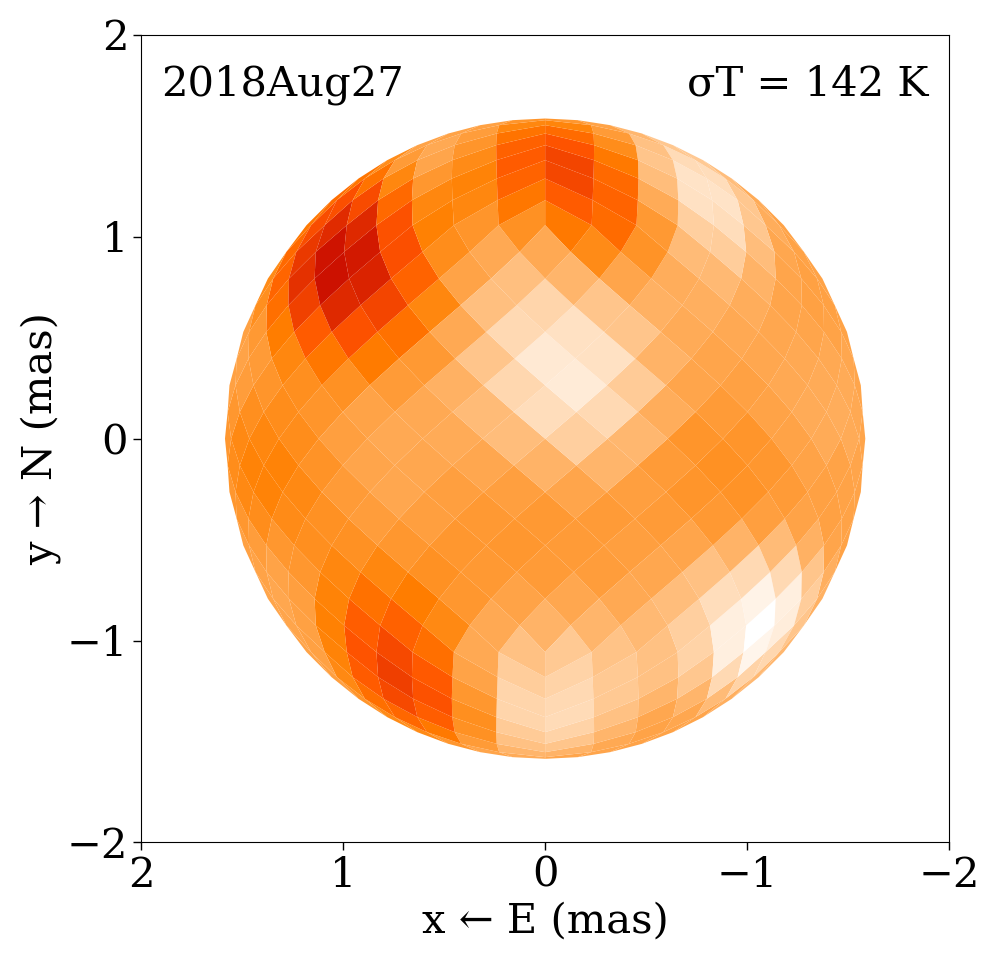}}
  \scalebox{0.158}{\includegraphics{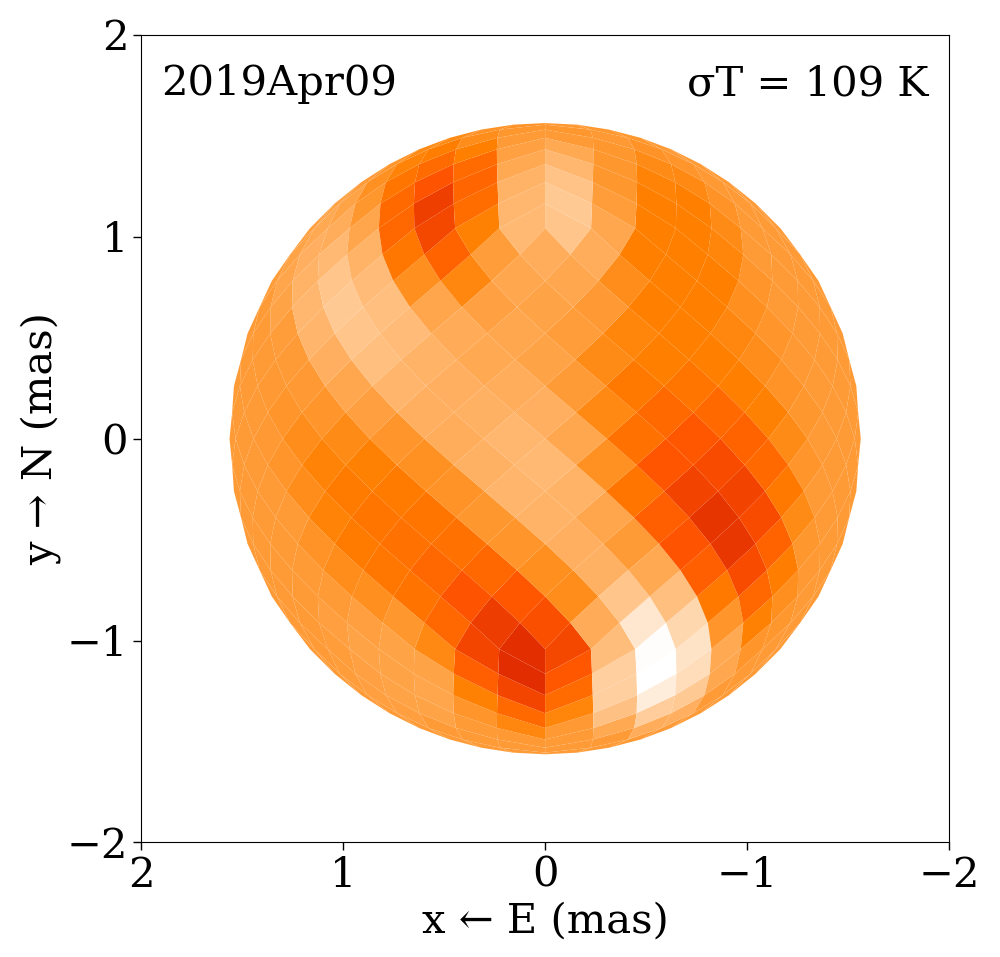}}
  \scalebox{0.158}{\includegraphics{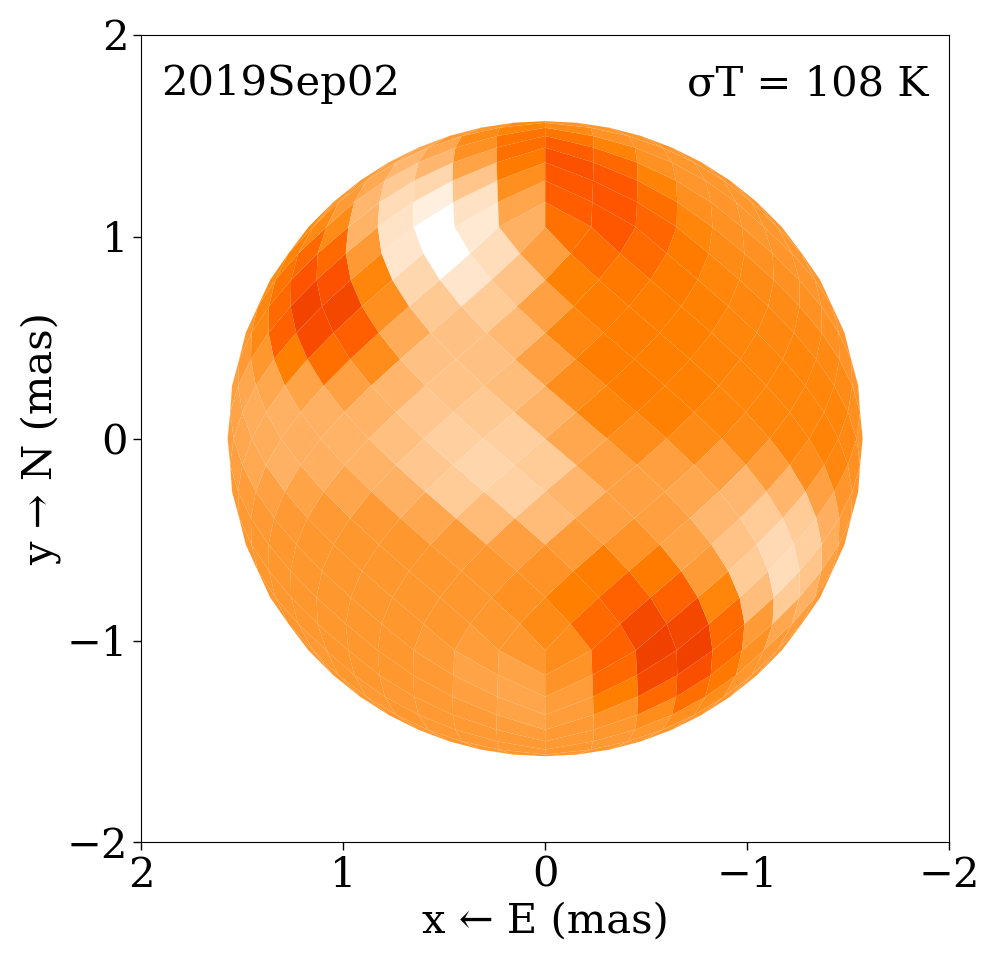}}
  \scalebox{0.158}{\includegraphics{rotir_polaris_2021Apr02_03_04_sigmaT.png}} \\
  \scalebox{0.158}{\includegraphics{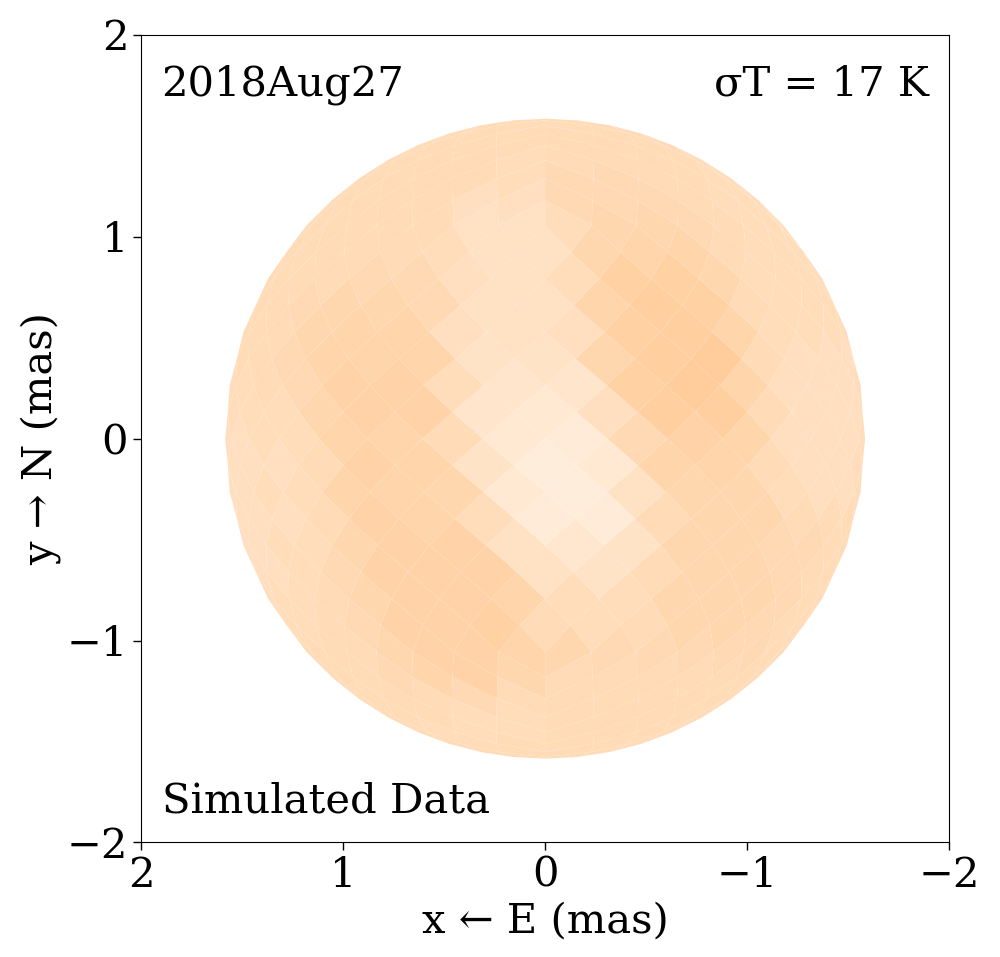}}
  \scalebox{0.158}{\includegraphics{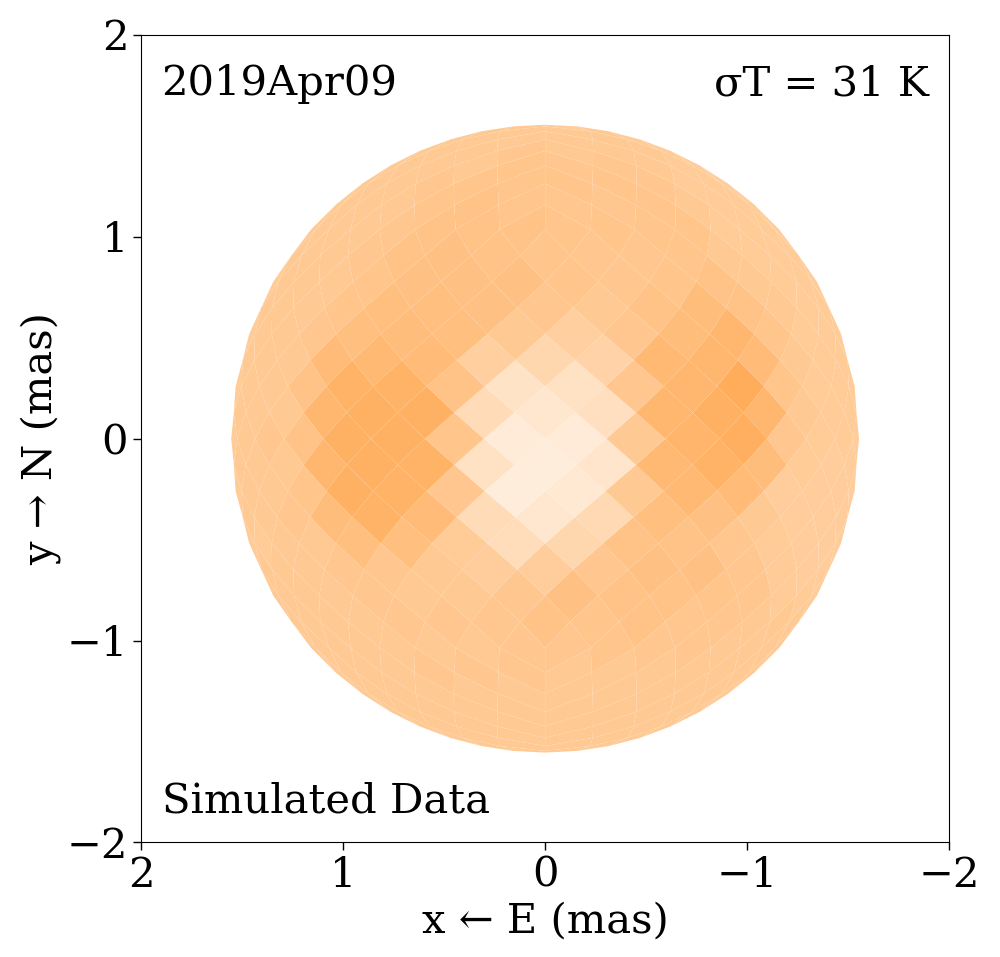}}
  \scalebox{0.158}{\includegraphics{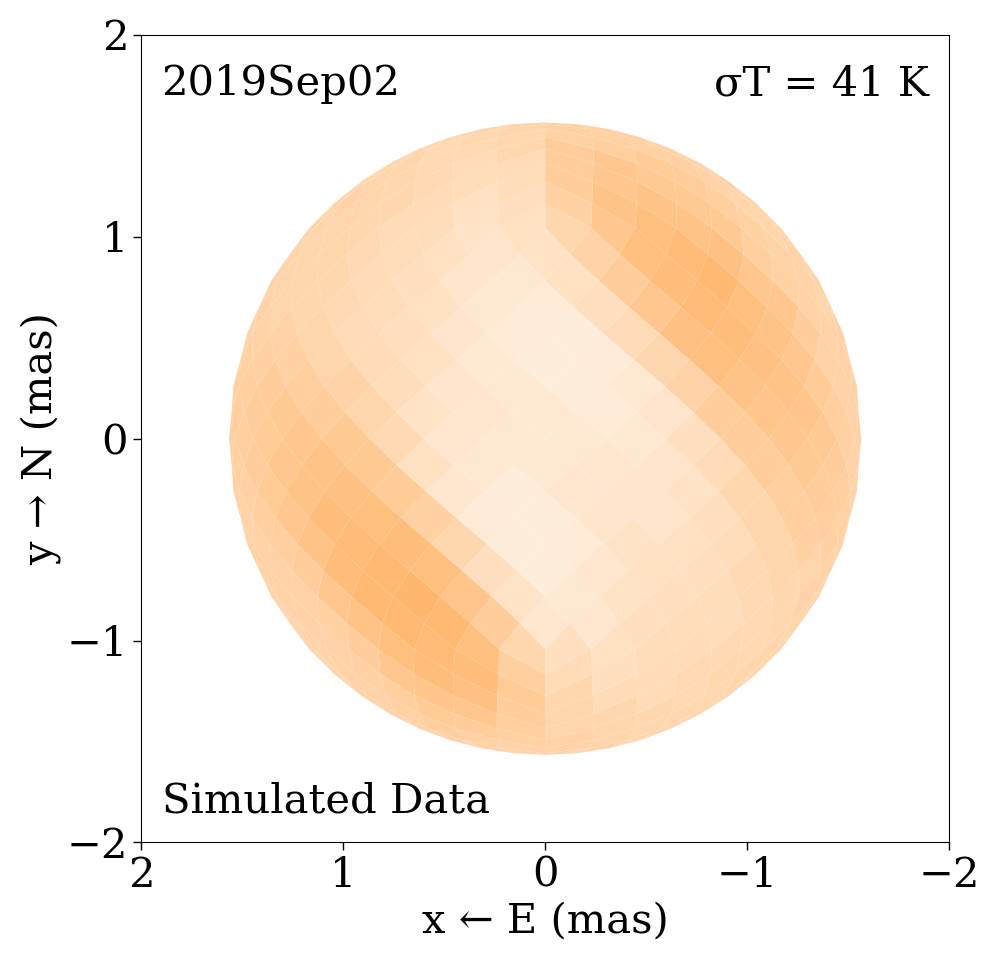}}
  \scalebox{0.158}{\includegraphics{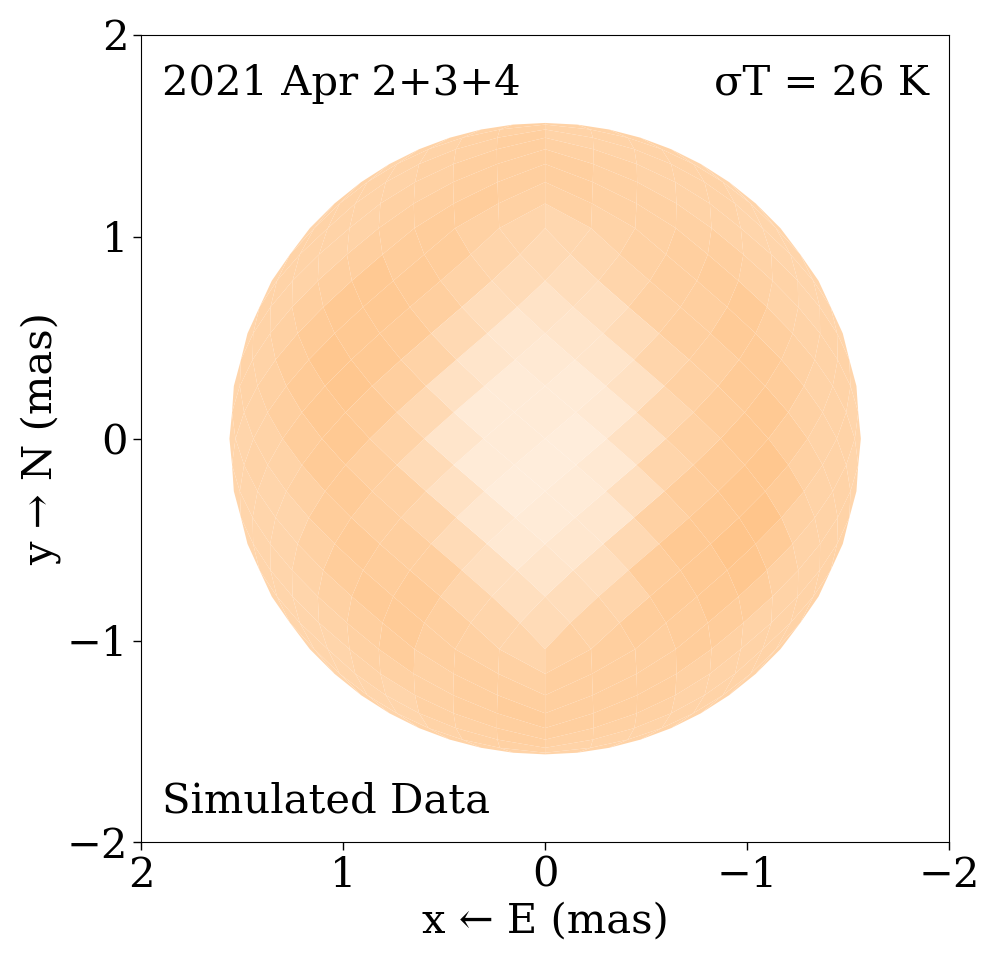}} \\
  \end{center}
  \caption{Surface images of Polaris Aa reconstructed from the CHARA data on four different nights. The top row shows the ($u, v$) coverage, the second row shows the Polaris images reconstructed using SURFING, the third row shows images reconstructed using ROTIR, and the bottom row shows images reconstructed from simulated data of a limb-darkened disk model without spots and with the same ($u, v$) sampling as the Polaris data. The simulated data highlights artifacts in the reconstruction from gaps in the ($u, v$) coverage. For the ROTIR images, the standard deviation in the surface temperature is listed in the upper right of each panel.  }
  \label{figure_images}
\end{figure*}

    Figure~\ref{figure_bestim} shows the image of Polaris Aa reconstructed with SURFING and ROTIR for the epoch with the most complete ($u,v$) coverage from UT 2021 Apr 2, 3, and 4. Both images show a large bright spot to the north of center.
 We show the SURFING and ROTIR images from four different nights and the corresponding ($u,v$) coverage in Figure~\ref{figure_images}. All of the images show spots on the surface, however, some of the surface features could be created by artifacts in the reconstruction process due to gaps in the ($u,v$) coverage or systematic errors in the visibility calibration.
The bottom row in  Figure~\ref{figure_images} shows reconstructed images from simulated data of a simple limb-darkened model without spots and with the same ($u, v$) sampling of Polaris to show which features are likely to be artifacts and which are likely to be spots on the surface
The contrast of the surface spots in the real data is larger than the contrast in the simulated limb-darkened disk data.
In addition to having much lower contrast, the spots in the simulated data are distributed in point symmetric patterns around the center of the star. The spots in the real data that do not follow these point-symmetric patterns are more likely to be true features on the stellar surface. Changing the simulated noise in the limb-darkened models changes slightly the shape and orientation of the artifacts in the simulated images, however the large-scale structure of the patterns remain consistent between iterations with different simulated noise.
The mirroring of bright and dark spots about the origin during some of the Polaris epochs is likely an artifact of the reconstruction process, however these features are indicative of an asymmetry in brightness distribution.
The interferometric data and observables extracted from the reconstructed images are shown in Appendix~\ref{section_appendix_data}.

\subsection{Binary Positions of Polaris Aa,Ab}

We used the data sets with 30-second integration times to search for the faint companion (Polaris Ab). The shorter integration time reduces time-smearing of the companion signal given the separations of 40--85 mas expected from the previous orbit fit. To account for the asymmetries on the surface of Polaris Aa, we first reconstructed images with SQUEEZE using all of the data from a given night. We then used a modified version of the IDL Binary Grid Search Procedure\footnote{\url{http://www.chara.gsu.edu/analysis-software/binary-grid-search}} \citep{schaefer16} where we extract visibilities and closure phases directly from the reconstructed image of Polaris Aa and analytically add in the companion. We started the grid search at the expected location of the binary based on the orbit computed by \citet{evans18} and searched within  $\Delta$RA = $\pm$20 mas and $\Delta$DEC = $\pm$20 mas from the  initial values.  We applied a correction for
bandwidth smearing to the binary fit.  The results from the binary grid search are shown in supplementary files.

We detected the Ab companion on five nights; these nights are indicated by an asterisk next to the UT date in Table~\ref{table_log}. On the nights with non-detections, the data quality was poor (e.g., 2016-11-18) or Polaris Ab was located outside of the interferometric field of view given by $\lambda^2 / \Delta\lambda$. The field of view corresponds to 50 mas in the low spectral resolution mode of MIRC and MIRC-X, so with an expected separation of $\sim$70 mas, we suspect that bandwidth smearing compromised the binary signal on UT 2018Aug27 and 2019Apr09. We were able to recover the detection of the companion on subsequent dates when we used the higher spectral resolution Prism102 and Grism190 modes with a field of view of 100 and 190 mas, respectively.

The average contrast of the companion was $\Delta H = 7.8 \pm 0.4$ mag, beyond the typical detection limits of the MIRC and MIRC-X instruments \citep{gallenne15}. To improve our confidence in the solutions, we also performed the binary grid search separately for each configuration observed on a given night. For the nights with high confidence solutions, the separations measured from each configuration separately were consistent with the results based on the full night of data. Additionally, the three successive nights in April 2021 produced consistent binary solutions. 

We used a bootstrap approach with 100 iterations to estimate the uncertainties based on the 67.5\% confidence ellipses for two parameters from the bootstrap distributions. These errors are added in quadrature with the uncertainties from the wavelength calibration. The best fit binary positions (corrected for wavelength calibration) and flux ratios ($f_{\rm Ab}/f_{\rm Aa}$) are listed in Table~\ref{table_astrometry_results}. The positions of Polaris Ab relative to Aa are given as the separation ($\rho$) and position angle measured east of north (PA). We also list the separations projected into right ascension ($\Delta\alpha$) and declination ($\Delta\delta$). The error ellipses are given by the semi-major axis ($\sigma_{\rm maj}$), semi-minor axis ($\sigma_{\rm min}$), and the position angle of the major axis ($\sigma_{\rm PA}$). The dates are given as heliocentric Julian day (HJD) and Julian year (JY). In Table~\ref{table_astrometry_results}, we also collected the {\it HST} astrometric measurements reported by \citet{evans08,evans18} in both polar and Cartesian coordinates. These papers reported the dates in Besselian years which we have converted to JY.

\subsubsection{Confirmation of binary positions with CANDID}

We confirmed the companion detections using the \emph{CANDID} tool\footnote{Available at \url{https://github.com/amerand/CANDID} or \url{https://github.com/agallenne/GUIcandid}} \citep[Companion Analysis and Non-Detection in Interferometric Data,][]{gallenne15}. \emph{CANDID} is a set of Python tools allowing us to search systematically for companions and estimate detection limits using all interferometric observables. We used the implemented bootstrap function to estimate the uncertainties on our previous parameters. In \emph{CANDID}, limb-darkening is parameterized with a power law function $I_\lambda (\mu) = I_\lambda (1) \mu^\epsilon $, and we chose to fix this parameter to 0.15. Bootstrapping is performed on the MJDs with our previous fitted parameters and using 1000 bootstrap samples. From the distribution, we took the median value and the maximum value between the 16th and 84th percentiles as uncertainty for the flux ratio and angular diameter. For the fitted astrometric position, the error ellipse is derived from the bootstrap sample (using a principal components analysis). The positions and uncertainties measured with CANDID were consistent with those determined from the IDL grid search procedure.


\begin{deluxetable*}{rrrrrrrrrr}
  \tablecaption{Relative astrometric position of the Polaris Ab companion.\label{table_astrometry_results}.}
  \tablewidth{0pt}
  \tablehead{
    \colhead{HJD $-$} &  \colhead{JY} & \colhead{$\rho$} & \colhead{PA} & \colhead{$\Delta \alpha$} & \colhead{$\Delta \delta$} & \colhead{$\sigma_{\rm maj}$} & \colhead{$\sigma_{\rm min}$} & \colhead{$\sigma_{\rm PA}$} & \colhead{$f_{\rm Ab}/f_{\rm Aa}$} \\
     \colhead{2,400,000} &  \colhead{ } & \colhead{(mas)} & \colhead{($^\circ$)} & \colhead{(mas)} & \colhead{(mas)} & \colhead{(mas)} & \colhead{(mas)} & \colhead{($^\circ$)} & \colhead{ } }
 \startdata
 \multicolumn{10}{c}{CHARA}\\
57643.735 & 2016.6974  &  33.008  &  71.696  &  31.338  &  10.366  &  5.757  &  1.159  &  18.49  & 0.00089 $\pm$ 0.05838 \\
58728.831 & 2019.6683  &  75.537  & 340.239  & -25.538  &  71.088  &  0.833  &  0.547  &  40.73  & 0.00072 $\pm$ 0.00744 \\
59306.890 & 2021.2509  &  84.507  & 315.900  & -58.810  &  60.686  &  1.200  &  0.663  &  94.49  & 0.00110 $\pm$ 0.00789 \\
59307.982 & 2021.2539  &  84.447  & 315.787  & -58.888  &  60.527  & 10.736  &  3.954  & 123.45  & 0.00070 $\pm$ 0.02214 \\
59308.785 & 2021.2561  &  84.232  & 315.669  & -58.861  &  60.252  &  0.880  &  0.804  & 135.12  & 0.00045 $\pm$ 0.00363 \\
\multicolumn{10}{c}{{\it Speckle}}\\
60237.913 & 2023.800   & 98.8     & 286.7    &  -94.6   &   28.4   &   5.3   &    4.2   &   0.0 &  \\
\multicolumn{10}{c}{{\it HST}}\\
53585.488 & 2005.5866  &  172.4  &  231.407  & -134.75  & -107.54  &   2.1   &    2.1   &   0.0 &  \\
53961.416 & 2006.6158  &  169.6  &  226.385  & -122.79  & -116.99  &   3.1   &    3.1   &   0.0 &  \\
54298.514 & 2007.5387  &  180.0  &  223.000  & -122.76  & -131.64  &  20.0   &   20.0   &   0.0 &  \\
55153.597 & 2009.8798  &  150.0  &  216.000  &  -88.17  & -121.35  &  20.0   &   20.0   &   0.0 &  \\
56835.302 & 2014.4841  &   85.0  &  175.000  &    7.41  &  -84.68  &  20.0   &   20.0   &   0.0 &  \\
\enddata

 \tablecomments{Positions in Table 3 are for J2000.}

\end{deluxetable*}

\section{Speckle  Observations of Polaris} \label{speck}

Polaris was observed using the technique of speckle interferometry at the
Astrophysical Research Consortium (ARC) 3.5-m telescope at Apache Point
Observatory (APO) on 2023 Oct 19 and 20. The speckle camera used was the
Differential Speckle Survey Instrument (DSSI, Horch et al. 2009), which
is currently a visitor instrument at APO. DSSI takes speckle patterns
in two filters simultaneously; for the Polaris observations, we used
filters with center wavelengths of 692 nm and 880 nm, and with FWHMs
of 40 and 50 nm respectively. On each night, a total of nine thousand
short-exposure images were recorded in each filter and stored in nine
FITS files, each of which consisted of a 1000-frame stack of 256x256-pixel
frames. The exposure time of individual frames was 40 ms, and the seeing
was approximately 0.9 arcseconds on both nights.

The reduction and calibration of DSSI speckle data at APO is described in
Davidson et al. (2024); briefly, we bias-subtract and form the spatial
frequency power spectrum of individual frames, co-adding these in the
Fourier domain to arrive at a summed power spectrum for each data file.
We then deconvolve by dividing by the power spectrum of a bright point
source observed in the same way. The result for a binary star is a pure
fringe pattern, which we then fit using a downhill simplex algorithm as
described in Horch et al (1997). The fringe spacing, orientation, and
depth of the fringe minima determine the separation, position angle,
and magnitude difference of the pair. In this analysis, we do not
determine the quadrant of the secondary star; there is a 180-degree
ambiguity in the position angle due to the fact that the power spectrum
has no phase information. For the 2023 October speckle run at APO, the
pixel scale and orientation were obtained using a combination of several
binaries with extremely well-determined orbits and slit mask observations,
as described in Davidson et al. (2024). These methods resulted in a pixel
scale with uncertainty of approximately 0.24\% and offset angles between
pixel axes and celestial coordinates that are uncertain at the 0.1-degree level.

Polaris presents a complication in the speckle analysis because it is a system
with a large magnitude difference that is observed at airmass of near 2.0 from
Apache Point. The airmass adds dispersion to the speckle patterns, stretching
speckles out on the image plane in a direction leading toward the zenith. This
then changes the shape of the power spectrum and could affect the fringe fitting
results. To have a good estimate of the speckle transfer function for a particular
DSSI observation, a bright unresolved star is typically observed near in time and
close in sky position to the star of interest. In the case of Polaris, since it
was not possible to have as close a match in sky position as we typically use,
we instead observed a bright unresolved star at low airmass, and then applied
a dispersion model to that observation so that it matched Polaris’ zenith angle
and azimuth exactly. This removed most of the effect of the dispersion, but a
further correction was applied by allowing for additional free parameters in
the fringe fit to account for any residual dispersion in the Fourier domain.
In this way, it was possible to determine relatively consistent relative
astrometry and photometry of the system for the sub-sample of the files
analyzed that had the lowest reduced-chi squared values from the fringe fitting procedure.

The measurements from the best sequences in the two nights of data resulted in
the following average relative position: $\Delta\alpha = -0\farcs0902 \pm 0\farcs0053$
and $\Delta\delta = +0\farcs0404 \pm 0\farcs0042$, for the equinox of the date of
observation (Julian year 2023.7999). These correspond to polar coordinates
$\theta = 294\fdg1 \pm 3\fdg7$ and $\rho = 0\farcs0988 \pm 0\farcs0052$.
Because of the dispersion complication noted above, we were not able to
derive a reliable measure of the magnitude difference between Polaris and its companion.

Given that the rest of the astrometric measurements in this work are
effectively referred to the equinox J2000, we applied rigorous precession
to our result above in order to reduce it to that same frame of reference.
The position angle correction is $-7\fdg4$. We obtained
$\Delta\alpha = -0\farcs0946 \pm 0\farcs0053$ and
$\Delta\delta = +0\farcs0284 \pm 0\farcs0042$ (J2000).






\section{Orbit Fitting} \label{section_orbit}

\subsection{Radial Velocities} \label{vr}

The orbital analysis depends on the combination of two datasets, the radial
velocities and relative astrometry. The radial velocity dataset is very large and
made up of measurements from many sources to cover multiple cycles of
the 30 year orbit of Polaris Aa-Ab. We used the complete set of 3,659
radial velocities with pulsation removed that are tabulated in Table 3
of \citet{torres23}\footnote{An error in the identification of some of the
  datasets in the online version of Table 3 of Torres (2023) was identified and corrected here}.
To those we added an additional 16 Hermes velocities
from the program VELOCE I (Anderson, et al 2024 in press). The
data had been prepared using the template approach detailed in
Anderson (2019).
Initially  we applied the radial velocity jitter and offset
terms in Table 5 of \citet{torres23} and the additional Hermes data.
(Jitter is as defined in Torres [2023], the uncertainty for each
dataset from the orbit fitting.)
After experimentation and preliminary
fits, the offsets and jitter were redetermined as listed in Table~\ref{vr_corr}.  The
sources and references are as listed in  Table 5 of \citet{torres23}. Sucessive
pairs of columns to the right provide the radial velocity offset and jitter for
the original orbit in  \citet{torres23}, and the solution for radial velocities and
CHARA and {\it HST} astrometry here.

\begin{deluxetable*}{lcccc}
		\tablecaption{\label{vr_corr} Corrections to Radial Velocity Sources}
		\tablewidth{0pt}
		\tablehead{
              \colhead{Source}        & \twocolhead{Torres 2023} &  \twocolhead{RV + Astrom}  \\
                  \colhead{}  \\
		  \colhead{} & \colhead{Offset} & \colhead{Jitter} & \colhead{Offset} &
                  \colhead{Jitter}  \\
		  \colhead{} & \colhead{km s$^{-1}$ } & \colhead{km s$^{-1}$} & \colhead{km s$^{-1}$ }
                  & \colhead{km s$^{-1}$ }  \\
                }
		\startdata
\hline
 1 Roemer 1965                         &      -   &  0.681    &      -  &   0.680   \\
 2 Hartmann 1901                       &   +1.39  &  0.51     &    +1.43  &  0.52    \\
 3 Kustner 1908                        &   +1.05 &   0.75     &    +0.94  &  0.83   \\
 4 Arellano Ferro 1983a + Kamper 1996  &   +1.027 &  0.524    &   +1.001  &  0.580      \\
 5 Dinshaw et al. 1989                 &  -11.829 &  0.469    &    -11.694 &  0.450  \\
 6 Gorynya et al. 1992                 &   +1.488 &  0.500    &   +1.375  &  0.498    \\
7 Hatzes \& Cochran 2000                &  -14.117 &  0.027    &  -14.242 &  0.027  \\
 8 Usenko et al. 2015                  &   +0.068 &  0.387    &    +0.100 &  0.409    \\
 9 Kamper \& Fernie 1998                &   -0.222 &  0.1061   &  -0.323  &  0.1065   \\
10 Eaton 2020                          &   +1.120 &  0.0444   &    +1.156 &  0.0398  \\
11 Lee et al. 2008                     &  -17.794 &  0.1103   &  -17.772 &  0.1124   \\
12 Bucke 2021                          &  -18.387 &  0.0976   &   -18.387 &  0.150    \\
13 Anderson 2019                       &   +0.551 &  0.0977   &     +0.622 &  0.1125  \\
14 Anderson 2024                       &    -     &   -       &   +0.440 &  0.103  \\
\enddata
	\end{deluxetable*}

\subsection{Orbit} \label{orb}

 The astrometry dataset has only 11 two-dimensional data points that cover only 16 years, or half the orbit. 

 Table~\ref{table_results} lists the orbital parameters from the spectroscopic radial velocity orbit computed by \citet{torres23}
 and Anderson et al. (2024) and a fit to only the visual orbit computed through a Newton Raphson technique using the IDL orbit fitting library\footnote{\url{http://www.chara.gsu.edu/analysis-software/orbfit-lib}} described by \citet{schaefer16}. The orbital parameters include the orbital period $P_\mathrm{orb}$, time of periastron passage $T_\mathrm{p}$, eccentricity $e$, argument of periastron with reference to the primary  $\omega$, primary radial velocity amplitude  $K_1$ , systemic velocity $\gamma$, longitude of ascending node $\Omega$
 for J2000, inclination $i$, and angular semi-major axis $a$. The eccentricity and $\omega_\mathrm{Aa}$ differ by 1.4\,$\sigma$ and 2.3\,$\sigma$, respectively, between the spectroscopic and visual orbits.

Our methods for fitting the radial velocities and astrometry simultaneously are described
in the subsections below. It was realized early on that the analysis approach is particularly
important since the orbit has a substantial eccentricity.  In a preliminary
 solution weighted solely by the errors on the datapoints in the two datasets,
the {\it HST} astrometry was systematically offset from the orbit.
The eccentricity and $\omega_\mathrm{Aa}$ are set essentially by the large number of
radial velocities.
We explored several approaches to combine the different datasets. The various 
weighting methods are described in more detail below with the orbital parameters
from each solution listed in Table~\ref{table_results}.






\subsection{Weights based on Measurement Errors} \label{section_error_weight}

To determine the orbital parameters, we simultaneously fitted the radial velocities (RVs) and astrometric position using a Markov chain Monte Carlo (MCMC) routine\footnote{With the Python package \emph{emcee} developed by \citet{foreman-mackey13}.}, whose  log-likelihood function is given as \citep{gallenne19a}
	\begin{displaymath}
	\log(\mathcal{L}) = - \dfrac{1}{2}\,\chi^2,\, \mathrm{with}\,\chi^2 = \chi^2_\mathrm{RV} + \chi^2_\mathrm{ast}.
	\end{displaymath}
	
	In our previous works on binary Cepheid systems with astrometric data \citep[see e.g.][]{gallenne19b,gallenne18}, both the pulsation and orbital motions of the Cepheids were fitted together with the astrometry. In this work, we decided to ignore the pulsation motion of the Cepheid by directly fitting the pulsation-corrected RVs from \citet{torres23}. In this case, our RV model is simply defined as
	\begin{displaymath}
	\chi^2_\mathrm{RV} =  \sum \dfrac{(V_1 - V_\mathrm{1m})^2}{\sigma_\mathrm{V_1}^2}, 
	\end{displaymath}
	in which $V_1$ and $\sigma_1$ denote the measured radial velocity and uncertainties for Polaris, and $V_\mathrm{1m}$ is the Keplerian velocity model characterized by Heintz (1978):
	\begin{eqnarray}
		V_\mathrm{1m} &=& \gamma + K_1\,[\cos(\omega + \nu) + e\cos{\omega}], \nonumber \\
		\tan \dfrac{\nu}{2} &= & \sqrt{\dfrac{1 + e}{1 - e}} \tan \dfrac{E}{2}, \nonumber \\
		E - e \sin E &=& \dfrac{2\pi (t - T_\mathrm{p})}{P_\mathrm{orb}}, \nonumber
	\end{eqnarray}
	where $\gamma$ is the systemic velocity, $e$ the eccentricity, $\omega$ the argument of periastron, $\nu$ the true anomaly,  $E$ the eccentric anomaly, $t$ the observing date, $P_\mathrm{orb}$ the orbital period, $T_\mathrm{p}$ the time of periastron passage, and $K_1$ the primary radial velocity amplitude. 
	
		$\chi^2_\mathrm{ast}$ measure for astrometry depends on the astrometric measurements  as
	\begin{eqnarray*}
		&\chi^2_\mathrm{ast} =& \chi^2_\mathrm{a} + \chi^2_\mathrm{b}, \\
		&\chi^2_\mathrm{a} =& \sum \frac{[ (\Delta \alpha - \Delta \alpha_\mathrm{m}) \sin \sigma_\mathrm{PA} + (\Delta \delta - \Delta \delta_\mathrm{m}) \cos \sigma_\mathrm{PA} ]^2}{\sigma^2_\mathrm{maj}}, \\
		&\chi^2_\mathrm{b} =& \sum \frac{[ -(\Delta \alpha - \Delta \alpha_\mathrm{m}) \cos \sigma_\mathrm{PA} + (\Delta \delta - \Delta \delta_\mathrm{m}) \sin \sigma_\mathrm{PA} ]^2}{\sigma^2_\mathrm{min}}, \\
	\end{eqnarray*}
	in which $(\Delta \alpha, \Delta \delta, \sigma_\mathrm{PA}, \sigma_\mathrm{maj}, \sigma_\mathrm{min})$ denote the relative astrometric measurements with the corresponding error ellipses, and $(\Delta \alpha_\mathrm{m}, \Delta \delta_\mathrm{m})$
       represent the astrometric model defined as:
	\begin{eqnarray*}
		\Delta \alpha_\mathrm{m} &=& r \,[ \sin \Omega \cos(\omega + \nu) + \cos i \cos \Omega \sin(\omega + \nu) ], \\
		\Delta \delta_\mathrm{m} &=& r \,[ \cos \Omega \cos(\omega + \nu) - \cos i \sin \Omega \sin(\omega + \nu) ], \\
		r &=& \dfrac{a (1 - e^2)}{1 + e\cos \nu},\\
	\end{eqnarray*}
	where $\Omega$ is the longitude of ascending node, $i$ the orbital inclination, and $a$ the angular semi-major axis. 
	
	As a starting point for our 100 MCMC walkers, we performed a least squares fit using orbital values from \citet{torres23} as first guesses. We then ran 100 initialization steps to well explore the parameter space and get settled into a stationary distribution. For all cases, the chain converged before 50 steps. Finally, we used the last position of the walkers to generate our full production run of 1000 steps, discarding the initial 50 steps. All the orbital elements are estimated from the distribution taking the median value and the maximum value between the 16th and 84th percentiles as uncertainty (although the distributions were roughly symmetrical).
	
	Polaris is a single-lined spectroscopic binary which does not allow the individual masses to be determined
        independently without additional information.
       To this end 
        we adopt the {\it Gaia} distance for Polaris B (above). In our MCMC procedure, we included the parallax measurement using a normal distribution centered on 7.3045\,mas with a standard deviation of 0.0178\,mas.
        Masses are then derived with the following equations:
	\begin{eqnarray*}
		M_\mathrm{T} &=& \dfrac{a^3}{P_\mathrm{orb}^2\,  \varpi^3} , \\
                M_2 &=& \dfrac{0.03357\, a^2\, K_1\, \sqrt{1 - e^2}} {\varpi^2\, P_\mathrm{orb} \sin{i}} \\
		M_1 &=& M_\mathrm{T} - M_2 \\
	\end{eqnarray*}
	with $M_\mathrm{T} = M_1 + M_2$ the total mass in $M_\odot$, $a$ and $\varpi$ in mas, $P_\mathrm{orb}$ in years, and $K_1$ in km\,s$^{-1}$.

        Initially, when fitting the orbit simultaneously to the radial velocities and astrometry, we weighted each measurement by their respective uncertainties.
        We performed a similar fit weighted by measurement uncertainties using the IDL orbit 
fitting library, but with uncertainties computed through a bootstrap technique.        
The orbital parameters from this fit are listed in the ``$\sigma$ Weights'' Column of Table~\ref{table_results}.
	

\subsection{Weights based on $N \sigma$} \label{section_Nweight}

To more evenly weight the two datasets, we can compute an average $\chi^2$ statistic multiplying the $\chi^2$ from each dataset by 1/N, where N is the number of data points \citep[][equation 5]{merand15}. The total $\chi^2$ for the simultaneous fit then becomes:
	\begin{displaymath}
	\chi^2 = \dfrac{1}{N_\mathrm{RV}}\,\chi^2_\mathrm{RV} + \dfrac{1}{N_\mathrm{ast}}\,\chi^2_\mathrm{ast}.
	\end{displaymath}
        where $N_\mathrm{RV}$ is the number of radial velocity measurements and $N_\mathrm{ast}$ is the number of astrometry data points. The orbital parameters are listed in the ``$N \sigma$ Weight'' Column of Table~\ref{table_results}. Uncertainties in the orbital parameters were determined from MCMC distributions.  


\subsection{Replicating the Astrometric Measurements} \label{section_replicate}

\begin{figure*}
  \centering
  \plottwo{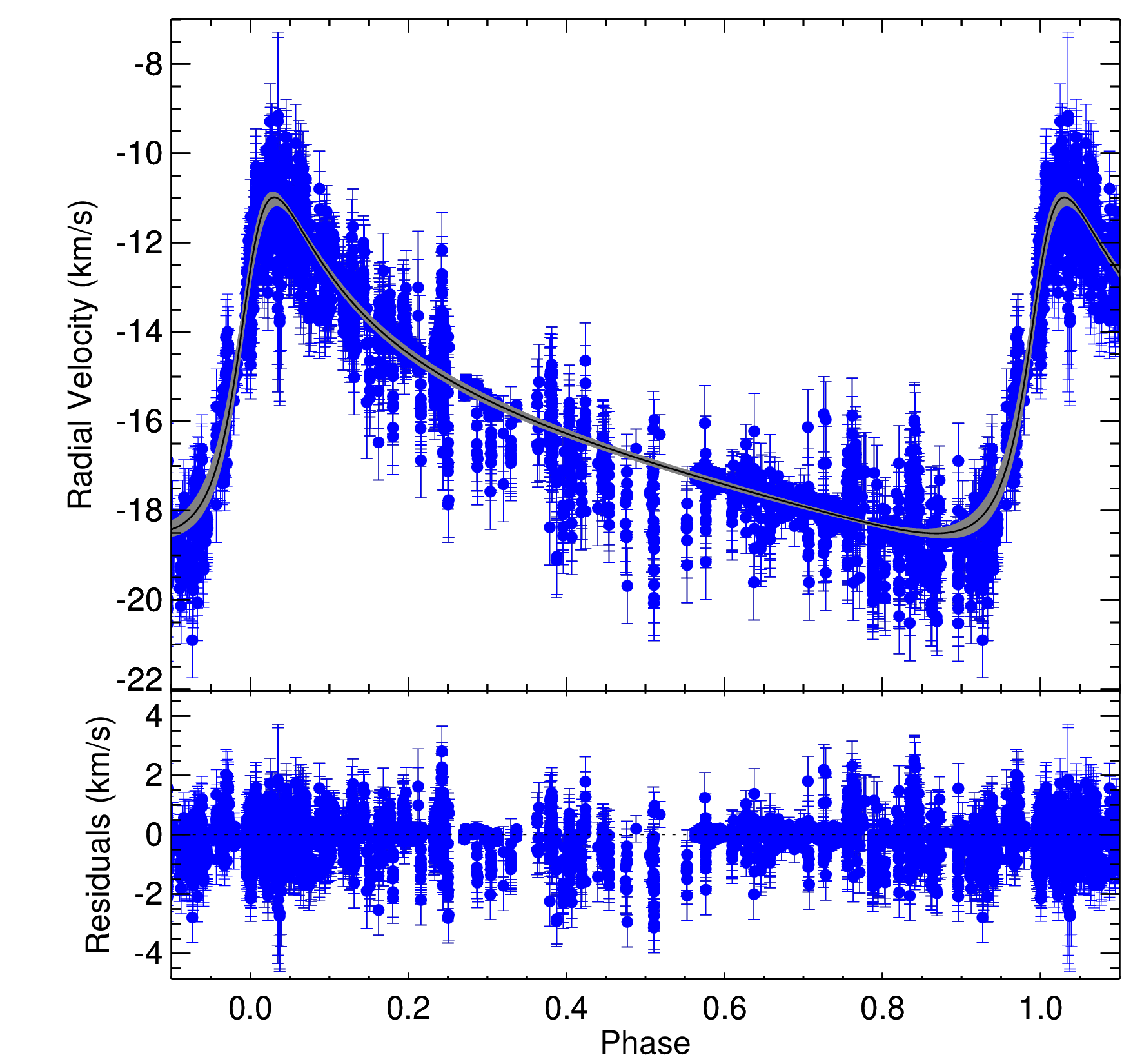}{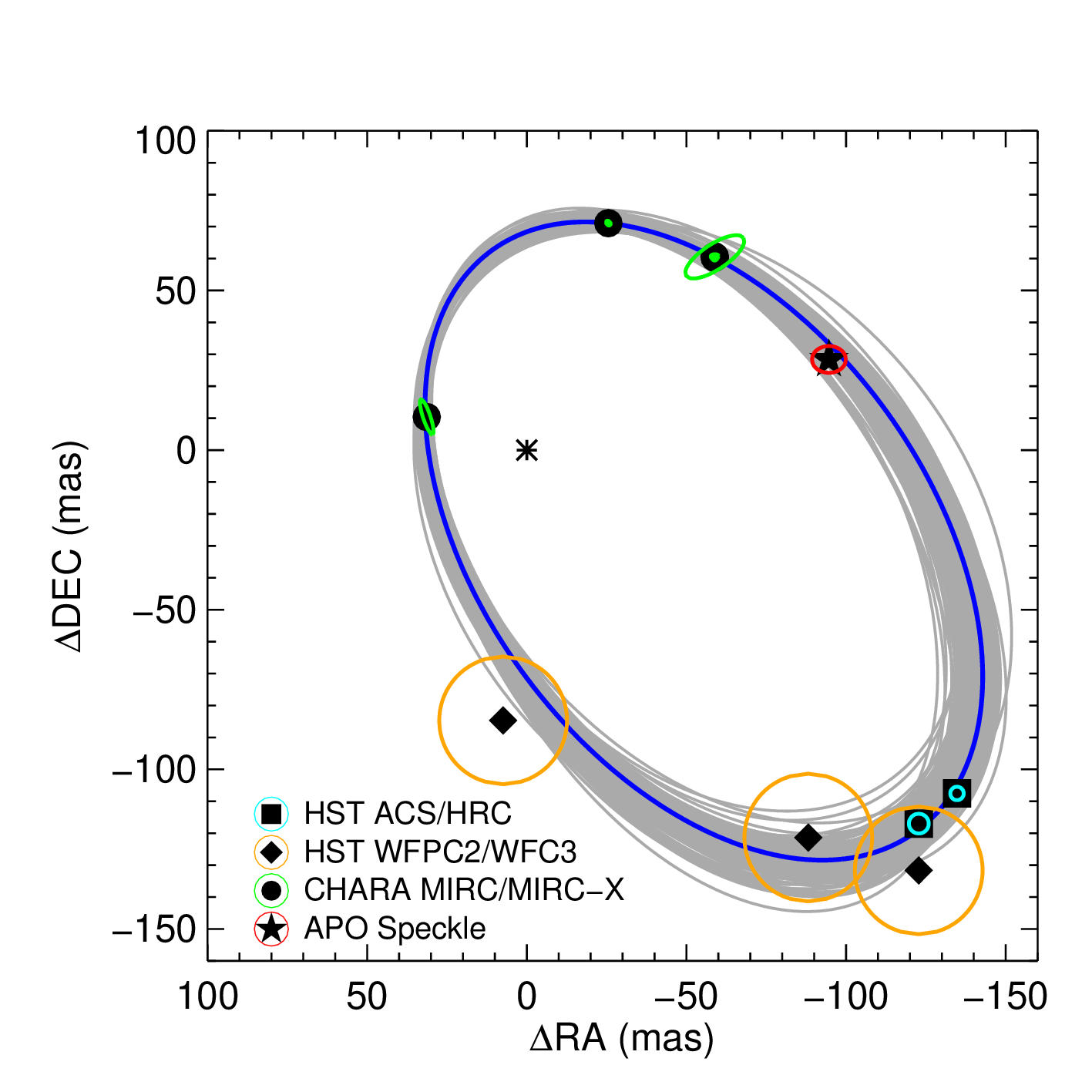}
\caption{Spectroscopic (left) and astrometric (right) orbital fit of Polaris.
    The symbols for the measurements by
    {\it HST} ACS, the {\it HST} WFPC2 and WFC3,  CHARA, and APS speckle are shown in the figure.  In addition to the best fit orbit, overplotted in gray is a sample of 1000 orbits selected at random while estimating the bootstrapped uncertainties.}
\label{orb.replic}
\end{figure*}

Another weighting scheme is motivated by the Synthetic Minority Over-sampling Technique (SMOTE) used in the machine learning community \citep{chawla02}.  In order to weight the two sets of measurements more evenly, we replicated each position measurement 167 times to give roughly even number of observables ($N$ = 167$\times$11$\times$2 = 3674). We applied random Gaussian uncertainties to the replicated measurements, creating a swarm of simulated observables for each data point. We then fit a simultaneous orbit  to the radial velocities and the replicated positions using the IDL orbit fitting library\footnote{\url{http://www.chara.gsu.edu/analysis-software/orbfit-lib}}. We followed a Monte Carlo bootstrap approach to compute uncertainties. For each bootstrap iteration, we randomly selected from the 11 astrometric measurements, with repetition. We then applied Gaussian uncertainties to the resampled measurements and replicated each data point as described above. We repeated this process 1,000 times and computed uncertainties in the orbital parameters from the standard deviation of the bootstrap distributions. The orbital parameters from this fit are listed in the ``Replicated'' column of Table~\ref{table_results}.  The spectroscopic and astrometric orbits are
shown in Figure~\ref{orb.replic}.
The orbital parameters and uncertainties are consistent with the 1/$N$ weighting method. The averaged $\chi^2$ in Section~\ref{section_Nweight} under-weights the radial velocities, while the replicated points over-weight the astrometry; but both methods provide similar results in the end.

In summary, all three weighting schemes produced similar masses within the errors.
 As stated in Section~\ref{orb} the solution based on measurement errors  (Section~\ref{section_error_weight}) the astrometric points from {\it HST} ACS/HRC are systematically displaced from the orbit solution.  Thus, the astrometry is poorly represented.  This is possibly due to systematic errors in the historical radial velocities, which are difficult to quantify, and which may on their own be distorting the shape of the orbit, and imposing that shape on the much less numerous astrometric measurements.   For this reason, a solution giving higher weight to the astrometry (Section~\ref{section_replicate}) is preferred
 with a mass of the Cepheid of 5.13  $\pm$ 0.28 (5\%)  $M_\odot$.





\begin{deluxetable}{lccccc}
  \tablecaption{Orbit Fitting Results.\label{table_results}.}
  \tablewidth{0pt}
  \tablehead{
    \colhead{Parameter} & \colhead{RV Only} & \colhead{Astrometry Only} & \colhead{$\sigma$ Weights} & \colhead{$N \sigma$ Weights} & \colhead{Replicated}  }
  \startdata
  $P_\mathrm{orb}$ (yr)			&  29.4330 $\pm$ 0.0079 &  29.49  $\pm$ 0.35    & 29.448  $\pm$   0.012  & 29.384 $\pm$ 0.017  &   29.416 $\pm$  0.028   \\
  $T_\mathrm{peri}$ (JY)		        &  2016.801 $\pm$ 0.011 & 2016.781 $\pm$ 0.049 & 2016.857 $\pm$ 0.012  & 2016.833 $\pm$ 0.017  &  2016.831 $\pm$  0.044  \\
  $e$					&  0.6195 $\pm$ 0.0015  & 0.6309 $\pm$ 0.0079  & 0.6290 $\pm$   0.0016  &  0.6336 $\pm$  0.0235  & 0.6354 $\pm$   0.0066    \\
  $\omega_\mathrm{Aa}$ ($^\circ$)		&  303.04 $\pm$ 0.34    & 305.22 $\pm$ 0.95    & 304.33 $\pm$    0.26    &  304.46 $\pm$  3.17  &  304.54  $\pm$  0.84     \\
  $K_1$ ($\mathrm{km~s^{-1}}$)		&  3.7409 $\pm$ 0.0075  &                     &  3.738  $\pm$  0.010    &  3.796 $\pm$ 0.311  &    3.762 $\pm$   0.025  \\
  $\gamma$	($\mathrm{km~s^{-1}}$)	&  -16.084 $\pm$ 0.025  &                      &  -16.0970  $\pm$   0.0047 & -16.110 $\pm$ 0.170   &   -16.1024 $\pm$  0.0084  \\
  $\Omega$	($^\circ$)		&                       & 202.16 $\pm$ 0.94    &  200.16 $\pm$   0.70  &  201.20  $\pm$ 2.59  & 201.28  $\pm$  1.18    \\
  $i$ ($^\circ$)				&                       & 127.82 $\pm$ 0.77   &  128.57 $\pm$  1.00   &  127.707 $\pm$ 3.191  &  127.57  $\pm$   1.21   \\
  $a$ (mas)				&                       & 129.07 $\pm$ 1.18    & 127.72  $\pm$  1.85  &  129.371 $\pm$ 5.432   & 129.55  $\pm$  2.05     \\
  Reference                             &  \citep{torres23} &  Section~\ref{orb} & Section~\ref{section_error_weight} & Section~\ref{section_Nweight} & Section~\ref{section_replicate}     \\
  \hline                                                                 
  $M_\mathrm{tot}$ ($M_\odot$)             &                       & 6.35 $\pm$ 0.23      &  6.16 $\pm$   0.28  &  6.4153 $\pm$ 0.7576   &  6.45 $\pm$   0.31   \\
  $M_\mathrm{Aa}$ ($M_\odot$)		&                       &                      &  4.87 $\pm$  0.26  &   5.0864  $\pm$ 0.6245   &   5.13 $\pm$    0.28   \\
  $M_\mathrm{Ab}$ ($M_\odot$)		&                       &                      &  1.295 $\pm$     0.019   &  1.3289 $\pm$   0.2702  &   1.316  $\pm$   0.028 \\
  \enddata
  \tablecomments{Masses computed assuming the {\it Gaia} distance of 136.90 $\pm$ 0.34 pc  \\}
\end{deluxetable}

\section{Discussion}

\subsection{The Cepheid Mass}

Approximately three quarters of the 30 year orbit of Polaris has now been observed with {\it HST}, CHARA, and
speckle measurements. The
resulting mass  of the Cepheid is  5.13  $\pm$ 0.28 (5\%)  $M_\odot$ which is larger than before the CHARA
and speckle observations were included, but
with smaller errors. 
The previous mass was  3.45 $\pm$ 0.75 $M_\odot$ (Evans et al. 2018; based on a slightly different
distance).

\subsection{Cepheid Mass Luminosity Relation}

Cepheids provide a quantitative test of evolutionary calculations. In the case of Polaris,
the mass derived here is combined with a luminosity to compare with theoretical
predictions.  For MW Cepheids in general, definitive luminosities will be produced by
{\it Gaia} in the release that includes orbital fits in the analysis.  For Polaris, we
use the distance to Polaris B as discussed in Section~\ref{intro.pol}.  Since Polaris B is
not a close binary, an orbit is not required in the {\it Gaia} solution.

\begin{figure*}[ht]
  \begin{center}
    \scalebox{0.5}{\includegraphics{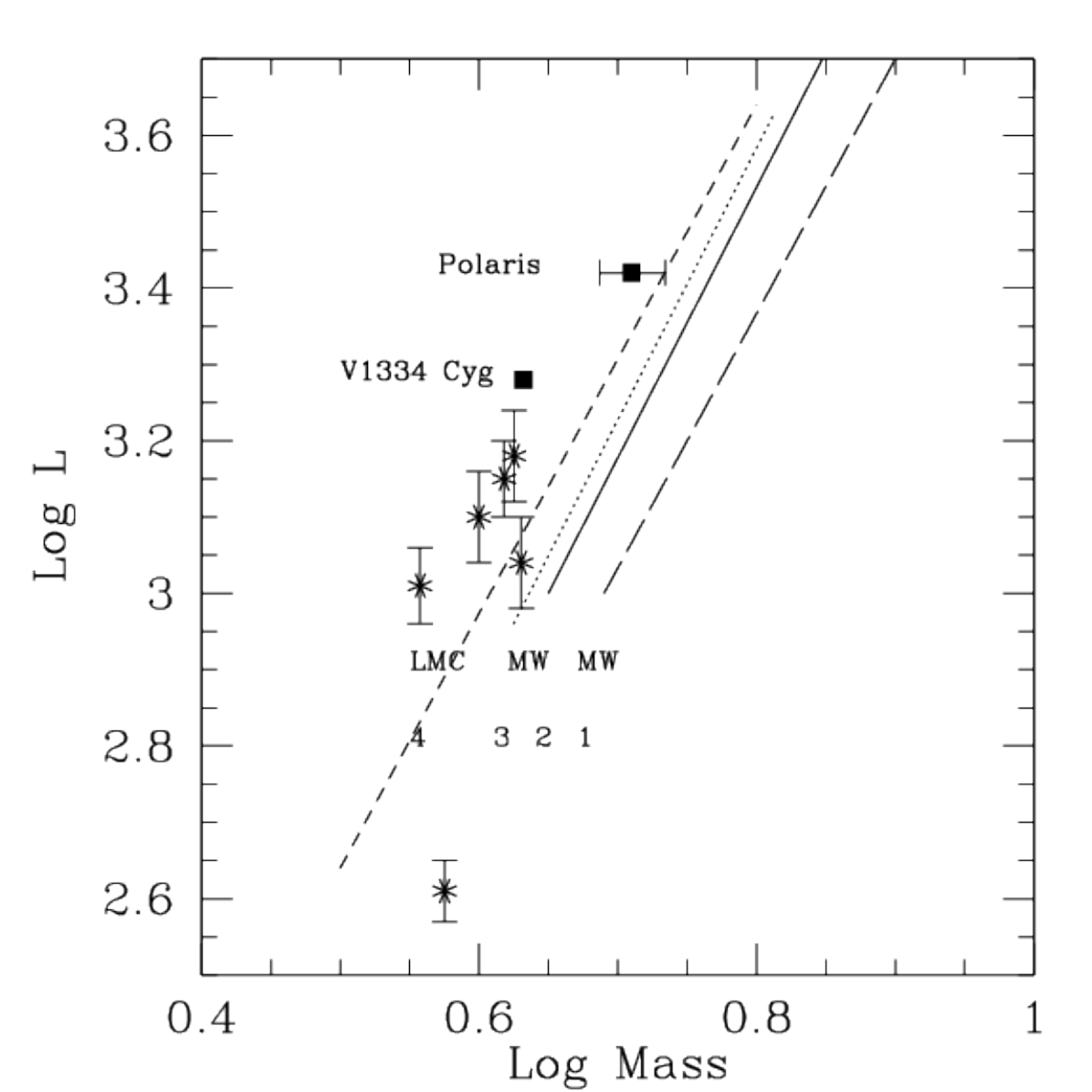}}
  \end{center}
  \caption{Cepheid mass-luminosity relation. The Milky Way Cepheids, Polaris (this work) and V1334 Cyg \citep{gallenne18}, are plotted as squares. LMC Cepheids \citep{pilecki21} are plotted as asterisks. Overplotted are predictions from evolutionary tracks (Bono et al. 2016; Anderson et al. 2014;
    Anderson et al. 2016) for (1) long dash: MW metallicity, no main sequence convective overshoot; (2) solid: MW metallicity, moderate convective overshoot; (3) dotted line: MW metallicity, small convective overshoot and rotation; (4) short dash line: LMC metallicity, moderate convective overshoot.  Mass and Luminosity are in solar units.}
  \label{figure_masslum}
\end{figure*}

Figure~\ref{figure_masslum} shows the Cepheid mass-luminosity relation with the location of Polaris
in comparison with the predictions from evolutionary calculations of several groups, which predict the
luminosity for a given mass.

 Figure~\ref{figure_masslum} includes V1334 Cyg, which has the most accurate Cepheid mass at present
\citep{gallenne18}. The solution also provides an accurate (3\%) distance to the system.
It is a good comparison for Polaris since it has similar characteristics.
It pulsates in the first overtone, and has a period (luminosity) similar to Polaris.  
Also shown in  Figure~\ref{figure_masslum} are masses for LMC Cepheids in
eclipsing binaries from Pilecki et al. (2021).  Luminosities for the MW Cepheids
are from the distance to Polaris (above) and V1334 Cyg.  Luminosities for the
LMC Cepheids are from   Pilecki et al. (2021).
The observed mass--luminosity combinations are compared with evolutionary
predictions from Bono et al. (2016) and Anderson et al. (2014).
The Bono tracks cover MW and LMC metallicities.  For the MW the relations are
shown for no main sequence convective core overshoot and moderate overshoot. The
Anderson relation is for main sequence rotation and moderate convective overshoot. 
The lower metallicity of the LMC stars results in a higher luminosity than for
solar abundance stars, as indicated by the models.

The determination of the mass of Polaris in this study is part of a series of
studies to measure of Cepheids.
Additional MW Cepheids are
to be added to  Figure~\ref{figure_masslum}.  However the first indication is
that in the examples so far (Polaris, V1334 Cyg, and the LMC Cepheids), the
luminosity is larger than  predicted by current evolutionary
calculations.


\subsection{Evolutionary Tracks}

\begin{figure*}[ht]
  \begin{center}
\scalebox{0.5}{\includegraphics{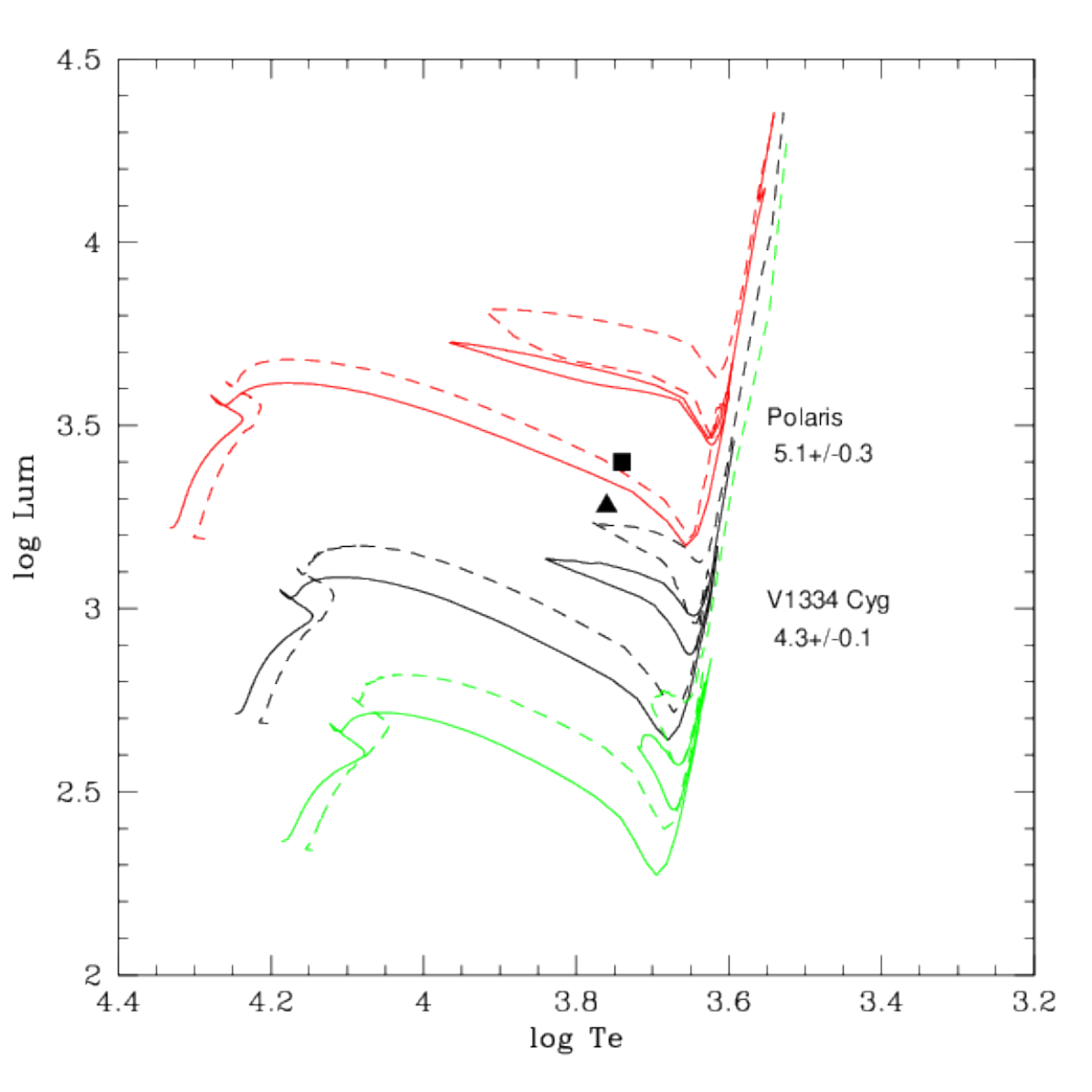}}
  \end{center}
  \caption{Polaris and V1334 Cyg compared with evolutionary tracks.
 For Polaris (filled square) and V1334 Cyg (filled triangle)
    the mass in  $M_\odot$ is listed next to the
    star name.  The evolutionary tracks  from Georgy et al. (2013) are for
    4, 5, and 7 $M_\odot$ in colors of green, black, and red respectively.
    Solid lines are for zero rotational velocity
    on the main sequence; dashed lines have 0.95 breakup velocity.
    Luminosity is in solar units; temperature
    is in Kelvin. 
  }
  \label{ev.model}
\end{figure*}

To explore the details of the calculation results, we provide Figure~\ref{ev.model}
updated from Figure 7 in Evans et al. (2018). Temperatures are from (B-V)$_0$, as
discussed in Evans et al. (2018).
Figure~\ref{ev.model} illustrates the location of Polaris and also V1334 Cyg with respect
to evolutionary calculations for 4, 5, and 7 $M_\odot$ from Georgy et al. (2013).  The
tracks contrast evolution for stars with zero rotational velocity on the main sequence
and those with 0.95 breakup velocity.  Both Polaris and V1334 Cyg are more luminous than
the 5  $M_\odot$ track in the blue loop, even for large main sequence rotation.   
For a comparison between codes, Figure 8 in Evans et al. (2018) 
 shows the location of Polaris compared with three sets of tracks
for a 5$M_\odot$ model:
Geneva (Georgy et al. 2013), MIST (Choi et al. 2016) and PARSEC
\footnote{\url{http://philrosenfield.github.io/padova\_tracks}} (based on
the Bressan et al.\ (2012) group).
All three sets of tracks are shown for zero initial main sequence rotation; for the
Geneva and MIST calculations, a track is also shown for substantial rotation. 
  In this figure, the MIST tracks (both with and without substantial
main sequence rotation) approximately coincide with the Georgy tracks with rotation.


One further criterion for the comparison between observed masses and evolutionary tracks is
that Cepheids are found in a region of the Luminosity--Temperature diagram (Figure~\ref{ev.model}) where
the blue loops penetrate the instability strip.  This was a motivation of the exploration of
the effect of rotation in main sequence progenitors (Anderson et al. 2014), since increasing the
main sequence core convective overshoot increases luminosity at the Cepheid phase, but truncates the
blue loops.  Evolutionary tracks for additional masses probe this further. 





%

\subsection{ Properties of Polaris}


This study provides astrometry from CHARA interferometry and  APO speckle observations
for the Polaris Aa and Ab system.
When combined with previous {\it HST} observations and radial velocities (Torres 2023), a
mass of  5.13 $\pm$ 0.28 $M_\odot$  is derived.  The observed luminosity for Polaris is brighter than
predicted by current evolutionary tracks.   The predicted luminosity depends on the rotation of the main sequence progenitor, but Fig~\ref{ev.model} shows that Polaris is at least 0.4 mag brighter than the predicted tracks including a small correction from the 5 $M_\odot$ tracks to the mass of Polaris (based on 0.5 critical breakup velocity [Anderson et al. 2014]).
This study is part of a series of studies of
Cepheid masses using interferometry as well as {\it Gaia}  astrometry to examine this
questions.

In the case of Polaris several of its properties  are unusual in the context of
Cepheids.  It pulsates in the first overtone, which may be linked with the
rapid period change. Overtone pulsators have more instability in their
pulsation periods, which can sometimes be interpreted as rapid period change. 
The variation in pulsation amplitude may also be related to its
pulsation mode.  The recent
discussion of velocities (Torres 2023) postulates that the ``glitch'' in pulsation
may occur at  periastron passage.  This  would add a new factor to the
interpretation of the observations.  It has been suggested for other Cepheids
with a ``phase jump'' that they are likely to be binary systems (Csorneyei et al. 2022).
In summary, while these properties are unusual for Cepheids, they exist in
other stars.  Thus Polaris fits in the framework of pulsating supergiants,
particularly if orbital motion is included.
While these characteristics complicate the interpretation of observations, all are found in
stars without abnormal evolution.  

The identification of star spots is consistent with several properties of Polaris.
It has a very low pulsation amplitude, which sets it apart from full amplitude
Cepheids.  This may mean that the atmosphere is  like that of a nonvariable
supergiant, which have often indicators of activity.  It is not clear how full
amplitude pulsation affects the atmosphere and magnetic field in pulsators, so
Polaris is an interesting test case.

Polaris has a couple of other characteristics which would be consistent with
magnetically related spots. 
Polarization has been found (Barron et al. 2022).  However, 
the polarization measurements look more like those of the nonvariable supergiant
$\alpha$ Per than other Cepheids (Grunhut et al. 2010),
but this may just reflect the very low pulsation amplitude.
Polaris has also been detected in X-rays (Evans et al. 2022).

Our identification of star spots opens the prospect of  determinating
a rotation period.  It also can explain why additional periodicities have been
difficult to find, since the distribution of spots is variable.
The  long period
($\simeq$120 day) variation found by Lee et al. (2008), for instance, might be a
rotation period. 40 and 60 d periods found by Anderson (2019) may be related. 

Another possible interpretation of surface features on Polaris is convective
supergranules as discussed by Schwarzschild (1975).

A final characteristic of the orbit of Polaris is the 
 high eccentricity.  This is  often found in  systems which have
 undergone three body interaction, which would be consistent with the
 Cepheid being a merger product from a former triple system, as
 suggested  by Evans, et al (2018).  However, high eccentricity is
 frequently found in long period systems (Shetye et al. 2024) so
 it does not require special conditions.

 \section{Summary and Future Work}
 The CHARA and APO observations of the Polaris Aa and Ab system have added to previous
 astrometric observations to cover approximately three quarters of the orbit.  The mass
of the Cepheid  which results from these data and the spectroscopic orbit is
5.13  $\pm$ 0.28 (5\%)  $M_\odot$.  The Cepheid is overluminous for this mass
according to current calculations.  Starspots have been identified in the
CHARA images, providing another tool to use to investigate the Cepheid.  
 
Further observations with CHARA and APO will provide additional coverage of the
orbit and add to the determination of the mass.


\begin{acknowledgments}

  This work is based upon observations obtained with the Georgia State University Center for High Angular Resolution Astronomy Array at Mount Wilson Observatory.  The CHARA Array is supported by the National Science Foundation under Grant No. AST-1636624 and AST-2034336.  Institutional support has been provided from the GSU College of Arts and Sciences and the GSU Office of the Vice President for Research and Economic Development. Time at the CHARA Array was granted through the NOIRLab community access program (NOIRLab PropID: 2018B-0039; PI: N. Evans).
PK acknowledges funding from the European Research Council (ERC) under the European Union's Horizon 2020 research and innovation program (project UniverScale, grant agreement 951549). SK acknowledges funding for MIRC-X  from the European Research Council (ERC) under the European Union's Horizon 2020 research and innovation programme (Starting Grant No. 639889 and Consolidated Grant No. 101003096). JDM acknowledges funding for the development of MIRC-X (NASA-XRP NNX16AD43G, NSF-AST 1909165). AG acknowledges the support of the Agencia Nacional de Investigaci\'on Cient\'ifica y Desarrollo (ANID) through the FONDECYT Regular grant 1241073.  RMR acknowledges funding from the Heising-Simons Foundation 51 Pegasi b Fellowship.
  Support was provided to NRE by the Chandra X-ray Center NASA Contract NAS8-03060.This research has received support from the European Research Council (ERC) under the European Union's Horizon 2020 research and innovation programme (Grant Agreement No. 947660). RIA is funded by the Swiss National Science Foundation through an Eccellenza Professorial Fellowship (award PCEFP2\_194638).
This research is based on observations made with the Mercator Telescope, operated on the island of La Palma by the Flemish Community, at the Spanish Observatorio del Roque de los Muchachos of the Instituto de Astrofisica de Canarias (observations from Anderson et al.). Hermes is supported by the Fund for Scientific Research of Flanders (FWO), Belgium, the Research Council of K.U. Leuven, Belgium, the Fonds National de la Recherche Scientifique (F.R.S.- FNRS), Belgium, the Royal Observatory of Belgium, the Observatoire de Gene\`eve, Switzerland, and the Th\"uringer Landessternwarte, Tautenburg, Germany.

  This work has made
use of data from the European Space Agency (ESA) mission {\it Gaia} (https://www.cosmos.esa.int/gaia), processed by
the {\it Gaia} Data Processing and Analysis Consortium (DPAC, https://www.cosmos.esa.int/web/gaia/dpac/consortium).
Funding for the DPAC has been provided by national institutions, in particular the institutions participating in the
{\it Gaia} Multilateral Agreement.
This research made use of services provided by the Jean-Marie
Mariotti Center (Aspro, SearchCal, and OIDB).
The SIMBAD database, and NASA’s Astrophysics Data System Bibliographic Services
were used in the preparation of this paper.

The software used in this project is available from  Squeeze [10.5281/zenodo.11643336] and ROTIR [10.5281/zenodo.11643362]]{https://doi.org/DOI}.

\end{acknowledgments}


\appendix

\section{Interferometric Data Plots - Imaging Results} \label{section_appendix_data}

\begin{figure*}[ht]
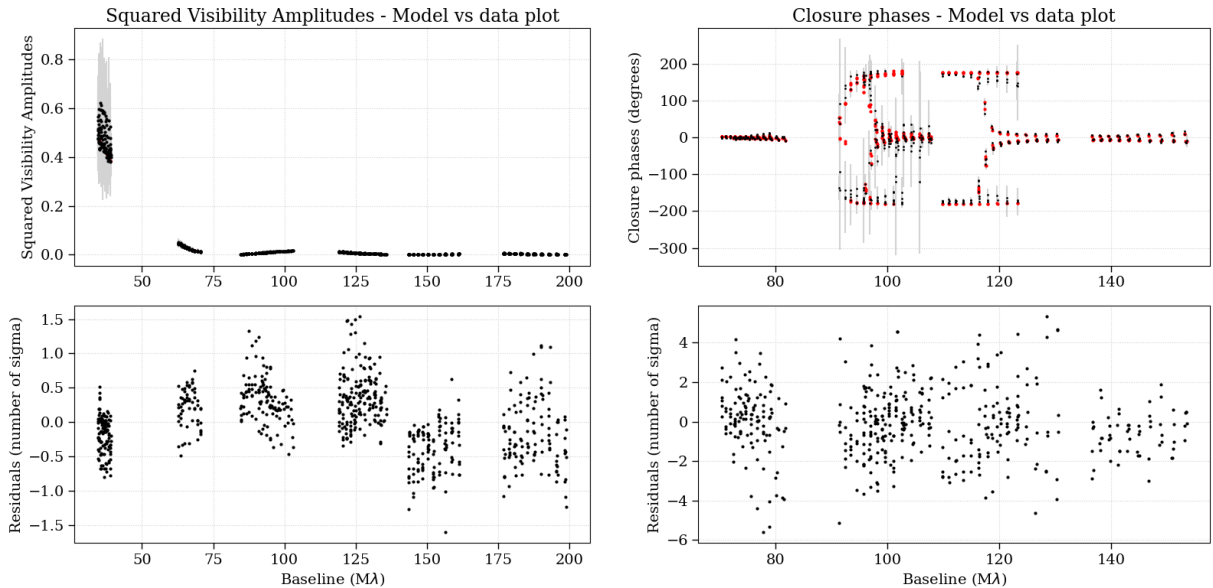

\figurenum{12.1}
  \begin{center}
    \scalebox{0.4}{\includegraphics{vis2_rotir_polaris_2018Aug27.png}}
    \scalebox{0.4}{\includegraphics{t3phi_rotir_polaris_2018Aug27.png}}
  \end{center}
  \caption{Visibilities and closure phases measured with CHARA MIRC-X and averaged over 10-minute observing sets for Polaris on UT 2018Aug27. The black symbols are the measured values while the red symbols are extracted from the ROTIR reconstructed image. The lower panels show the residuals between the data and the image.  \\
The complete set of figure set (4 images) is available in the online journal. 
  }
  \label{figure_vis2cp_2018aug}
\end{figure*}

Figure Set~\ref{figure_vis2cp_2018aug} shows the visibilities and closure phases measured with CHARA MIRC-X averaged over 10-minute observing sets. The data (black circles) and observables extracted from the reconstructed images (red circles) are plotted in the top panels of each figure. The bottom panels show the residuals between the data and the image.

\figsetstart
\figsetnum{12}
\figsettitle{Imaging Results}

\figsetgrpstart
\figsetgrpnum{12.1}
\figsetgrptitle{2018Aug27_im}
\figsetplot{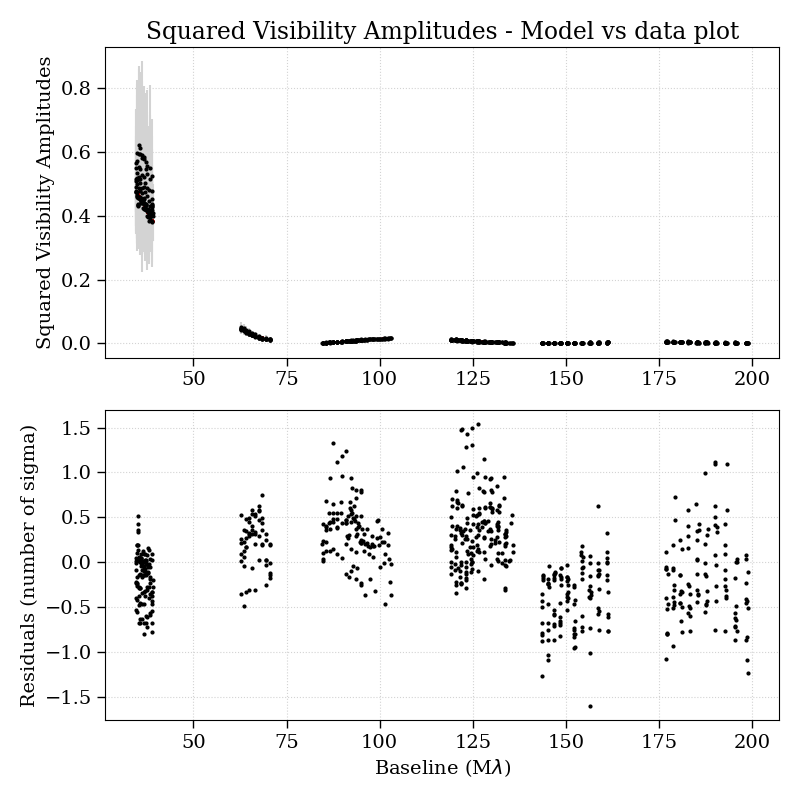}
\figsetplot{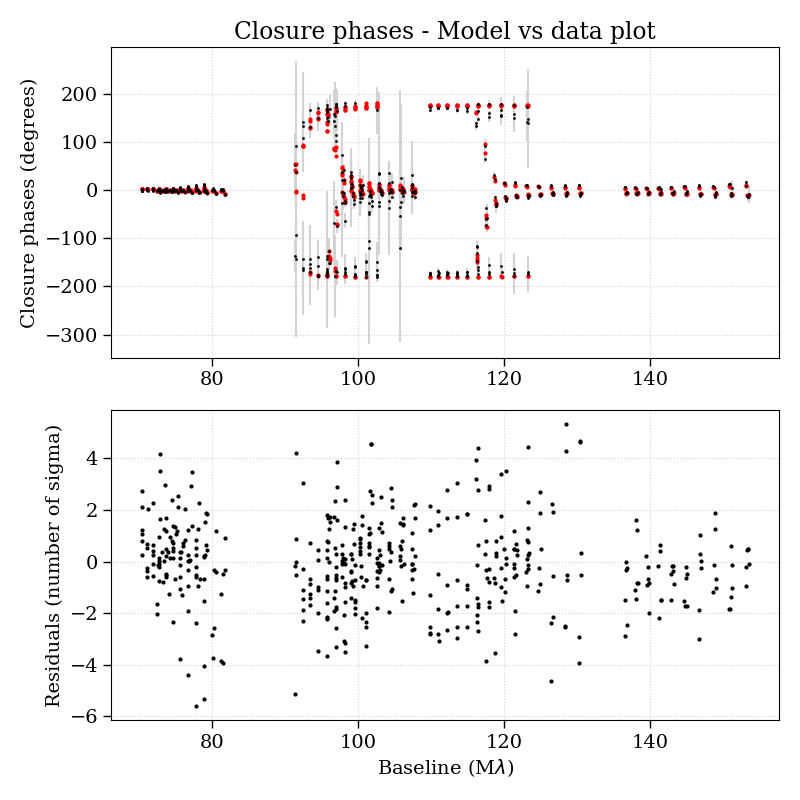}
\figsetgrpnote{Visibilities and closure phases measured with CHARA MIRC-X and averaged over 10-minute observing sets for Polaris on UT 2018Aug27. The black symbols are the measured values while the red symbols are extracted from the ROTIR reconstructed image. The lower panels show the residuals between the data and the image. \\
The complete set of figure set (4 images) is available in the online journal.}
\figsetgrpend

\figsetgrpstart
\figsetgrpnum{12.2}
\figsetgrptitle{2019Apr09_im}
\figsetplot{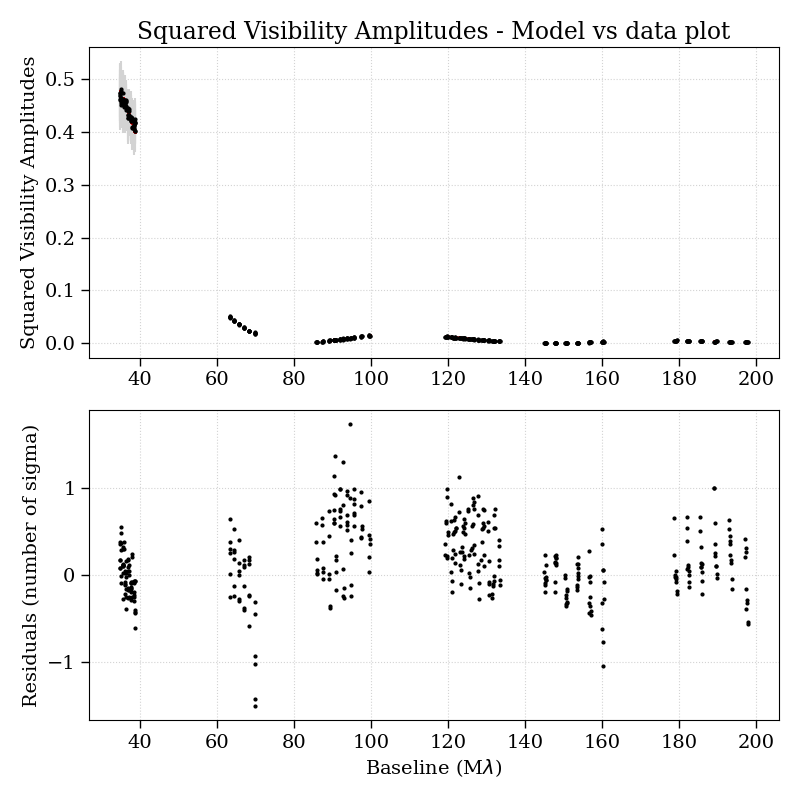}
\figsetplot{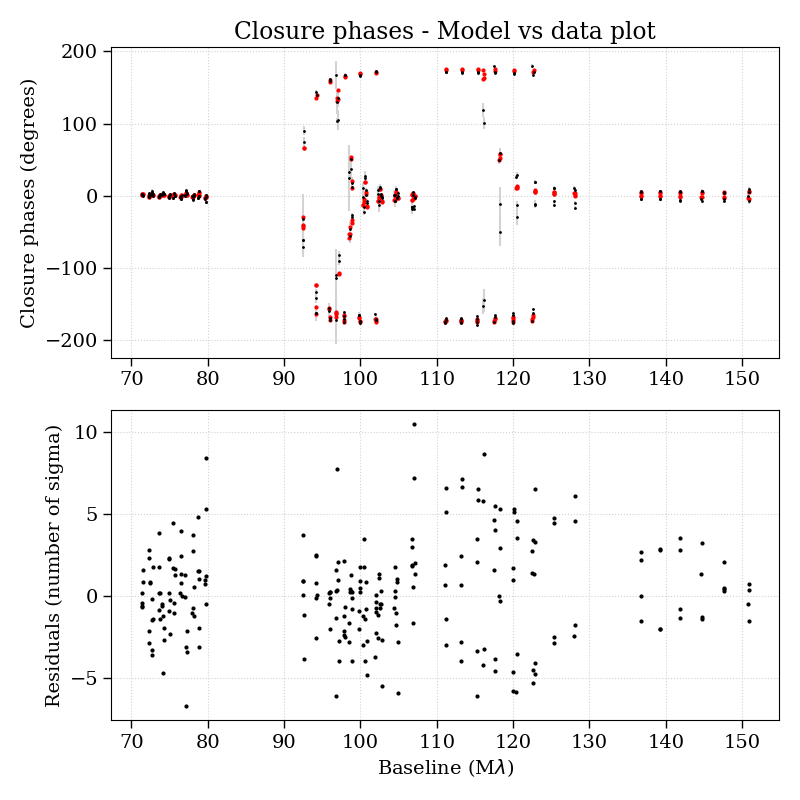}
\figsetgrpnote{Visibilities and closure phases measured with CHARA MIRC-X and averaged over 10-minute observing sets for Polaris on UT 2019Apr09. The black symbols are the measured values while the red symbols are extracted from the ROTIR reconstructed image. The lower panels show the residuals between the data and the image.}
\figsetgrpend

\figsetgrpstart
\figsetgrpnum{12.3}
\figsetgrptitle{2019Sep02_im}
\figsetplot{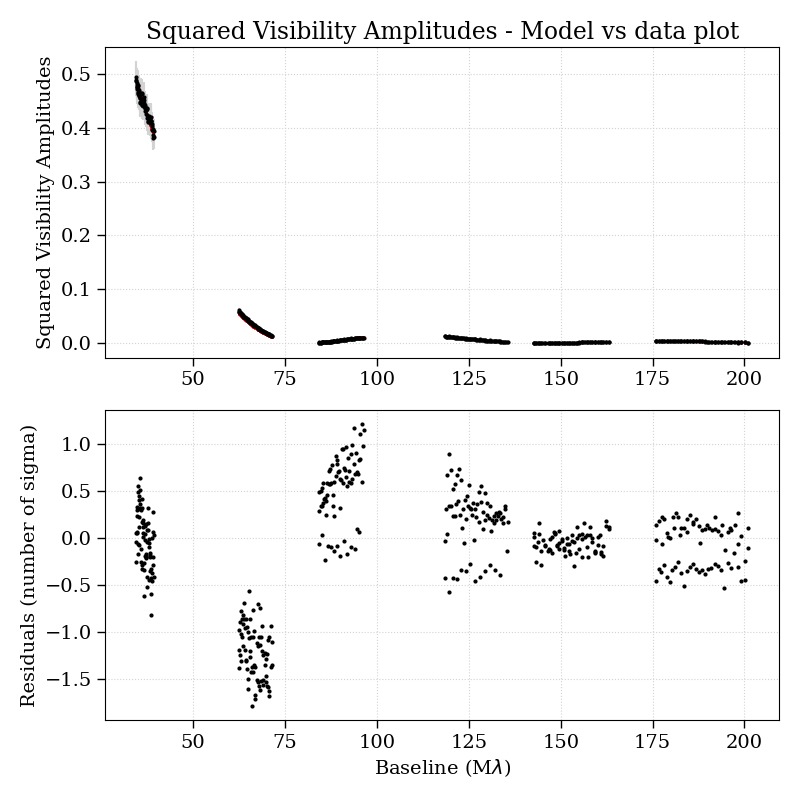}
\figsetplot{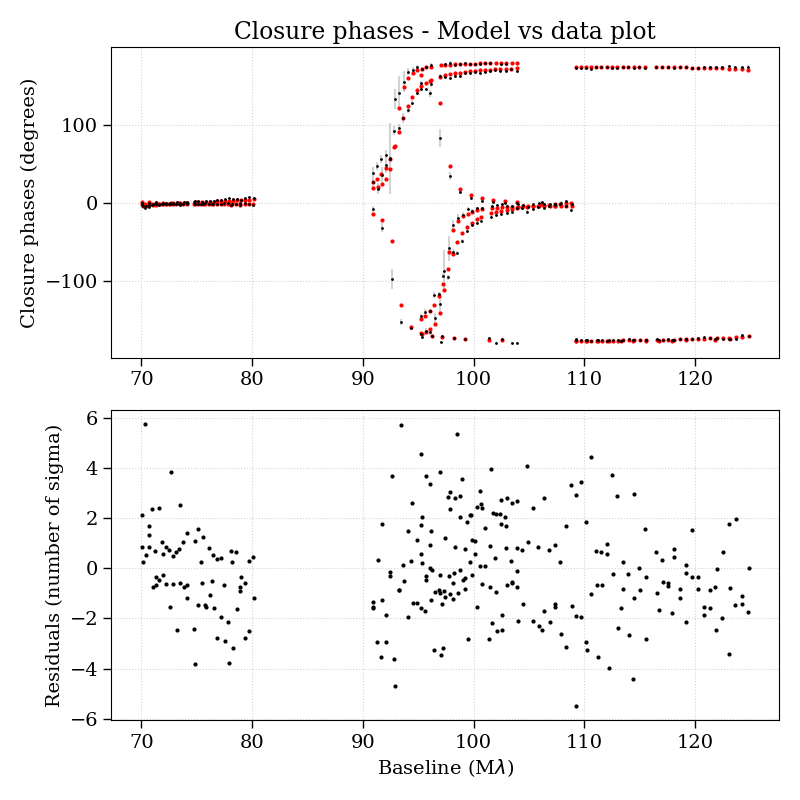}
\figsetgrpnote{Visibilities and closure phases measured with CHARA MIRC-X and averaged over 10-minute observing sets for Polaris on UT 2019Sep02. The black symbols are the measured values while the red symbols are extracted from the ROTIR reconstructed image. The lower panels show the residuals between the data and the image.}
\figsetgrpend

\figsetgrpstart
\figsetgrpnum{12.4}
\figsetgrptitle{2021Apr02_03_04_im}
\figsetplot{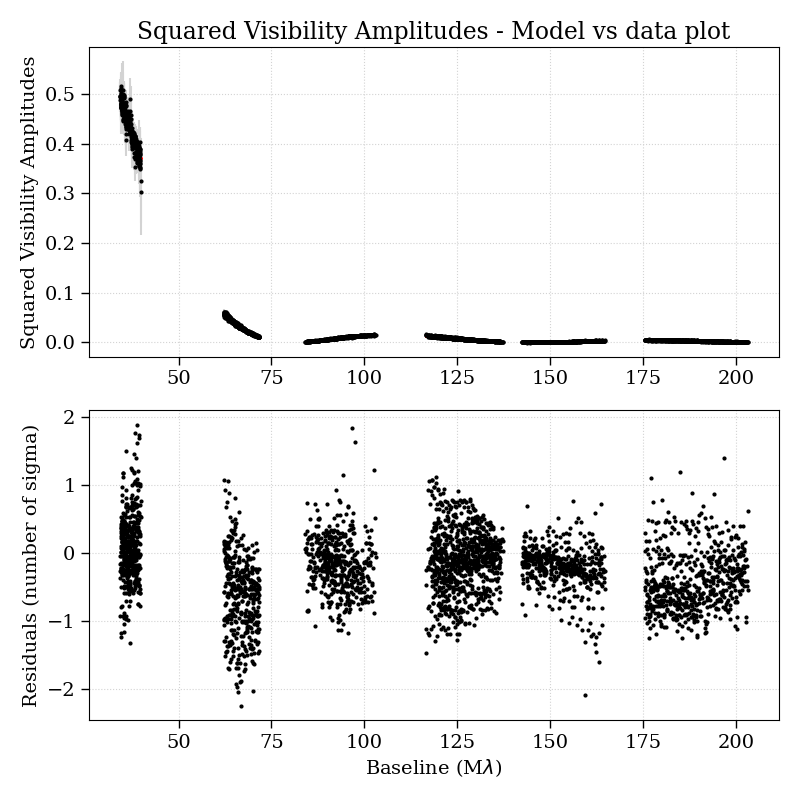}
\figsetplot{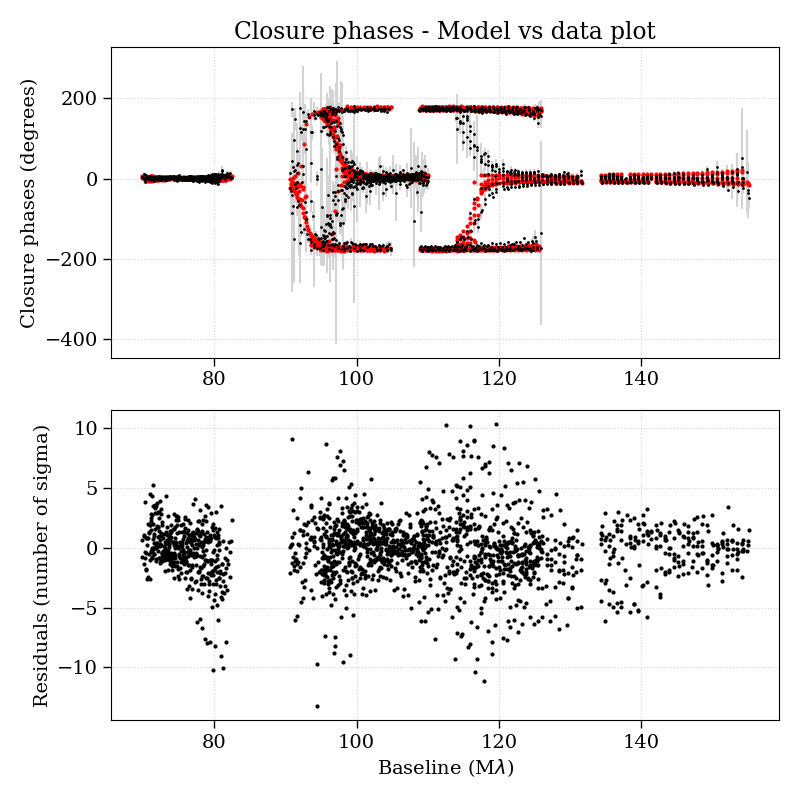}
\figsetgrpnote{Visibilities and closure phases measured with CHARA MIRC-X and averaged over 10-minute observing sets for Polaris on UT 2021Apr02, 2021Apr03, and 2021Apr04. The black symbols are the measured values while the red symbols are extracted from the ROTIR reconstructed image. The lower panels show the residuals between the data and the image.}
\figsetgrpend

\figsetend

\begin{figure*}[ht]
\figurenum{12.2}
  \begin{center}
    \scalebox{0.4}{\includegraphics{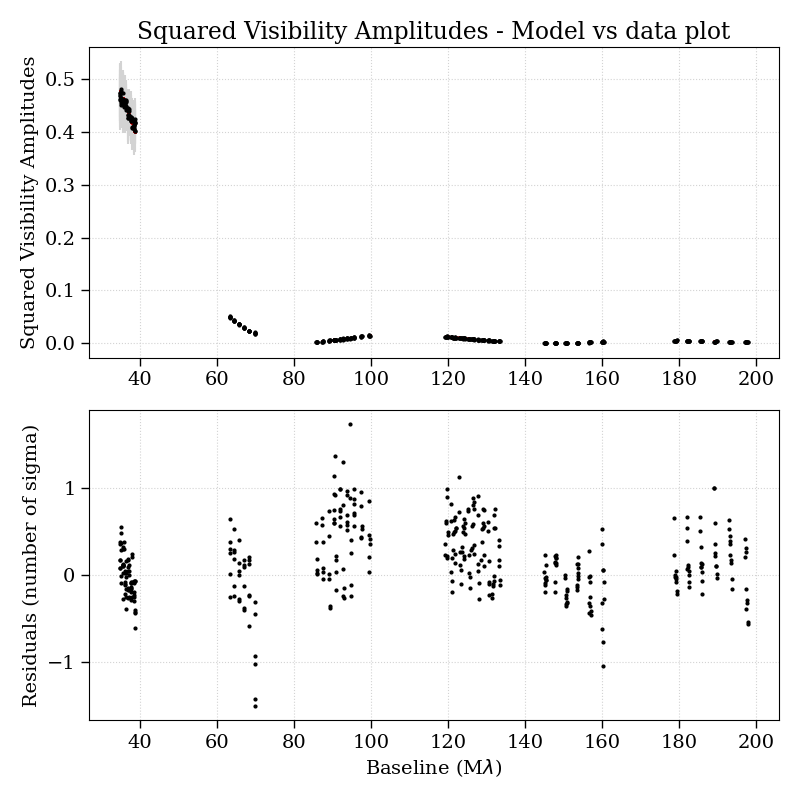}}
    \scalebox{0.4}{\includegraphics{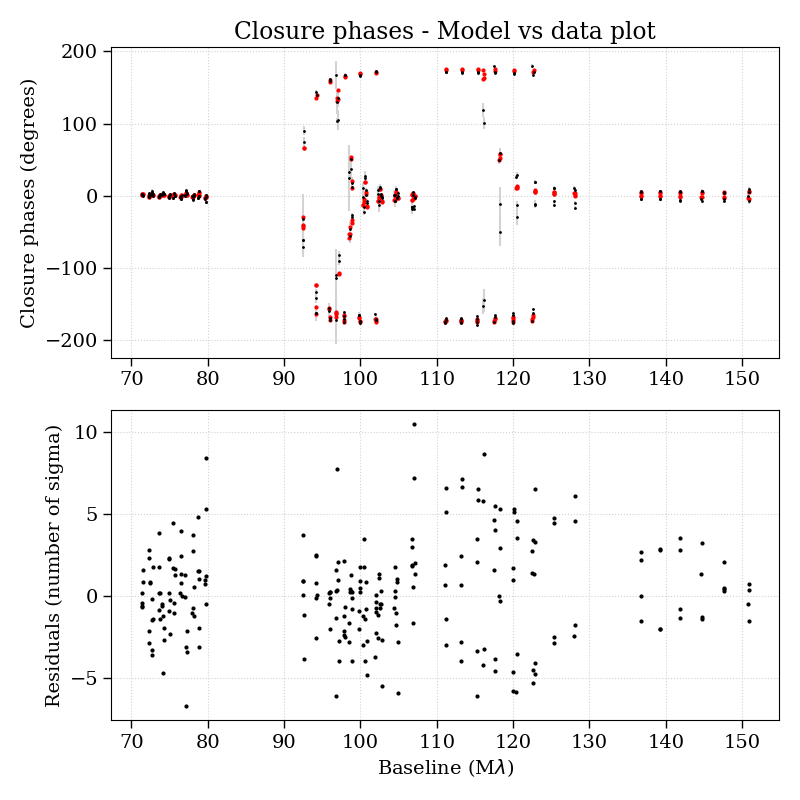}}
  \end{center}
  \caption{Visibilities and closure phases measured with CHARA MIRC-X and averaged over 10-minute observing sets for Polaris on UT 2019Apr09. The black symbols are the measured values while the red symbols are extracted from the ROTIR reconstructed image. The lower panels show the residuals between the data and the image.  }
  \label{figure_vis2cp_2019apr}
\end{figure*}

\begin{figure*}[ht]
\figurenum{12.3}
  \begin{center}
    \scalebox{0.4}{\includegraphics{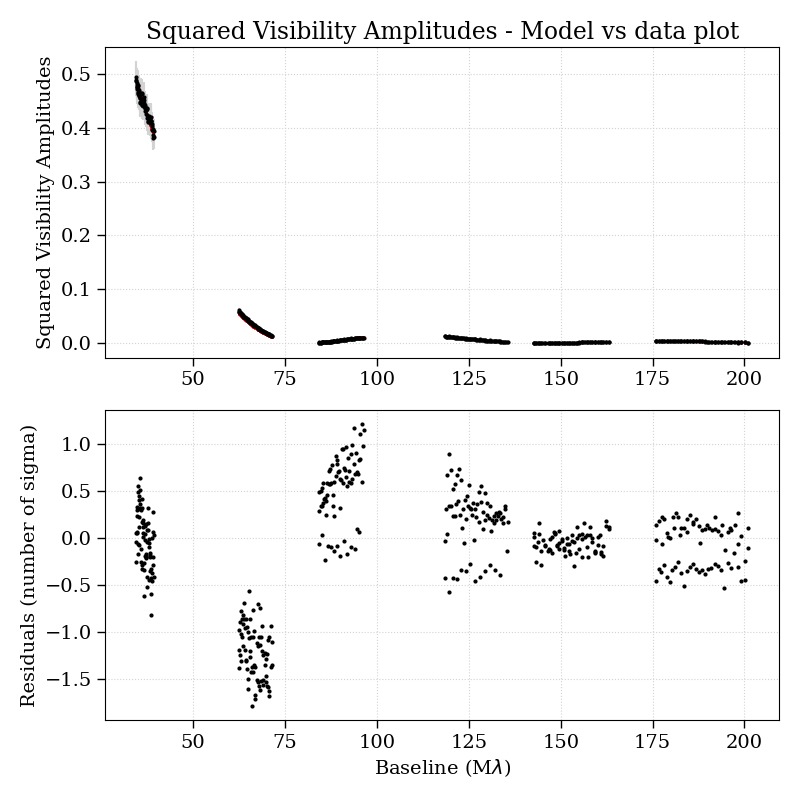}}
    \scalebox{0.4}{\includegraphics{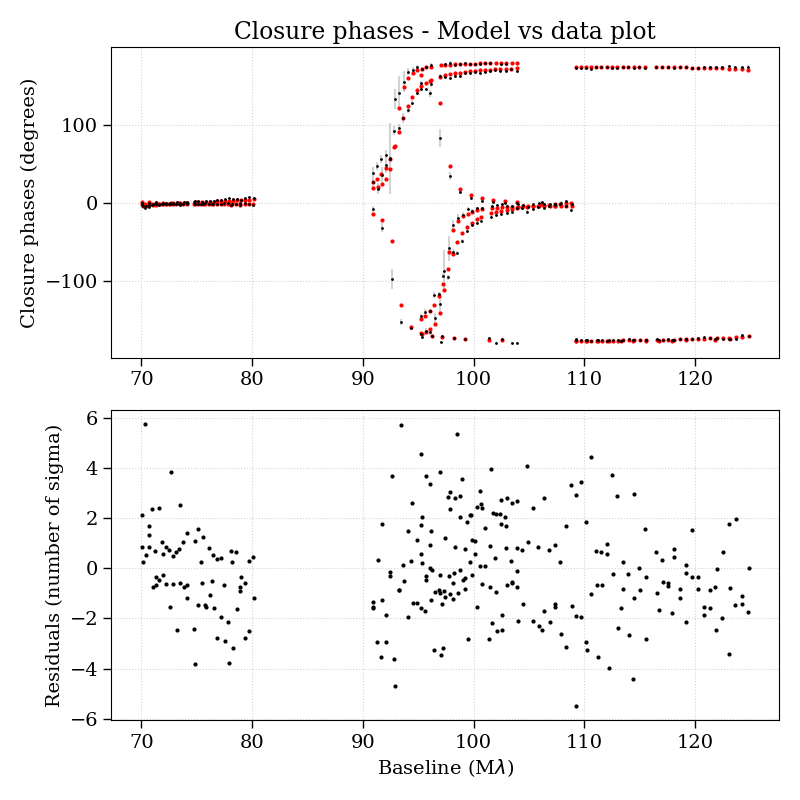}}
  \end{center}
  \caption{Visibilities and closure phases measured with CHARA MIRC-X and averaged over 10-minute observing sets for Polaris on UT 2019Sep02. The black symbols are the measured values while the red symbols are extracted from the ROTIR reconstructed image. The lower panels show the residuals between the data and the image.  }
  \label{figure_vis2cp_2019Sep}
\end{figure*}
            
\begin{figure*}[ht]
\figurenum{12.4}
  \begin{center}
    \scalebox{0.4}{\includegraphics{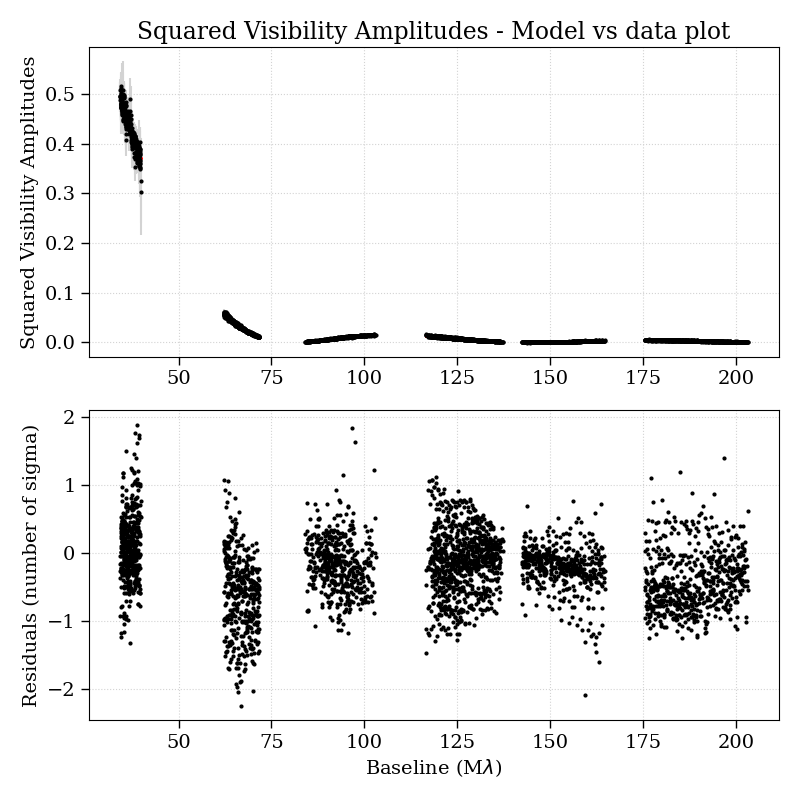}}
    \scalebox{0.4}{\includegraphics{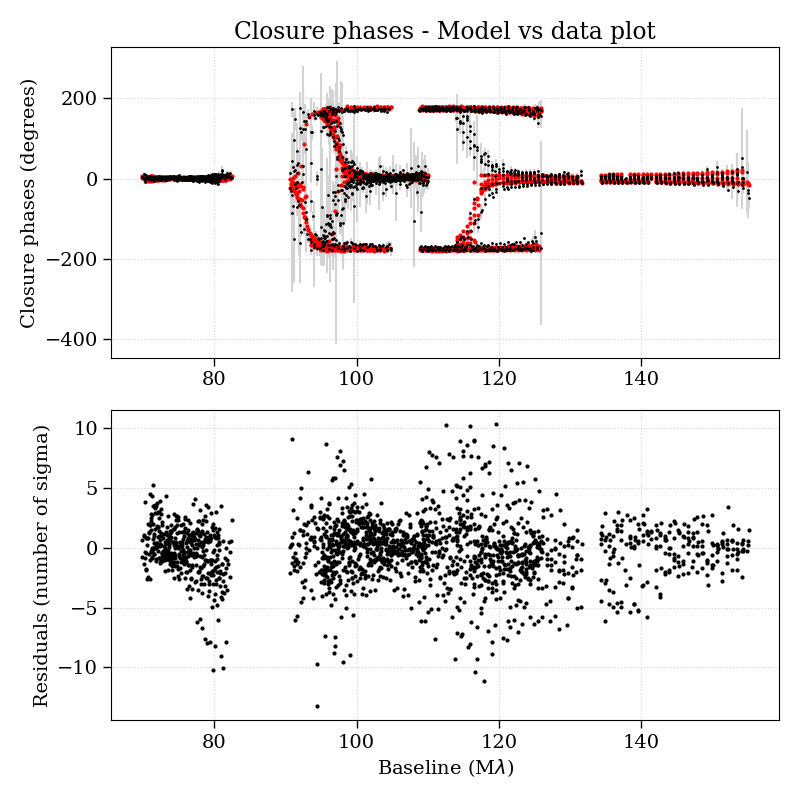}}
  \end{center}
  \caption{Visibilities and closure phases measured with CHARA MIRC-X and averaged over 10-minute observing sets for Polaris on UT 2021Apr02, 2021Apr03, and 2021Apr04. The black symbols are the measured values while the red symbols are extracted from the ROTIR reconstructed image. The lower panels show the residuals between the data and the image.  }
  \label{figure_vis2cp_2021apr}
\end{figure*}

\section{Interferometric Data Plots - Binary Fits} \label{section_appendix_fits}

\begin{figure}
\figurenum{13.1}
  \begin{center}
    \scalebox{0.12}{\includegraphics{oiprep_MIRC_L2.Polaris.2016Sep12_tcoh0075ms_GHS_30sec_2023Apr21.XCHAN.cal.SPLIT_uv_lam.png}}
    \scalebox{0.11}{\includegraphics{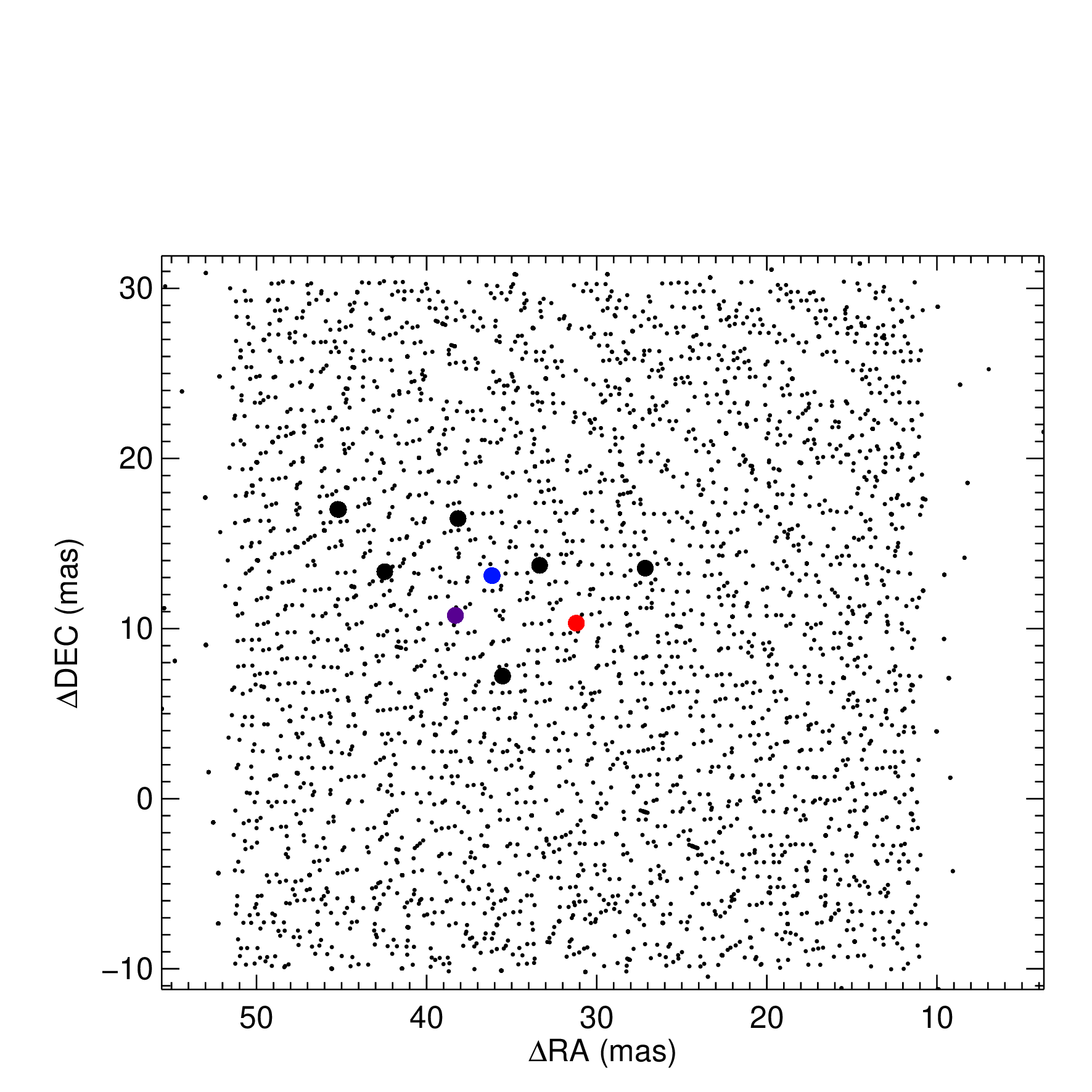}} \\
    \scalebox{0.1}{\includegraphics{oiprep_MIRC_L2.Polaris.2016Sep12_tcoh0075ms_GHS_30sec_2023Apr21.XCHAN.cal.SPLIT_logvis2_big.png}}
    \scalebox{0.1}{\includegraphics{oiprep_MIRC_L2.Polaris.2016Sep12_tcoh0075ms_GHS_30sec_2023Apr21.XCHAN.cal.SPLIT_t3_bline.png}}
    \caption{Binary fit for Polaris on UT 2016Sep12. Top Row: ($u,v$) coverage (left) and $\chi^2$ map from the binary grid search (right). In the $\chi^2$ map, the red, orange, yellow, green, blue, purple, large black, and small black symbols correspond to solutions within $\Delta \chi^2$ = 1, 4, 9, 16, 25, 36, 49, and $>$50 from the minimum $\chi^2$. Bottom row: The filled black circles show the squared visibilities (left) and closure phases (right) measured with MIRC-X using the 30 second integration time. The red crosses show the best-fit binary model. \\
The complete set of figure set (8 images) is available in the online journal. 
     }
\label{figure_2016Sep12_bin}
  \end{center}
\end{figure}

Figure Set~\ref{figure_2016Sep12_bin} shows the results from the binary fits to the CHARA MIRC-X interferometry using the 30 second integration time. Each figure shows the ($u,v$) coverage, the $\chi^2$ map from the binary grid search, the visibilities, and the closure phases. The $\chi^2$ maps are centered at the predicted location based on the updated orbit fit. The nights with reliable detections show a clear minimum in the $\chi^2$ indicated by the colored circles. The nights with unreliable binary fits show more ambiguity in the $\chi^2$ maps.


\figsetstart
\figsetnum{13}
\figsettitle{Binary Fits}

\figsetgrpstart
\figsetgrpnum{13.1}
\figsetgrptitle{2018Aug27_bin}
\figsetplot{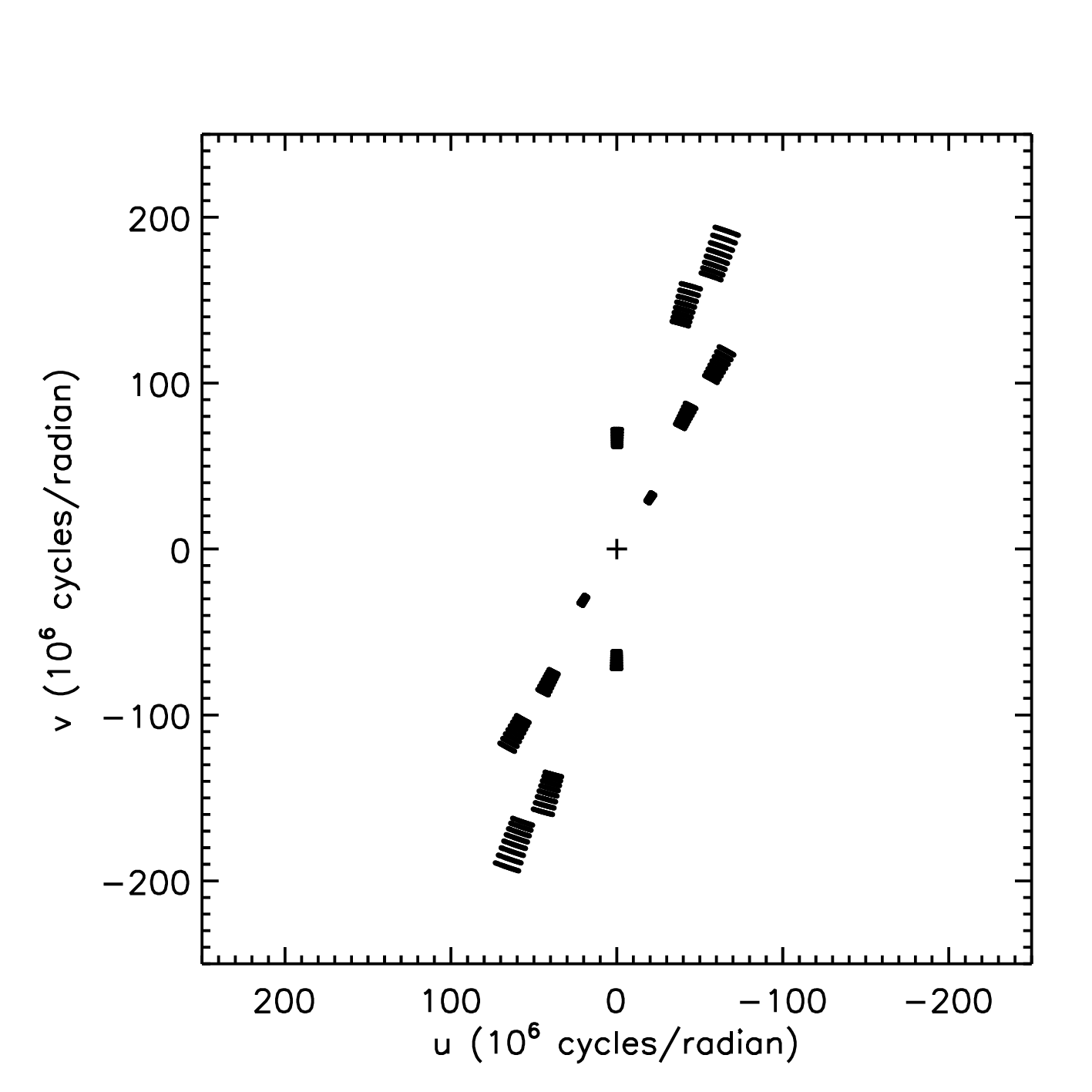}
\figsetplot{chi2_radec_oiprep_MIRC_L2.Polaris.2016Sep12}
\figsetplot{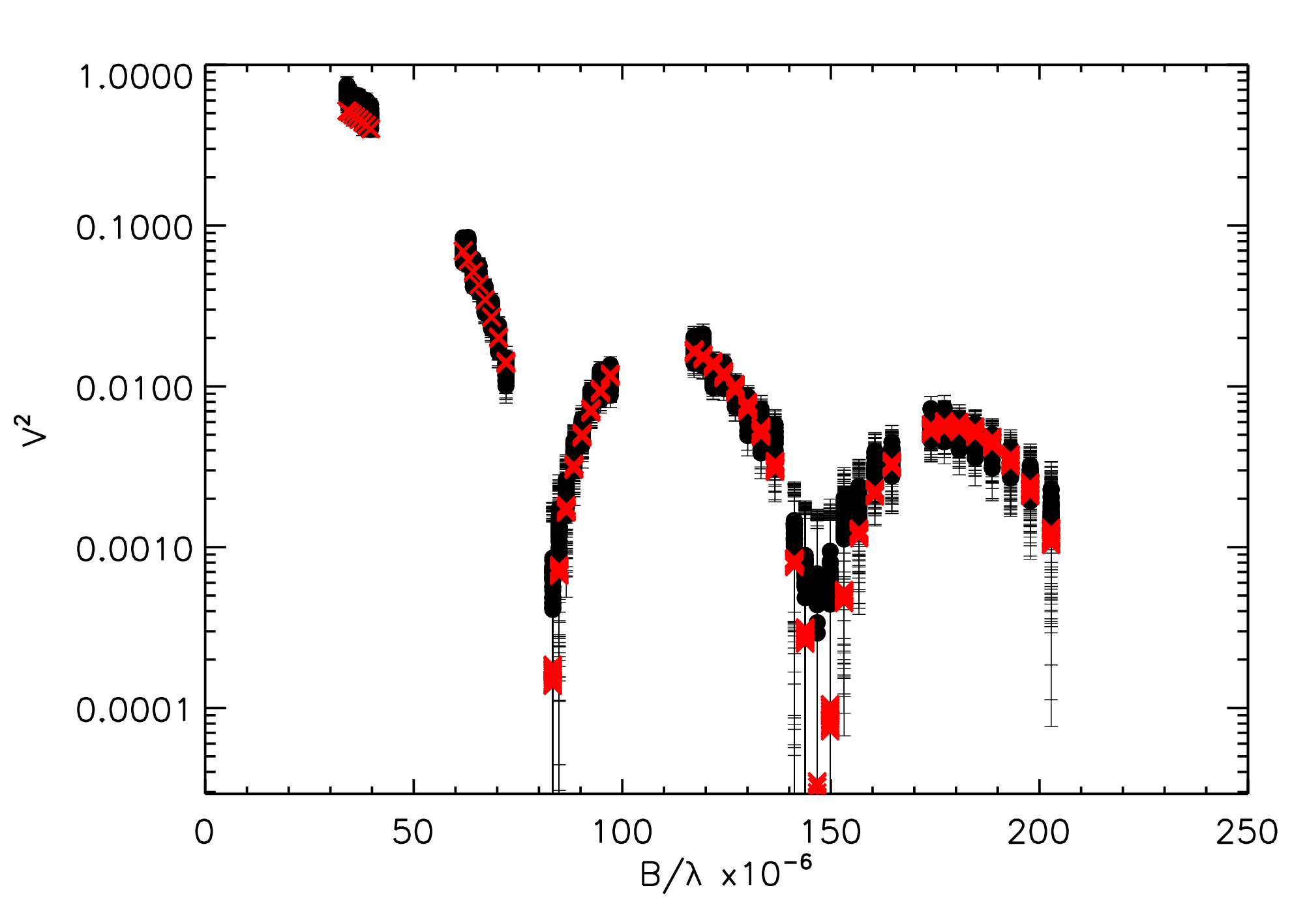}
\figsetplot{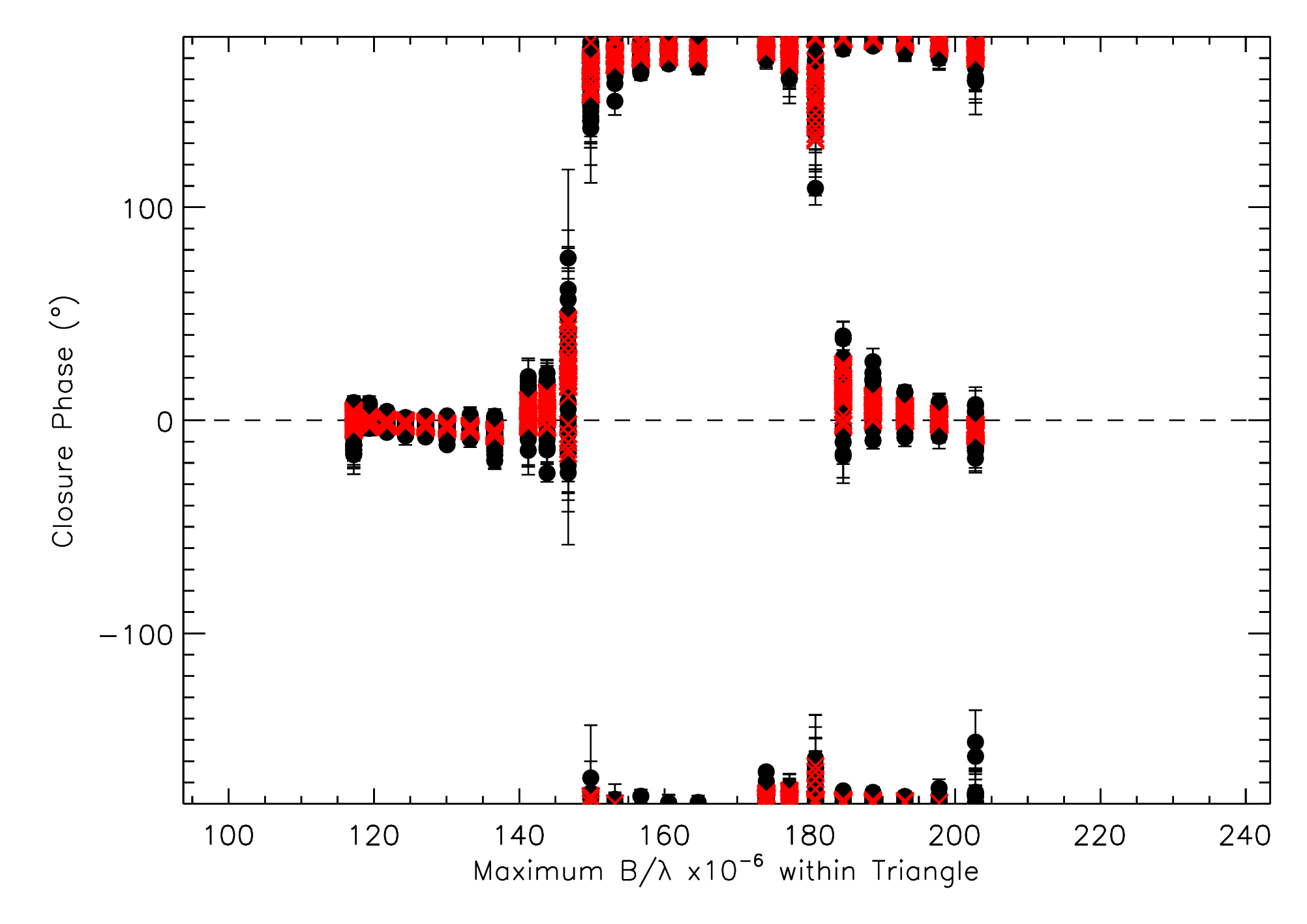}
\figsetgrpnote{Binary fit for Polaris on UT 2016Sep12. Top Row: ($u,v$) coverage (left) and $\chi^2$ map from the binary grid search (right). In the $\chi^2$ map, the red, orange, yellow, green, blue, purple, large black, and small black symbols correspond to solutions within $\Delta \chi^2$ = 1, 4, 9, 16, 25, 36, 49, and $>$50 from the minimum $\chi^2$. Bottom row: The filled black circles show the squared visibilities (left) and closure phases (right) measured with MIRC-X using the 30 second integration time. The red crosses show the best-fit binary model. }
\figsetgrpend

\figsetgrpstart
\figsetgrpnum{13.2}
\figsetgrptitle{2016Nov18_bin}
\figsetplot{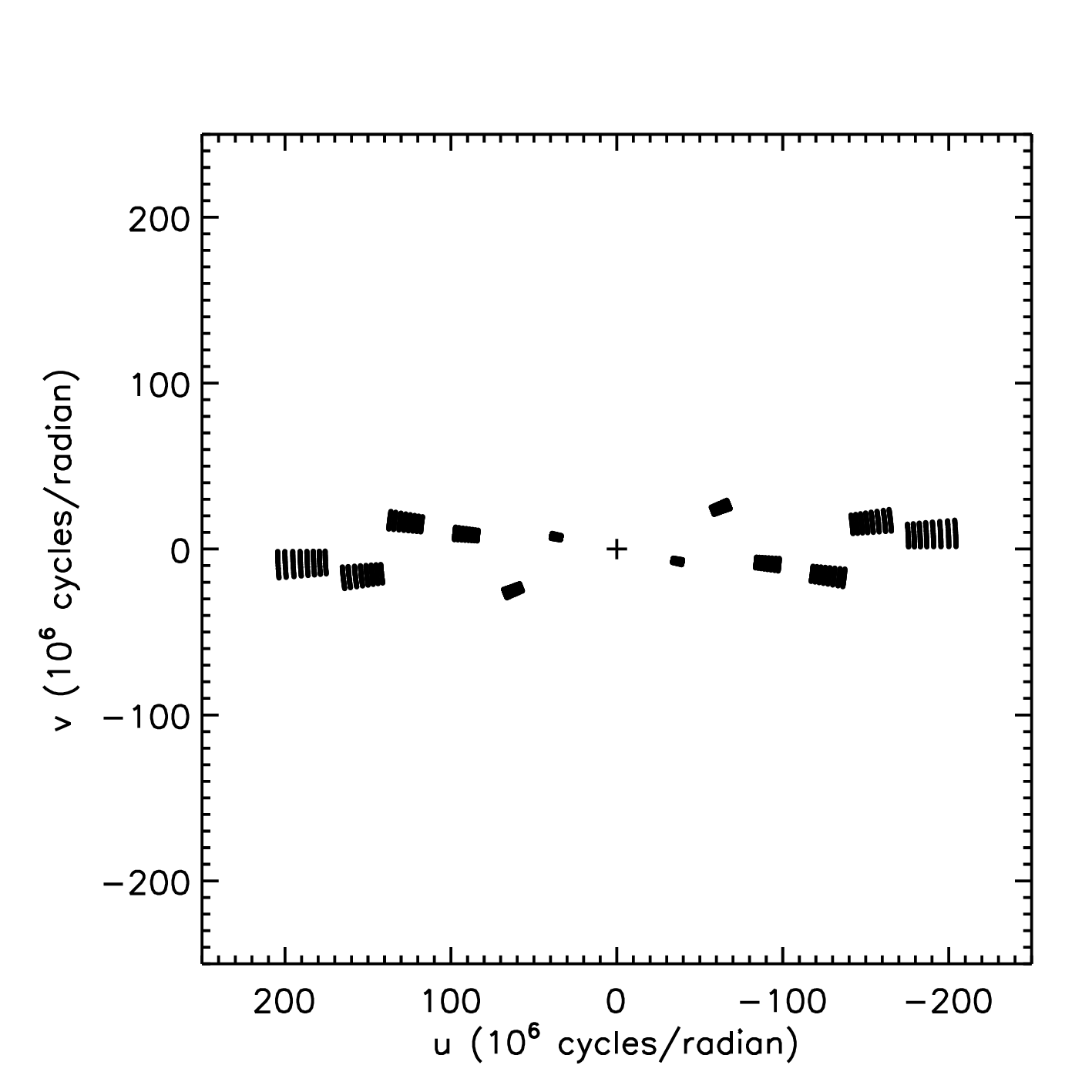}
\figsetplot{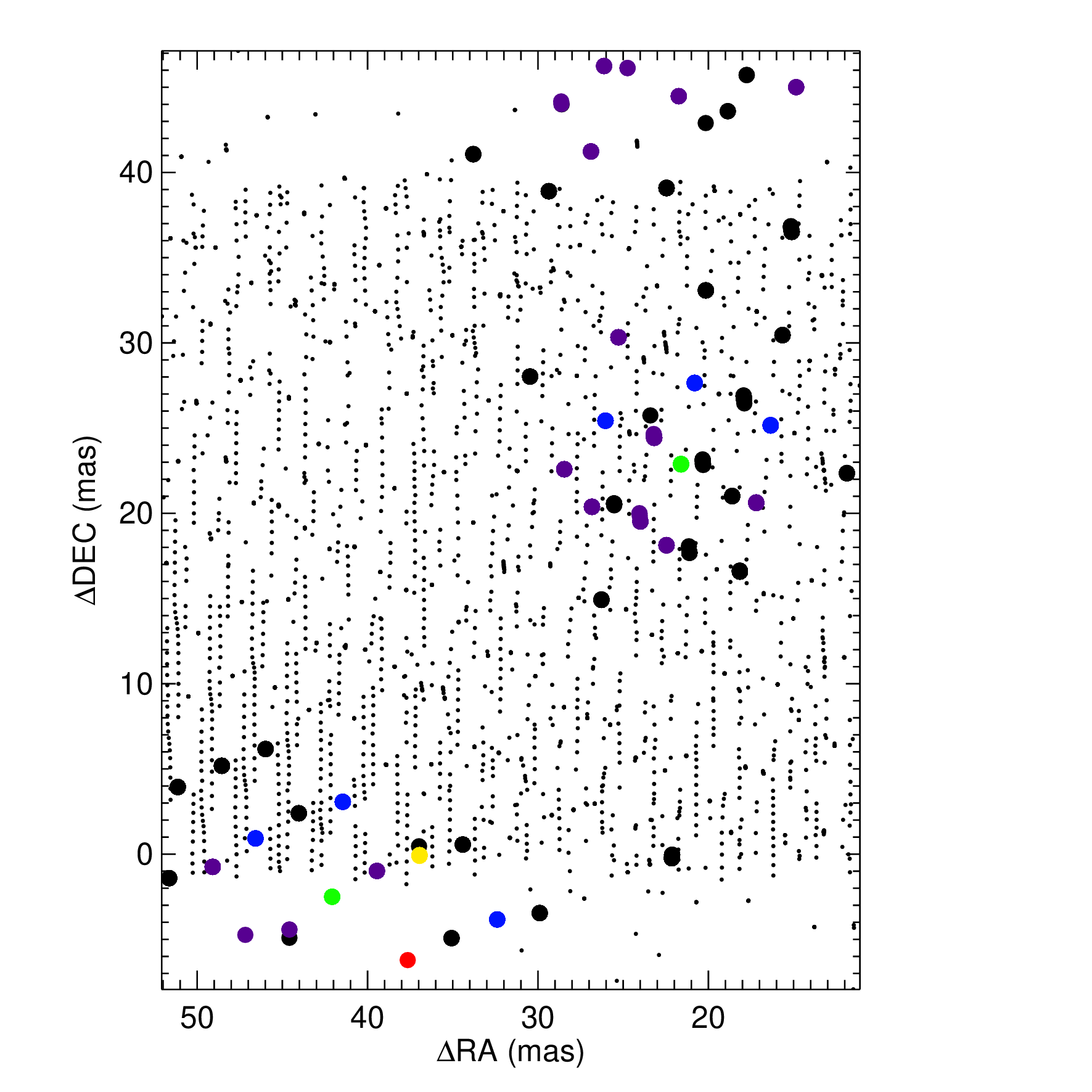}
\figsetplot{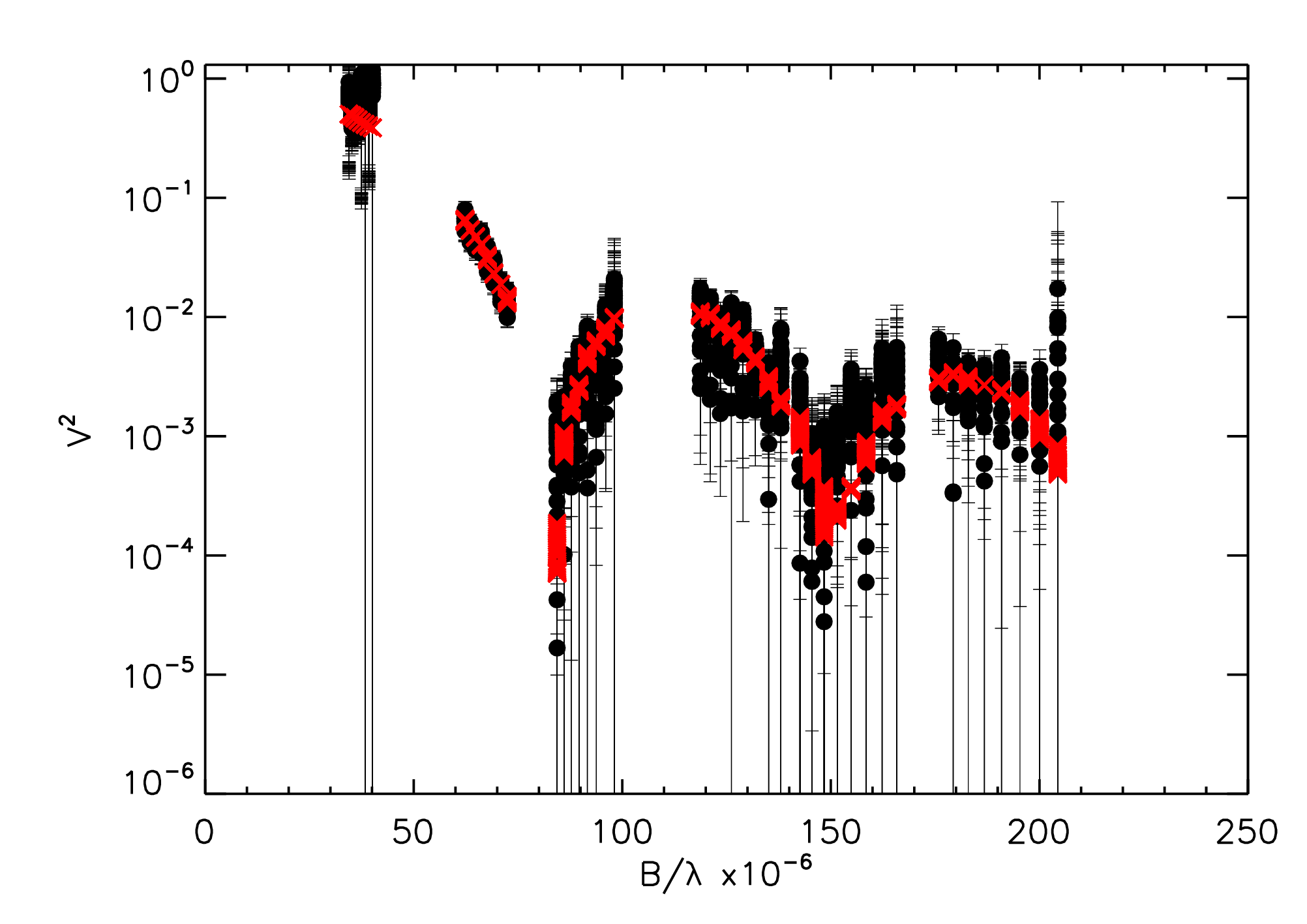}
\figsetplot{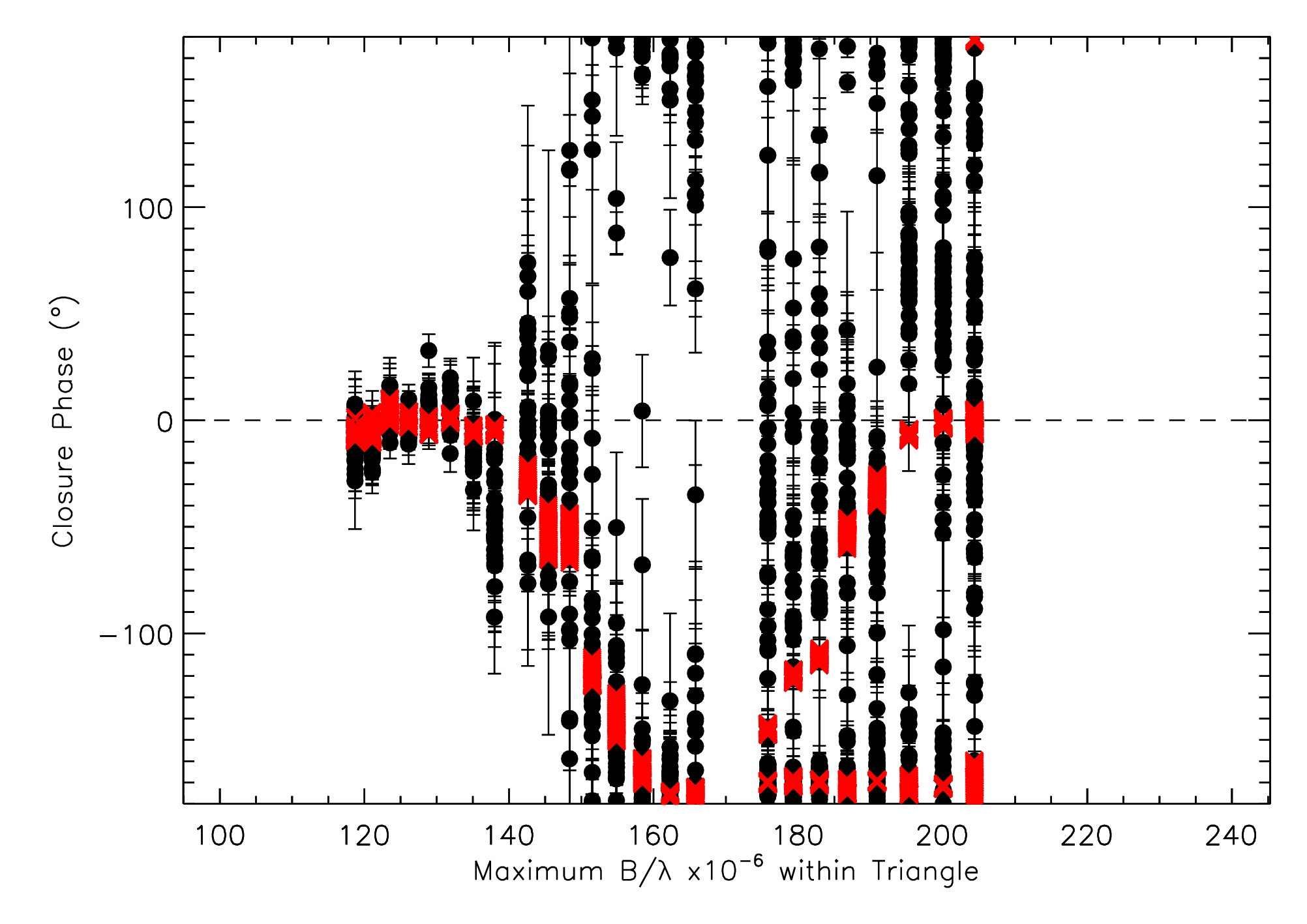}
\figsetgrpnote{Binary fit for Polaris on UT 2016Nov18. Top Row: ($u,v$) coverage (left) and $\chi^2$ map from the binary grid search (right). In the $\chi^2$ map, the red, orange, yellow, green, blue, purple, large black, and small black symbols correspond to solutions within $\Delta \chi^2$ = 1, 4, 9, 16, 25, 36, 49, and $>$50 from the minimum $\chi^2$. Bottom row: The filled black circles show the squared visibilities (left) and closure phases (right) measured with MIRC-X using the 30 second integration time. The red crosses show the best-fit binary model.}
\figsetgrpend

\figsetgrpstart
\figsetgrpnum{13.3}
\figsetgrptitle{2018Aug27_bin}
\figsetplot{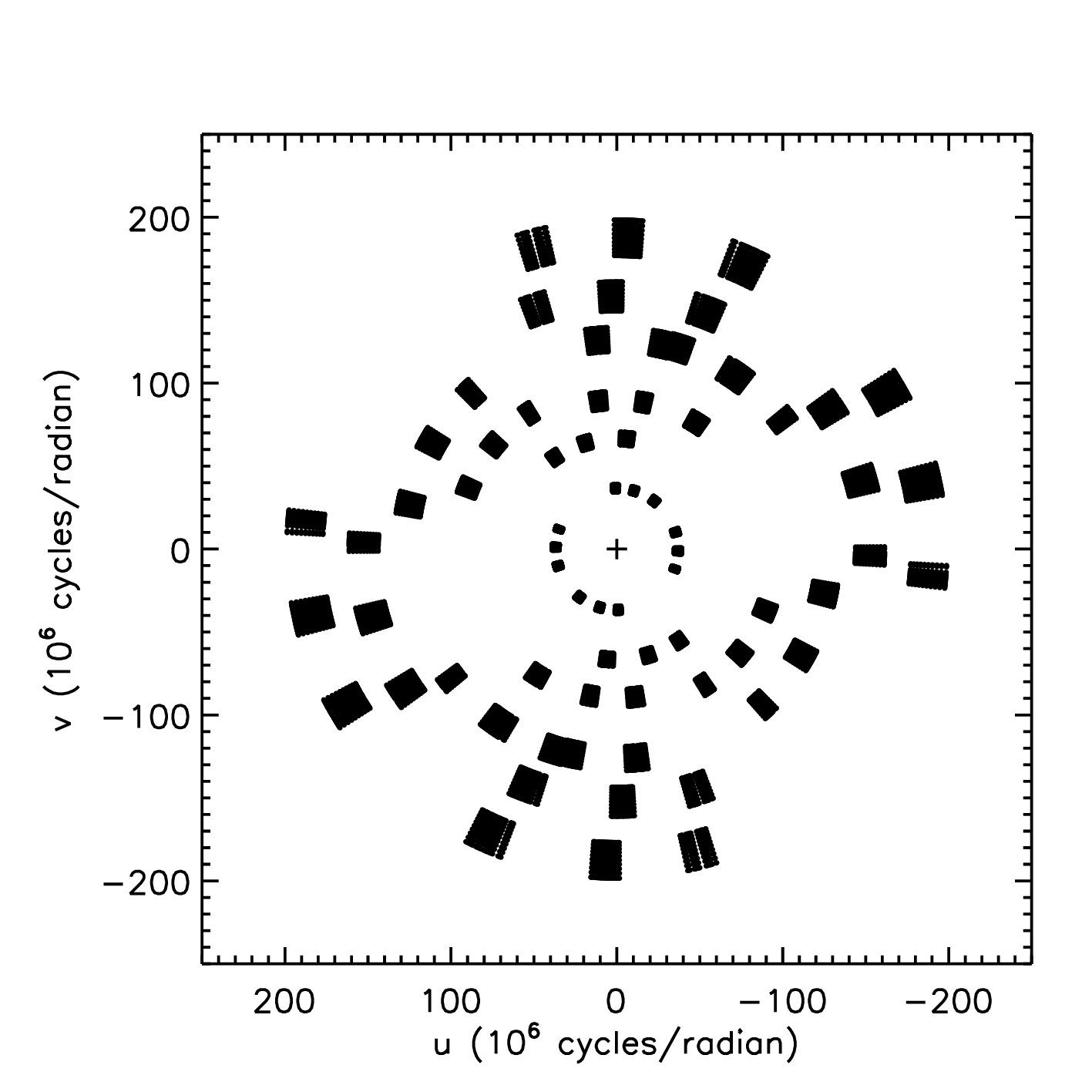}
\figsetplot{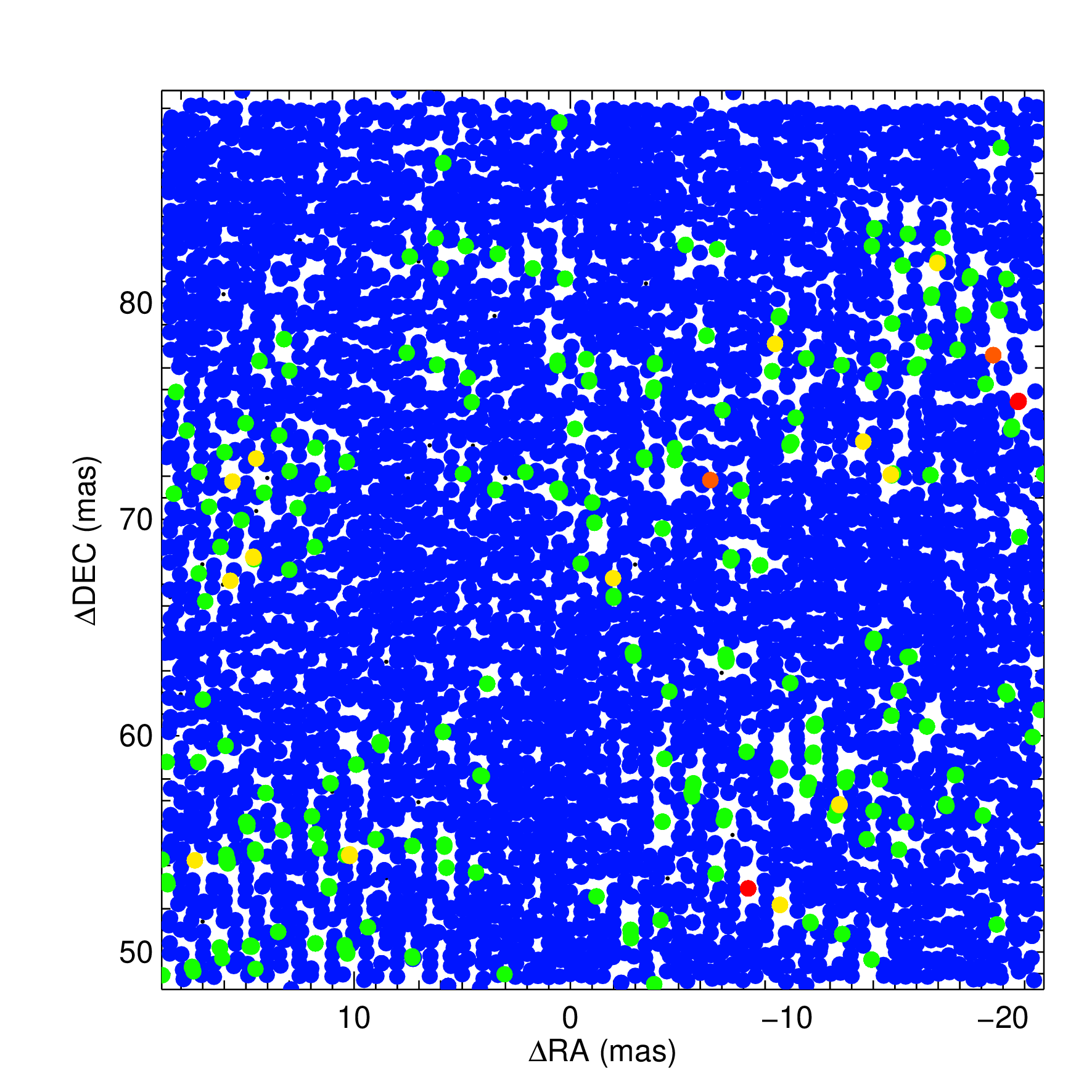}
\figsetplot{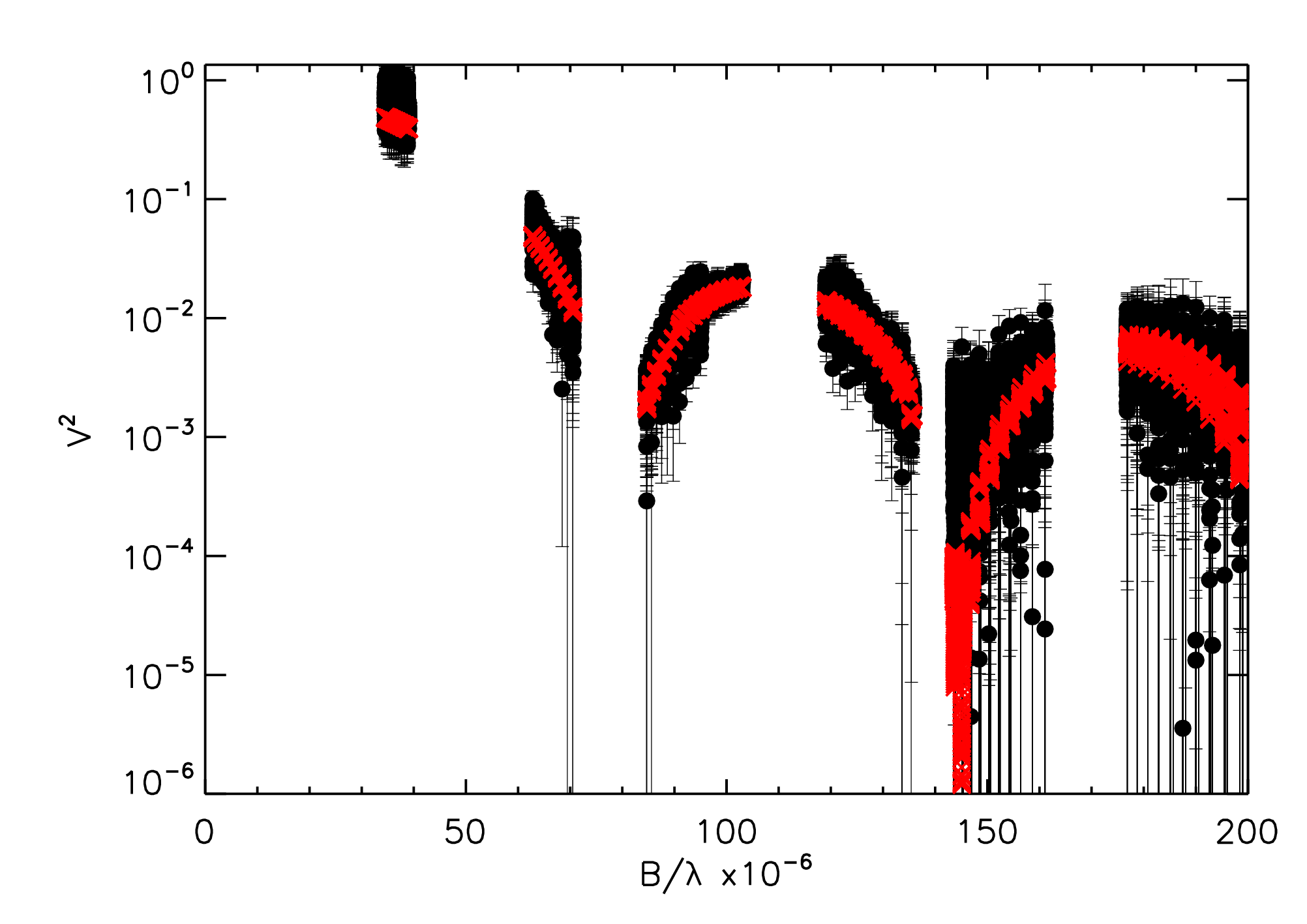}
\figsetplot{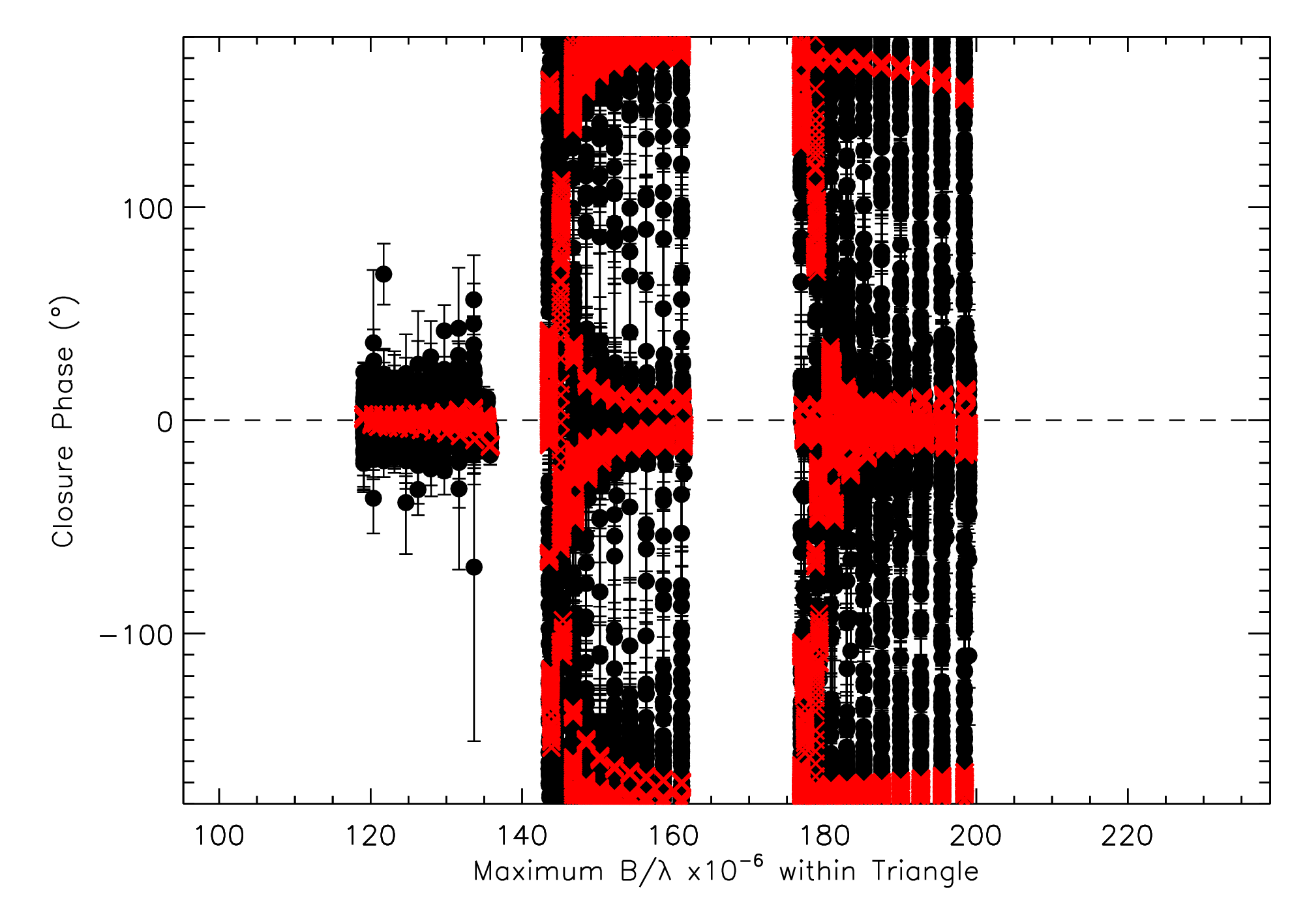}
\figsetgrpnote{Binary fit for Polaris on UT 2018Aug27. Top Row: ($u,v$) coverage (left) and $\chi^2$ map from the binary grid search (right). In the $\chi^2$ map, the red, orange, yellow, green, blue, purple, large black, and small black symbols correspond to solutions within $\Delta \chi^2$ = 1, 4, 9, 16, 25, 36, 49, and $>$50 from the minimum $\chi^2$. Bottom row: The filled black circles show the squared visibilities (left) and closure phases (right) measured with MIRC-X using the 30 second integration time. The red crosses show the best-fit binary model.}
\figsetgrpend

\figsetgrpstart
\figsetgrpnum{13.4}
\figsetgrptitle{2019Apr09_bin}
\figsetplot{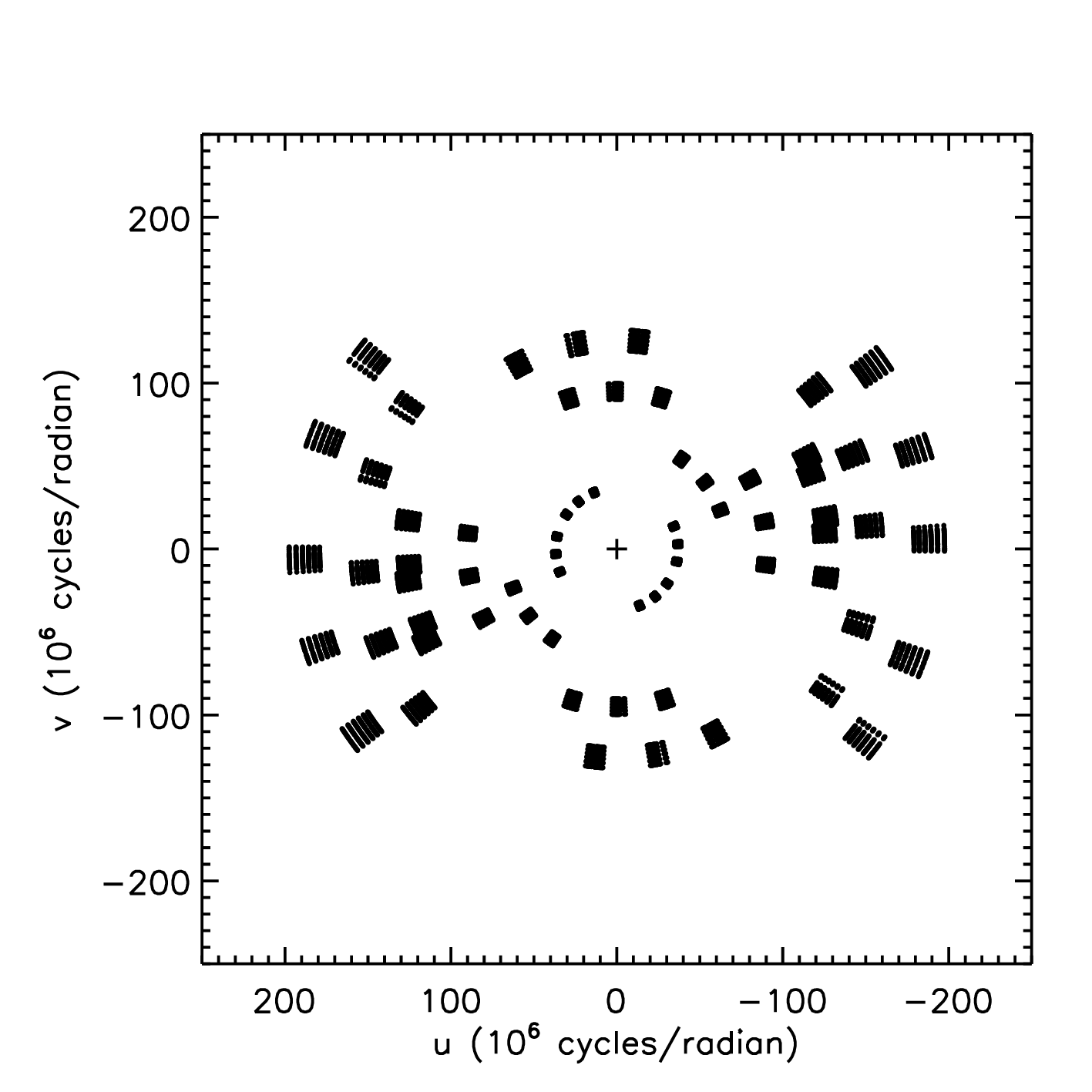}
\figsetplot{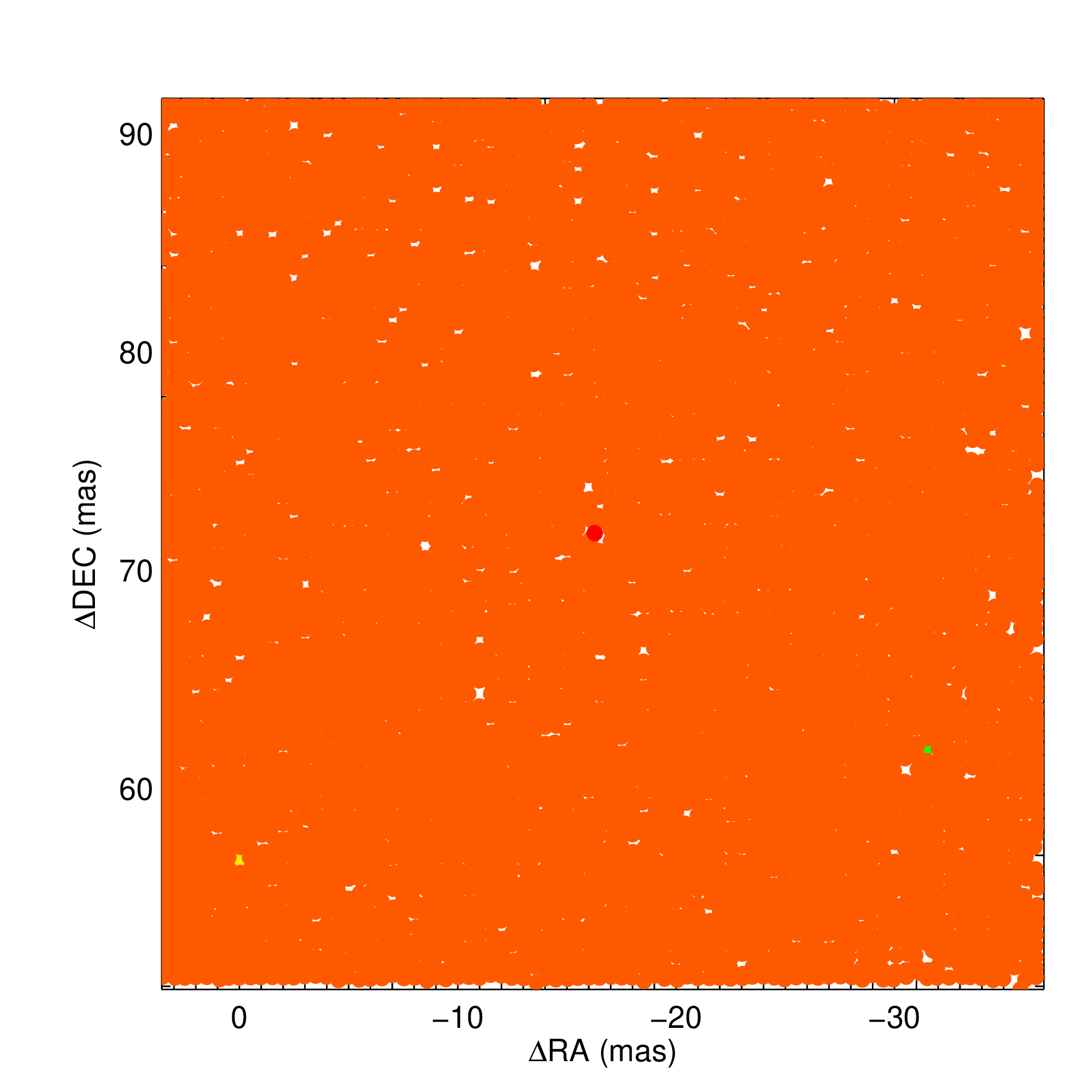}
\figsetplot{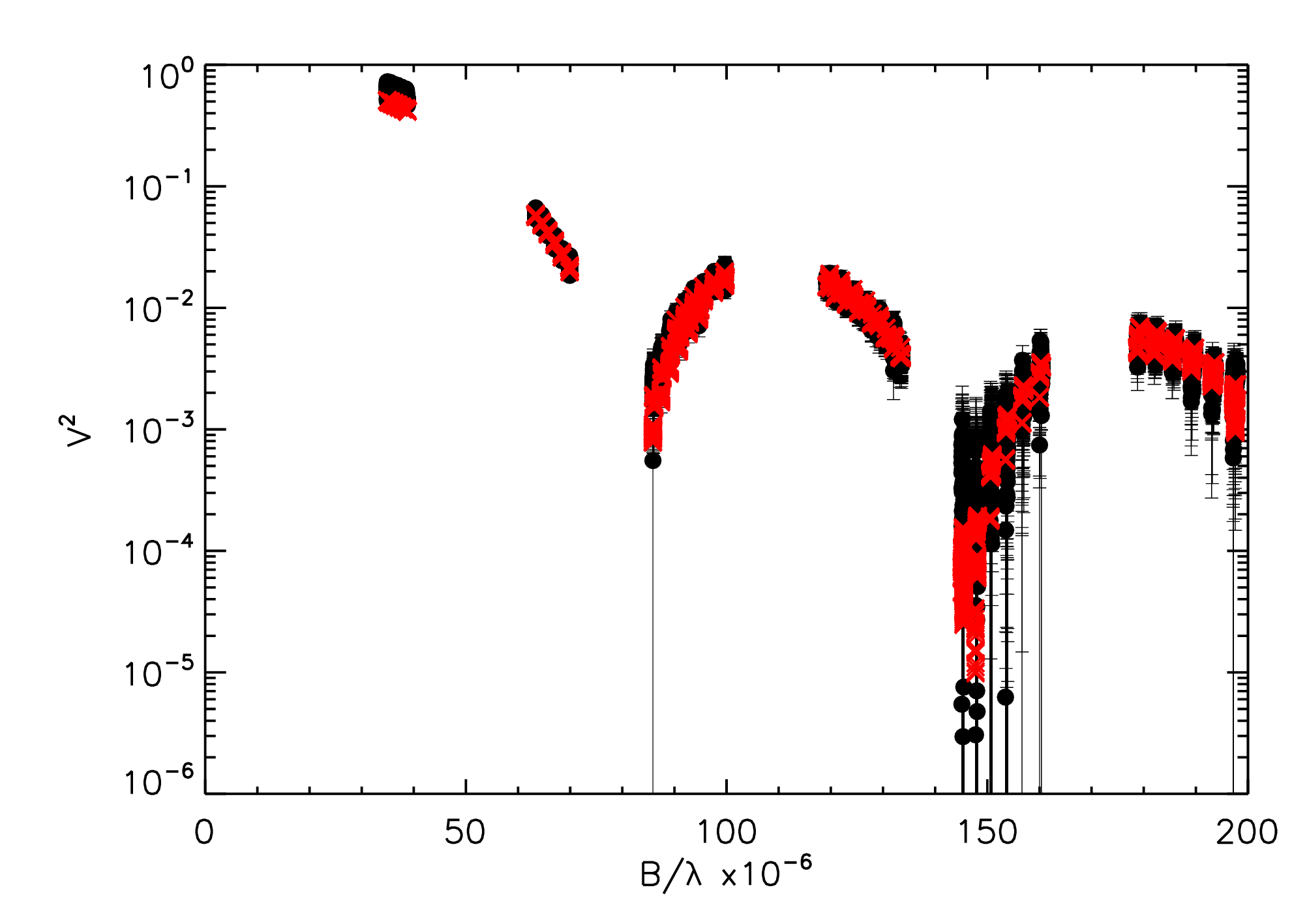}
\figsetplot{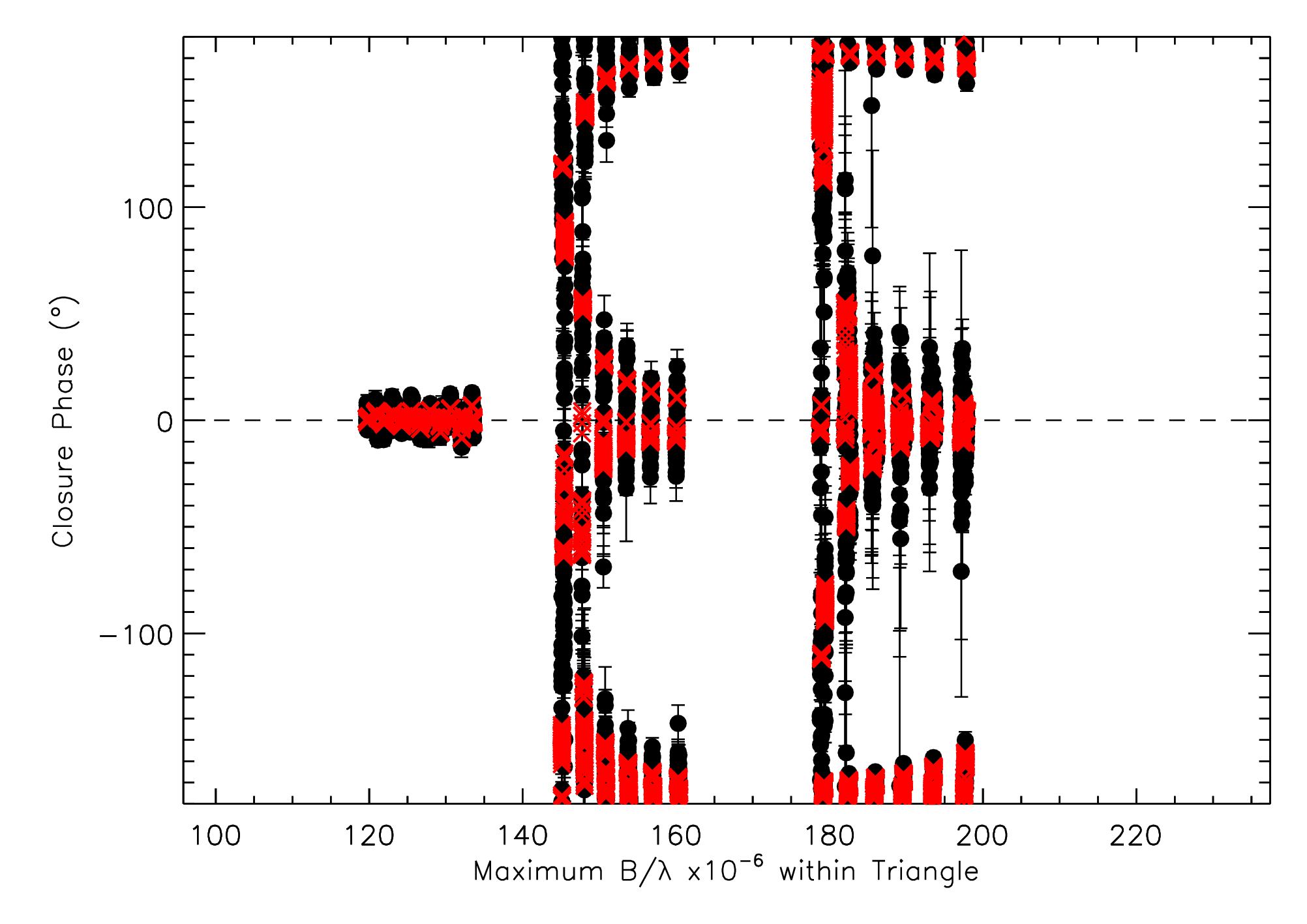}
\figsetgrpnote{Binary fit for Polaris on UT 2019Apr09. Top Row: ($u,v$) coverage (left) and $\chi^2$ map from the binary grid search (right). In the $\chi^2$ map, the red, orange, yellow, green, blue, purple, large black, and small black symbols correspond to solutions within $\Delta \chi^2$ = 1, 4, 9, 16, 25, 36, 49, and $>$50 from the minimum $\chi^2$. Bottom row: The filled black circles show the squared visibilities (left) and closure phases (right) measured with MIRC-X using the 30 second integration time. The red crosses show the best-fit binary model.}
\figsetgrpend

\figsetgrpstart
\figsetgrpnum{13.5}
\figsetgrptitle{2019Sep02_bin}
\figsetplot{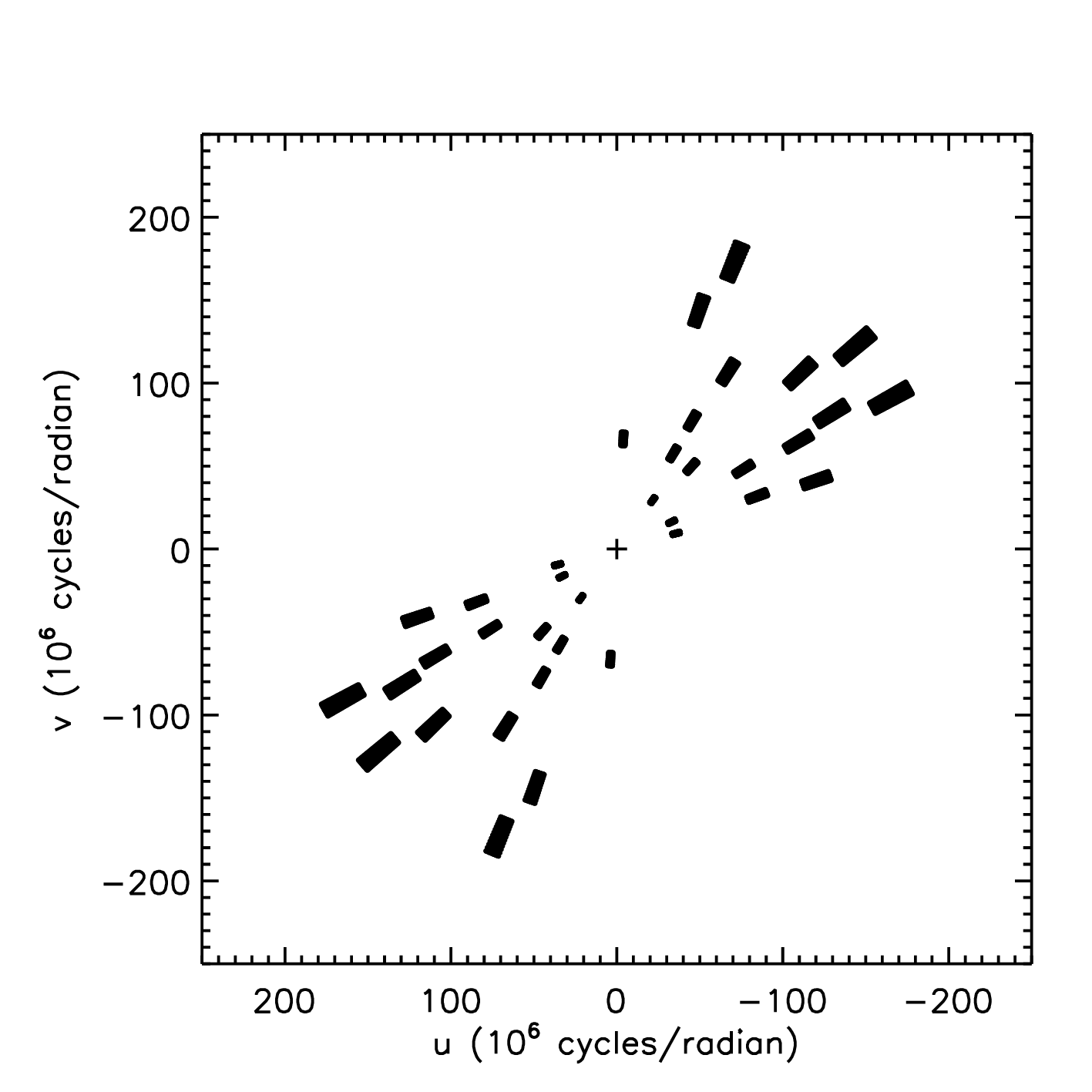}
\figsetplot{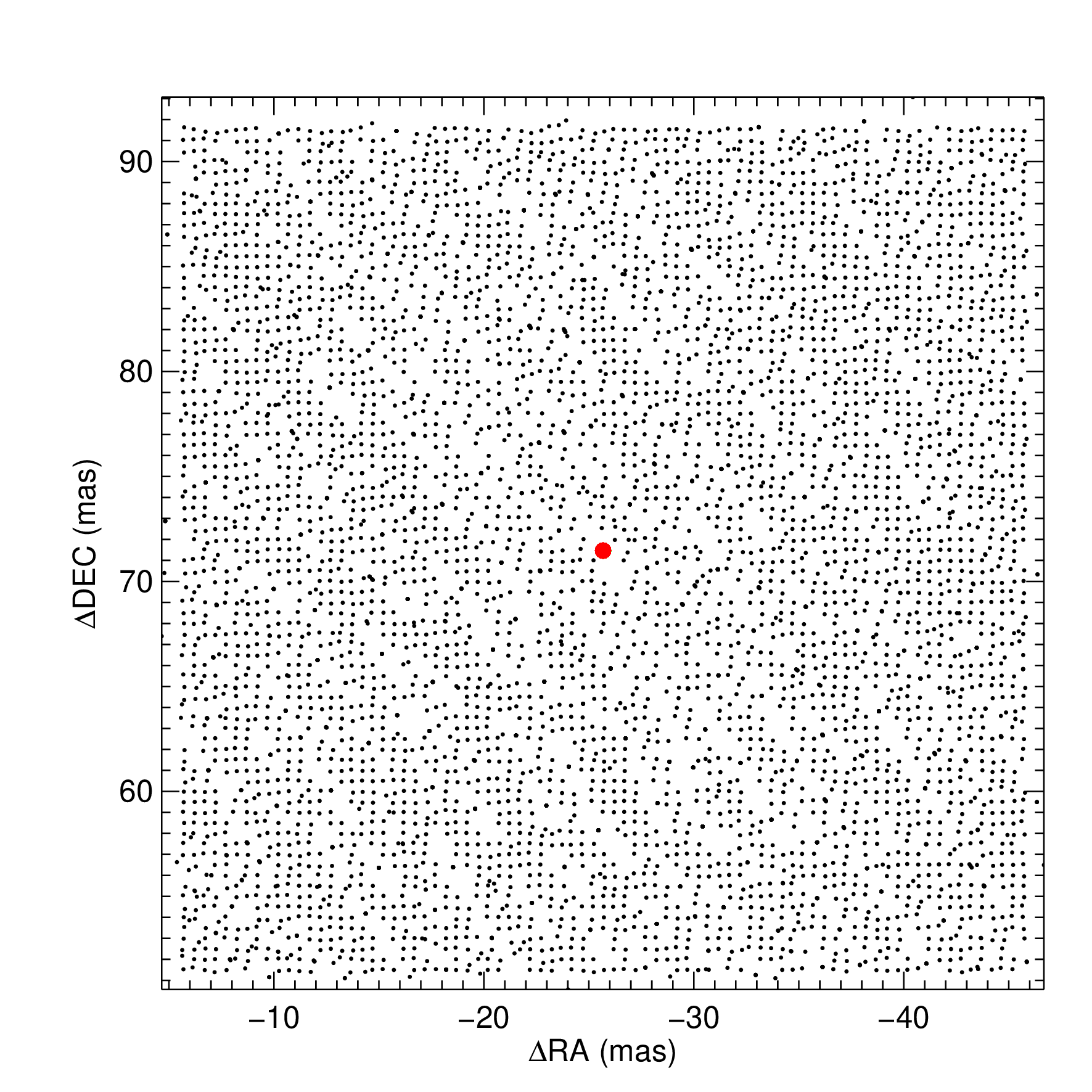}
\figsetplot{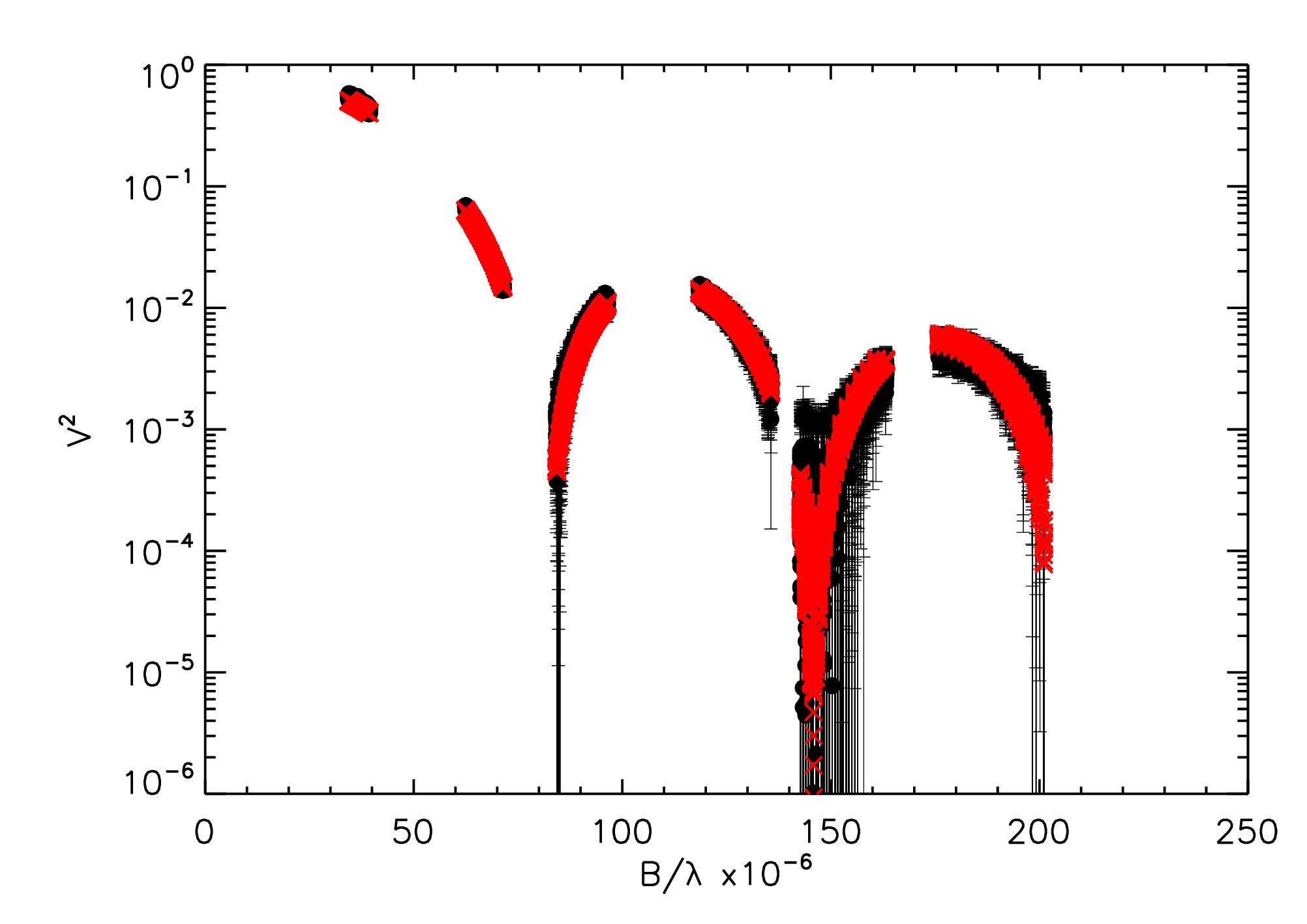}
\figsetplot{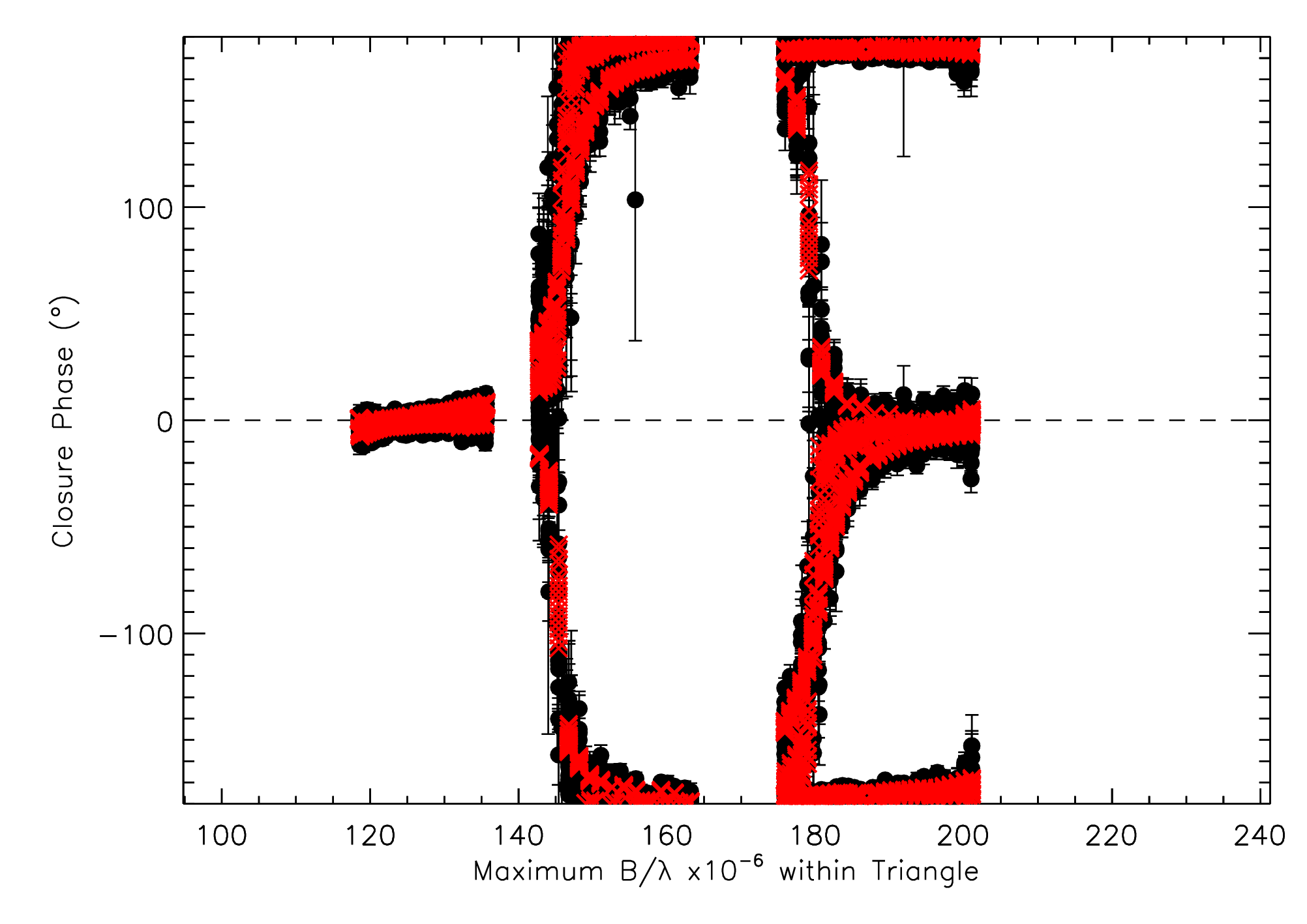}
\figsetgrpnote{Binary fit for Polaris on UT 2019Sep02. Top Row: ($u,v$) coverage (left) and $\chi^2$ map from the binary grid search (right). In the $\chi^2$ map, the red, orange, yellow, green, blue, purple, large black, and small black symbols correspond to solutions within $\Delta \chi^2$ = 1, 4, 9, 16, 25, 36, 49, and $>$50 from the minimum $\chi^2$. Bottom row: The filled black circles show the squared visibilities (left) and closure phases (right) measured with MIRC-X using the 30 second integration time. The red crosses show the best-fit binary model.}
\figsetgrpend

\figsetgrpstart
\figsetgrpnum{13.6}
\figsetgrptitle{2021Apr02_bin}
\figsetplot{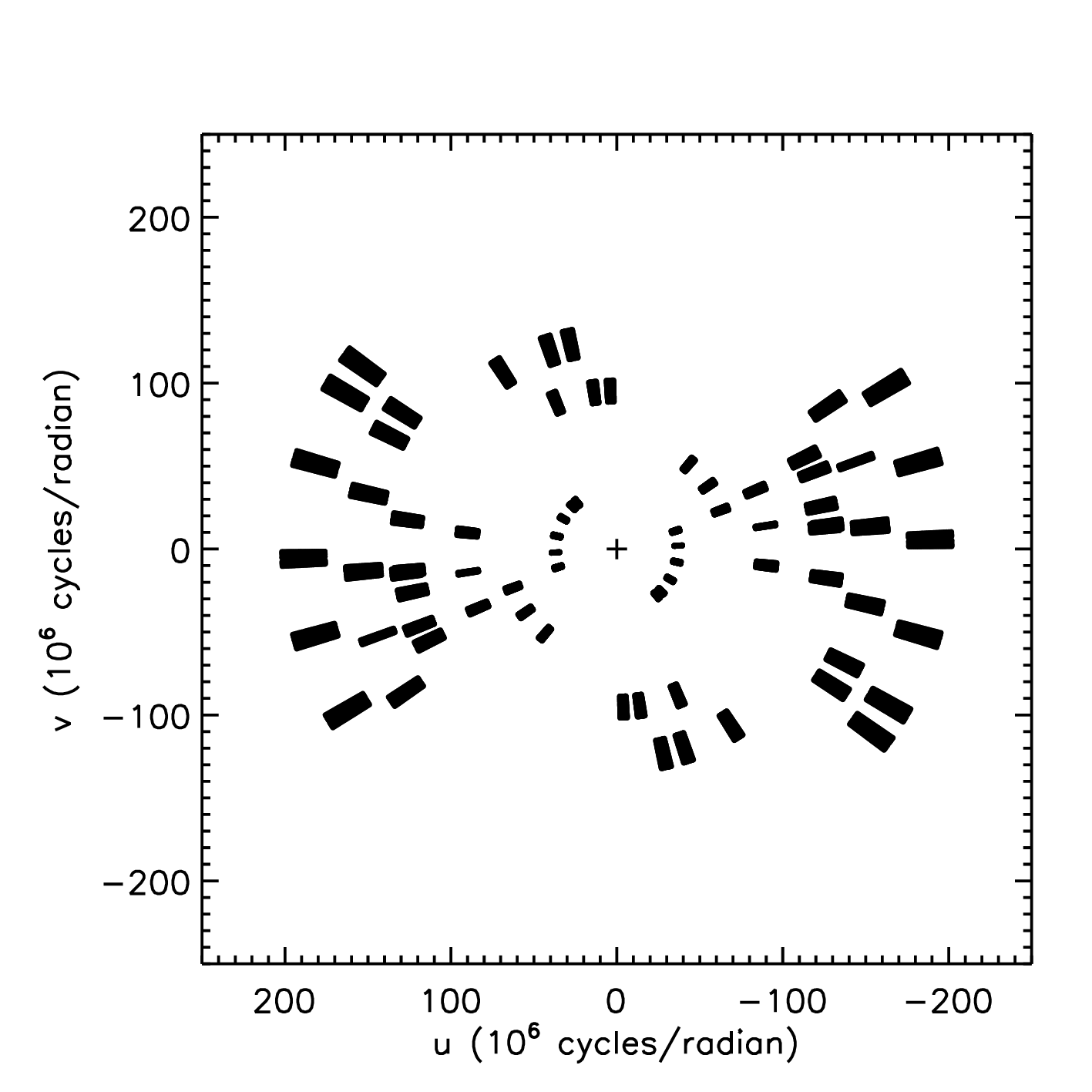}
\figsetplot{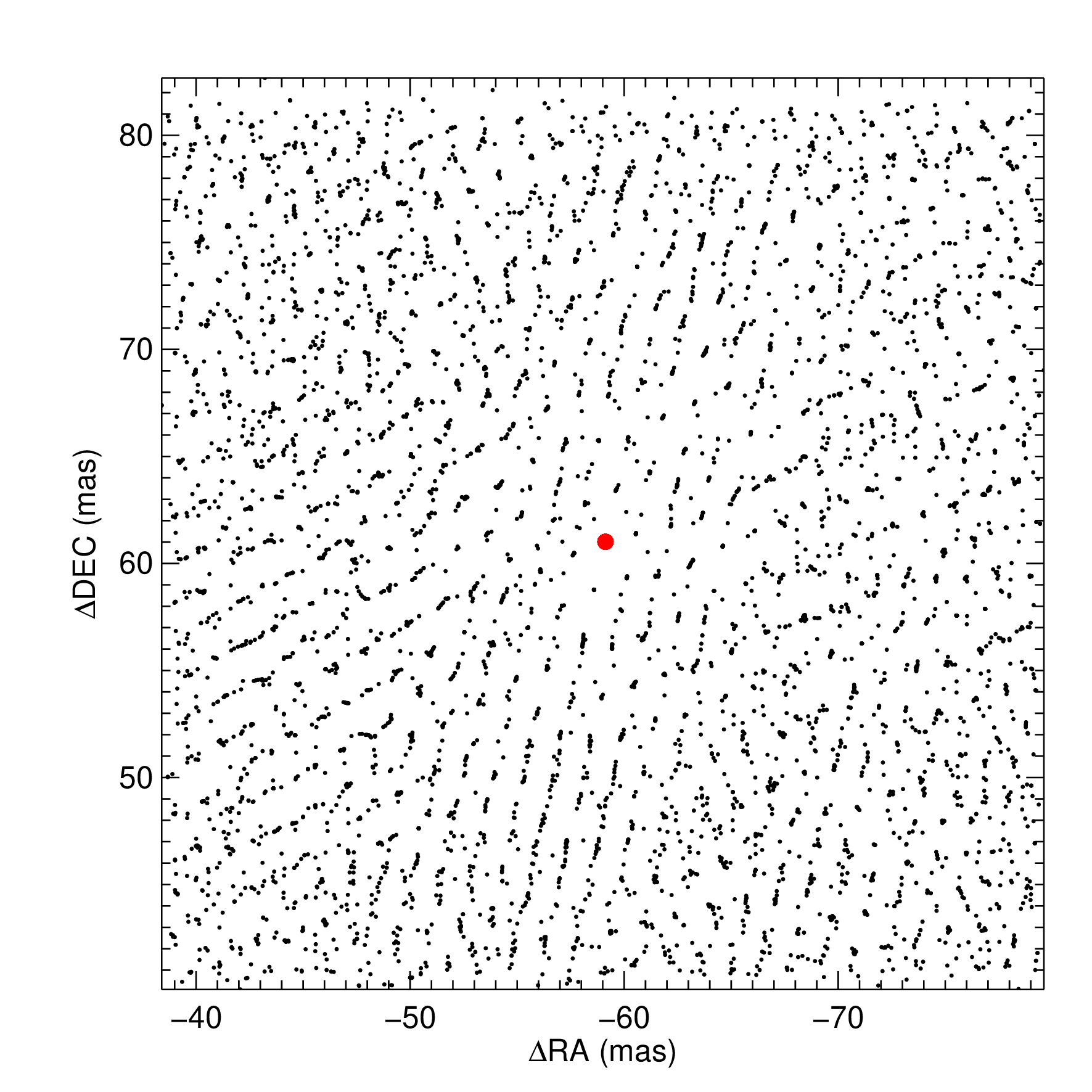}
\figsetplot{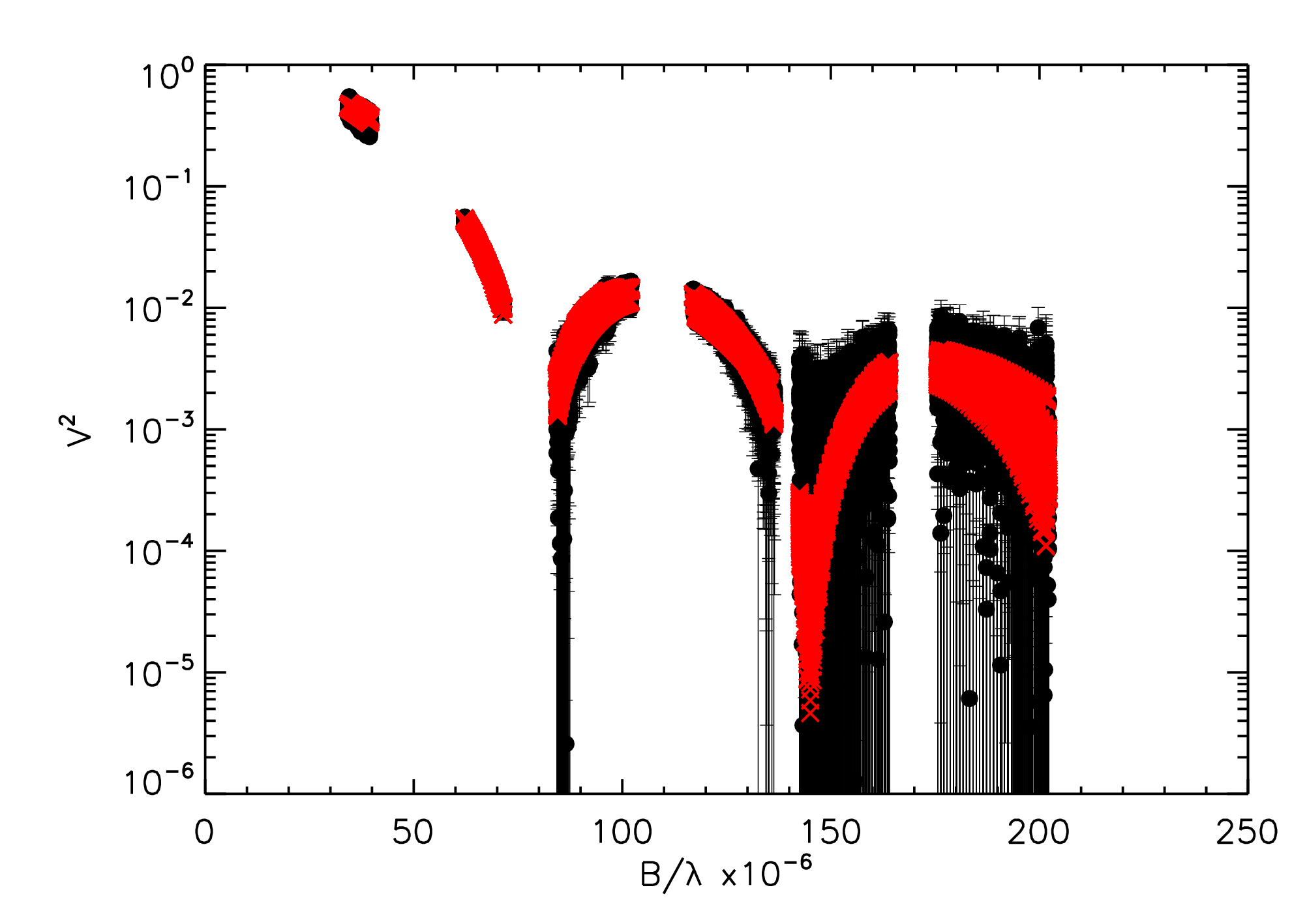}
\figsetplot{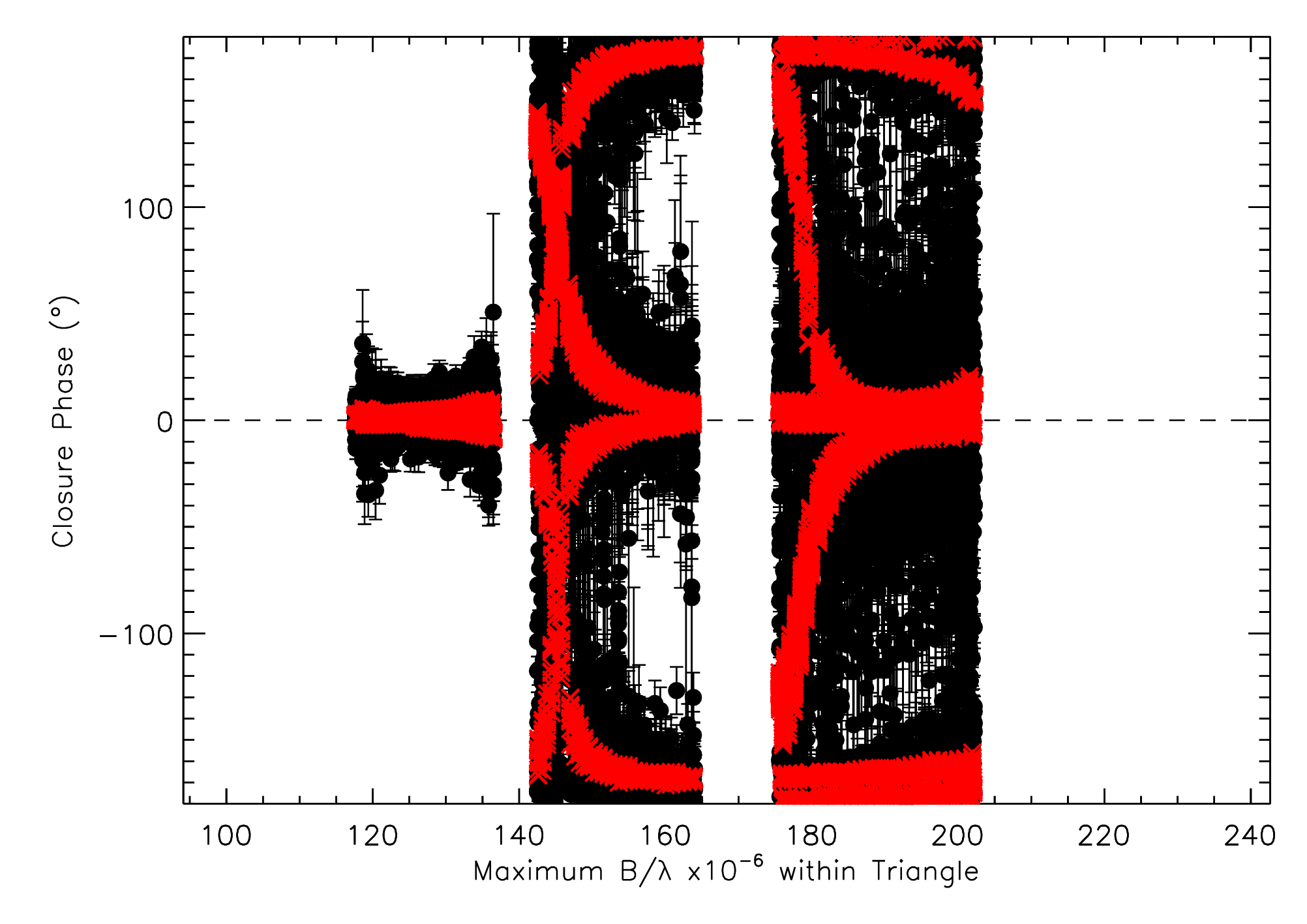}
\figsetgrpnote{Binary fit for Polaris on UT 2021Apr02. Top Row: ($u,v$) coverage (left) and $\chi^2$ map from the binary grid search (right). In the $\chi^2$ map, the red, orange, yellow, green, blue, purple, large black, and small black symbols correspond to solutions within $\Delta \chi^2$ = 1, 4, 9, 16, 25, 36, 49, and $>$50 from the minimum $\chi^2$. Bottom row: The filled black circles show the squared visibilities (left) and closure phases (right) measured with MIRC-X using the 30 second integration time. The red crosses show the best-fit binary model.}
\figsetgrpend

\figsetgrpstart
\figsetgrpnum{13.7}
\figsetgrptitle{2021Apr03_bin}
\figsetplot{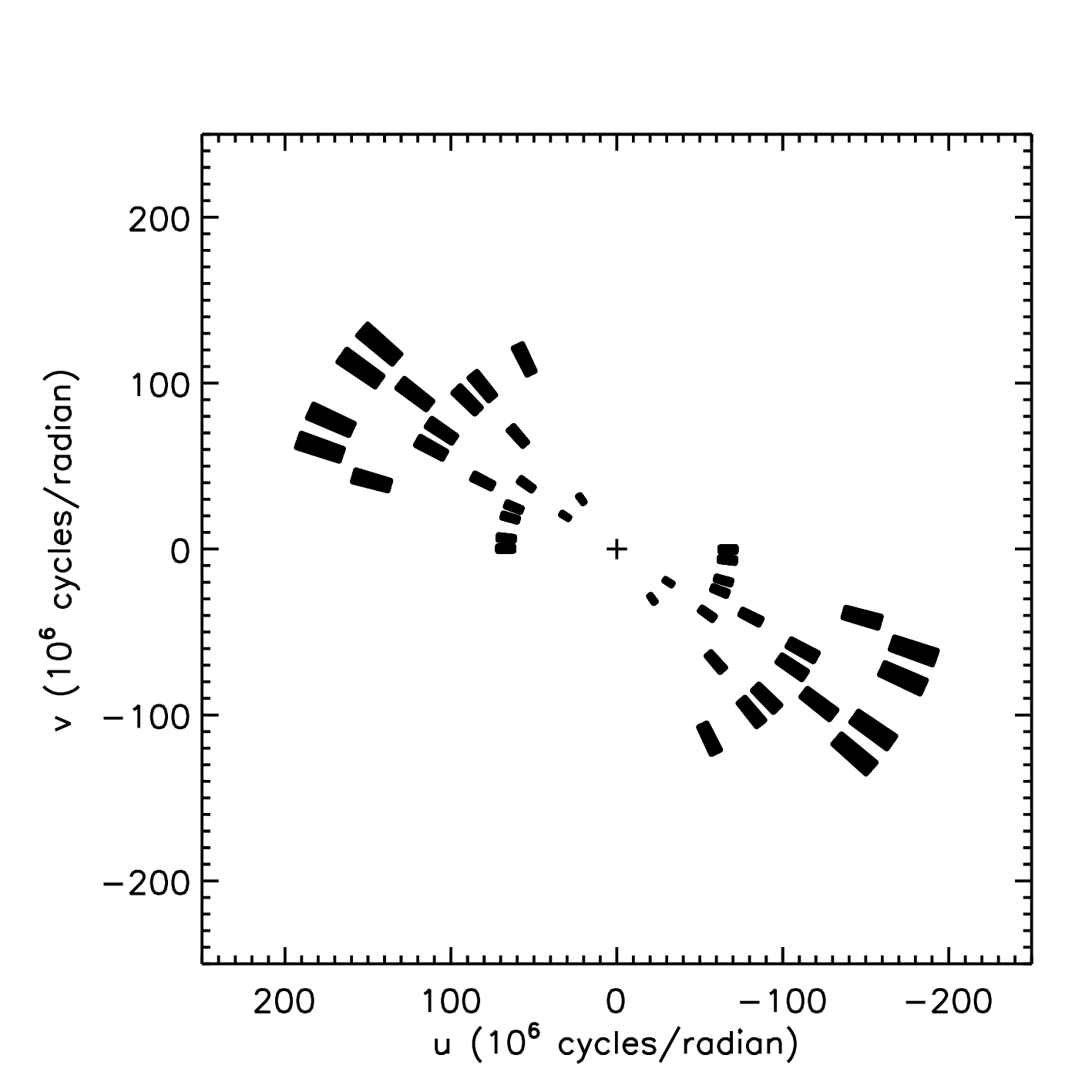}
\figsetplot{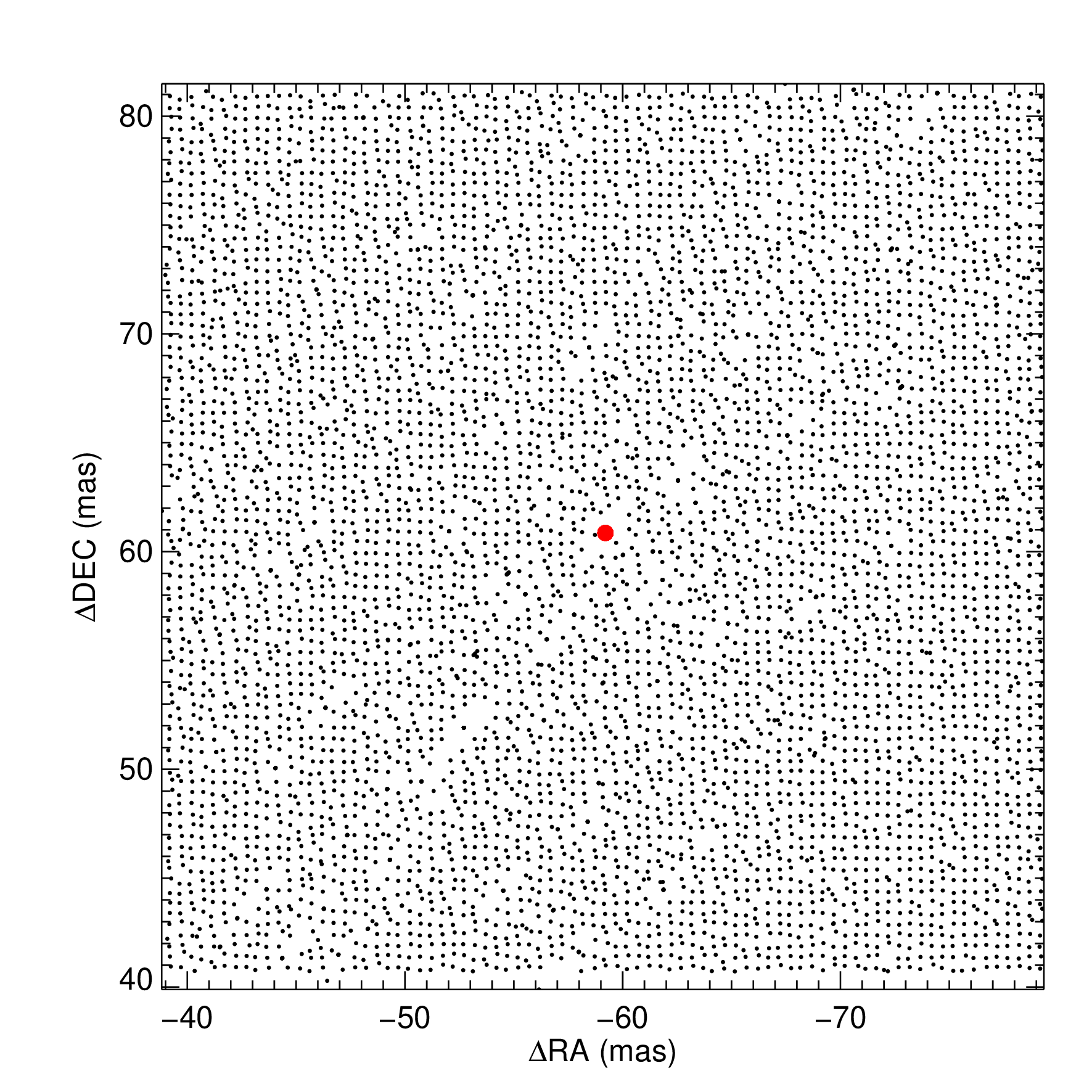}
\figsetplot{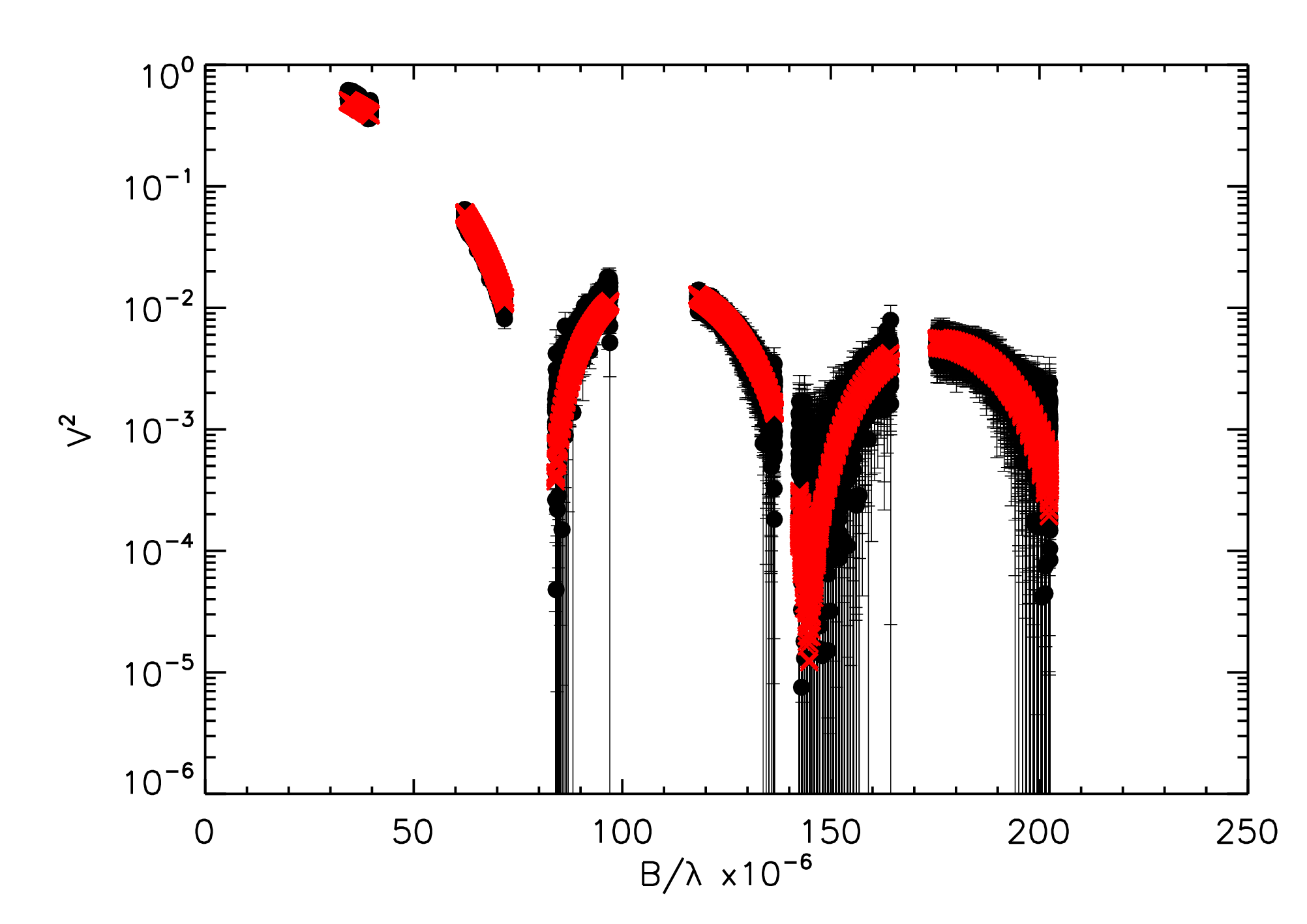}
\figsetplot{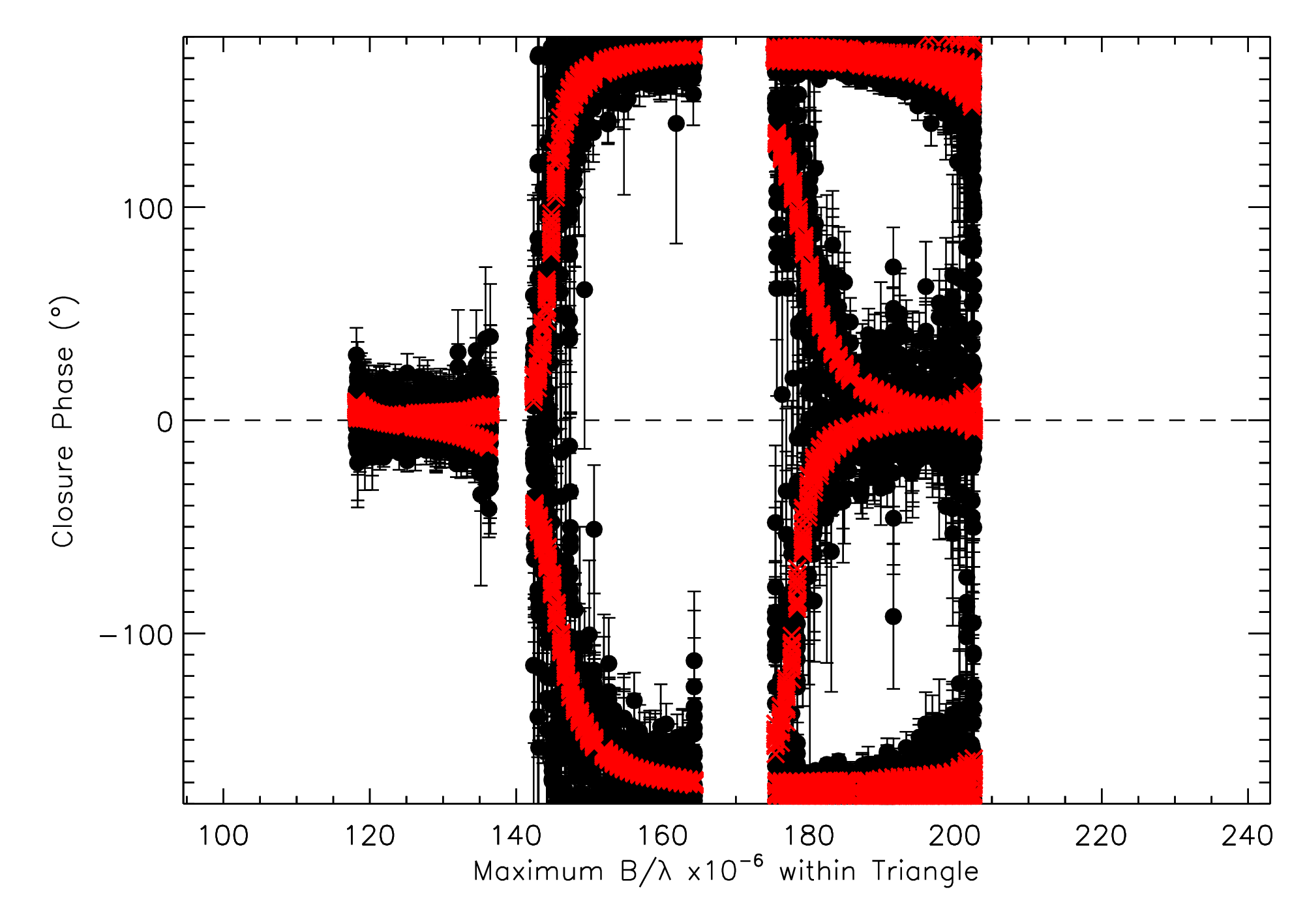}
\figsetgrpnote{Binary fit for Polaris on UT 2021Apr03. Top Row: ($u,v$) coverage (left) and $\chi^2$ map from the binary grid search (right). In the $\chi^2$ map, the red, orange, yellow, green, blue, purple, large black, and small black symbols correspond to solutions within $\Delta \chi^2$ = 1, 4, 9, 16, 25, 36, 49, and $>$50 from the minimum $\chi^2$. Bottom row: The filled black circles show the squared visibilities (left) and closure phases (right) measured with MIRC-X using the 30 second integration time. The red crosses show the best-fit binary model.}
\figsetgrpend

\figsetgrpstart
\figsetgrpnum{13.8}
\figsetgrptitle{2021Apr04_bin}
\figsetplot{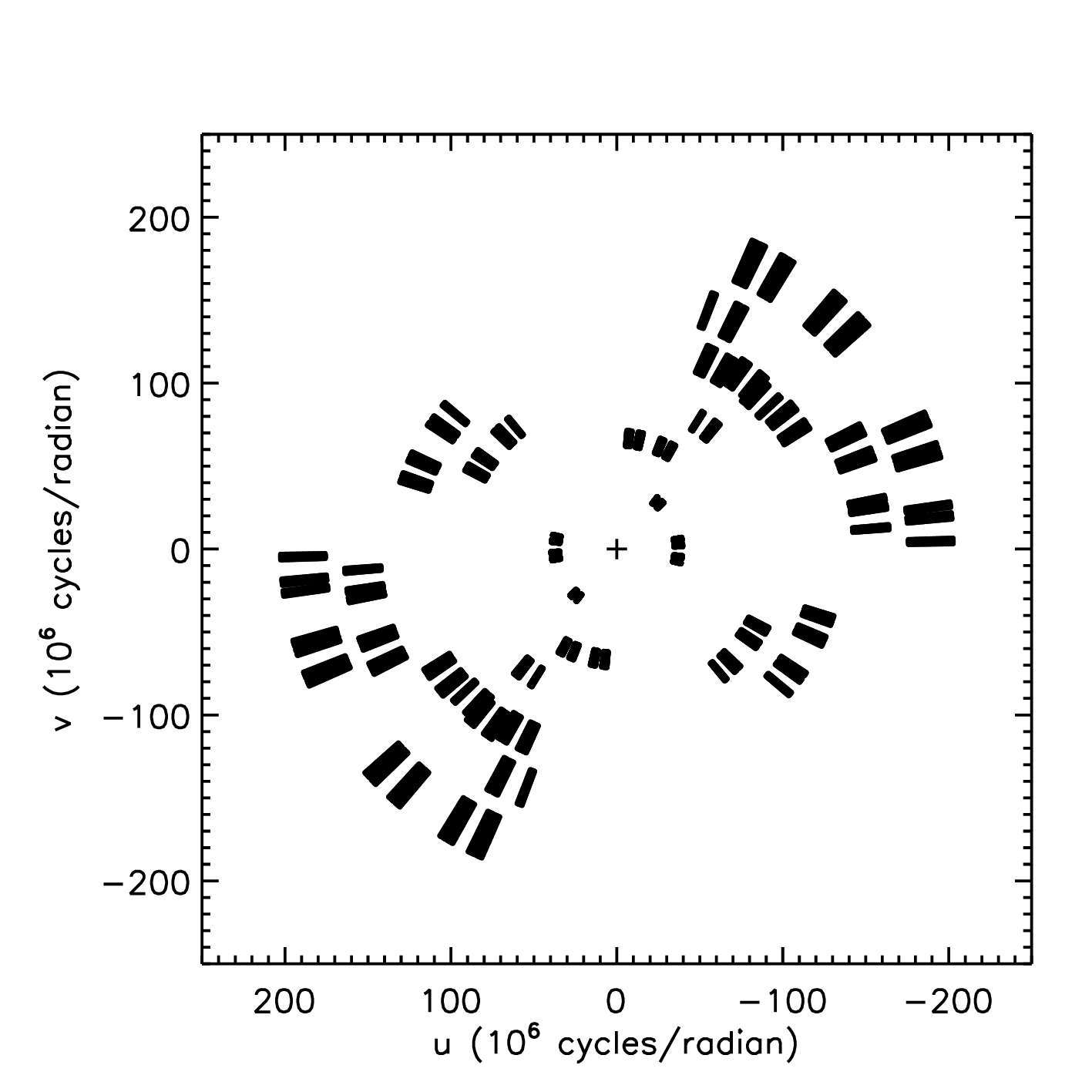}
\figsetplot{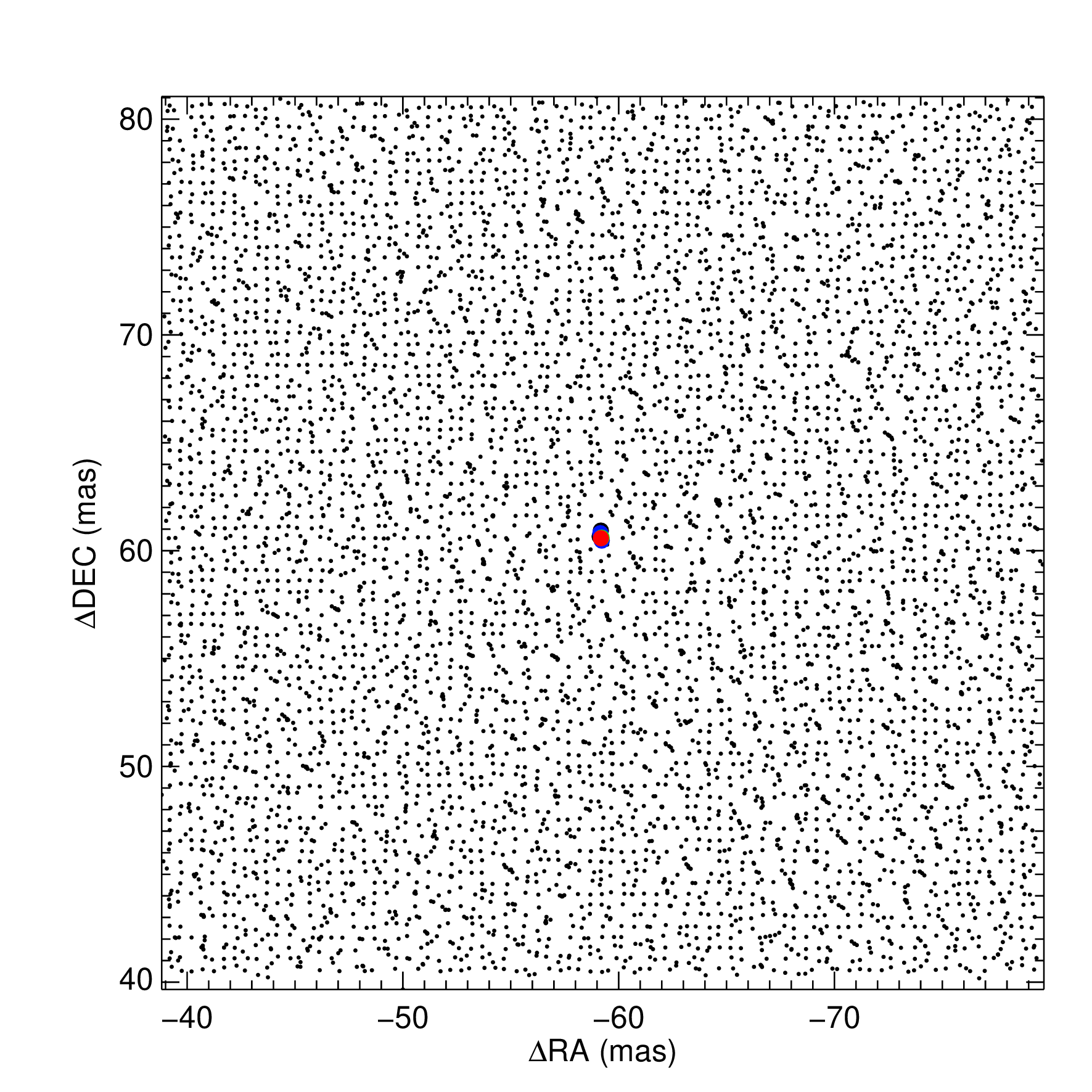}
\figsetplot{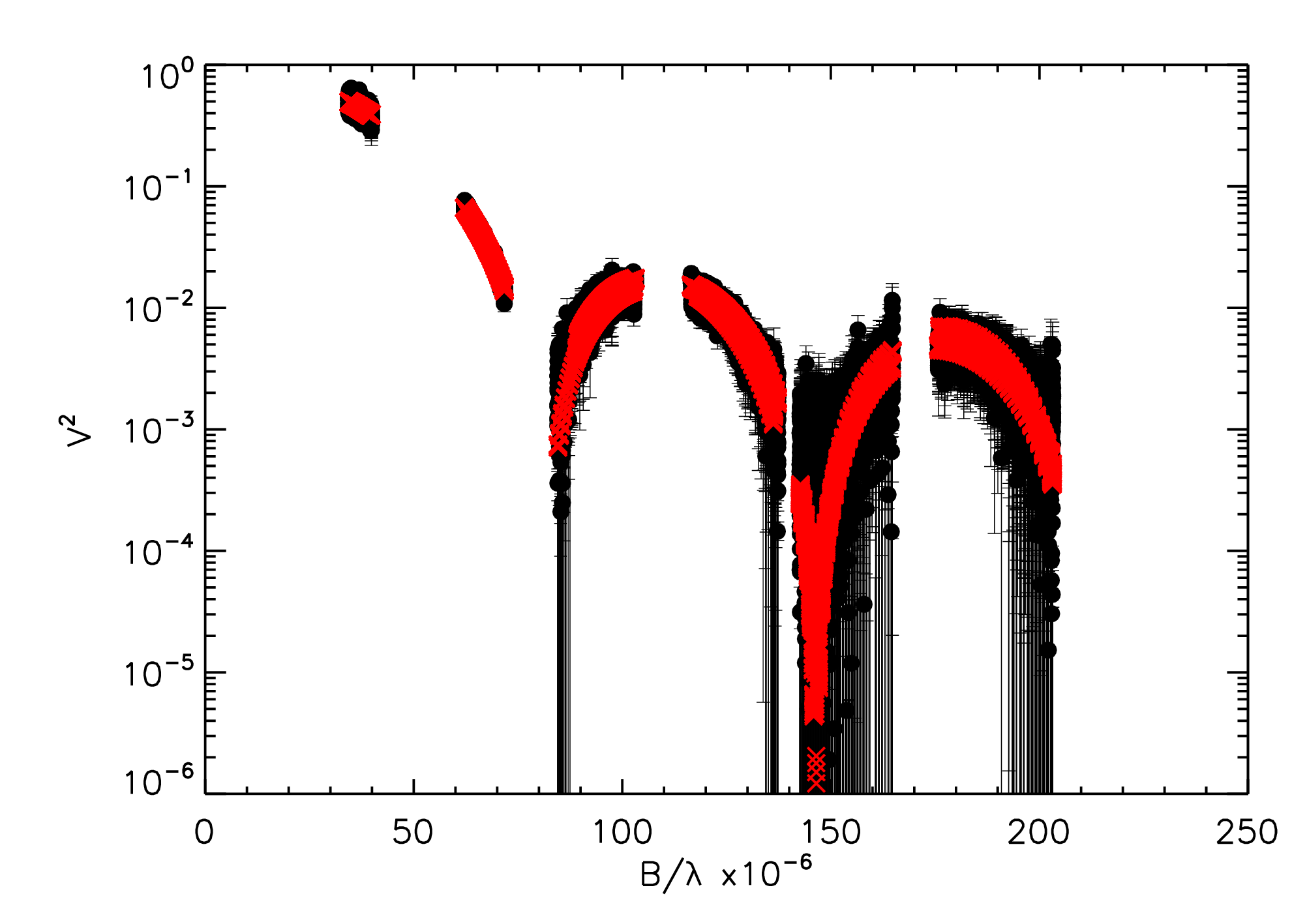}
\figsetplot{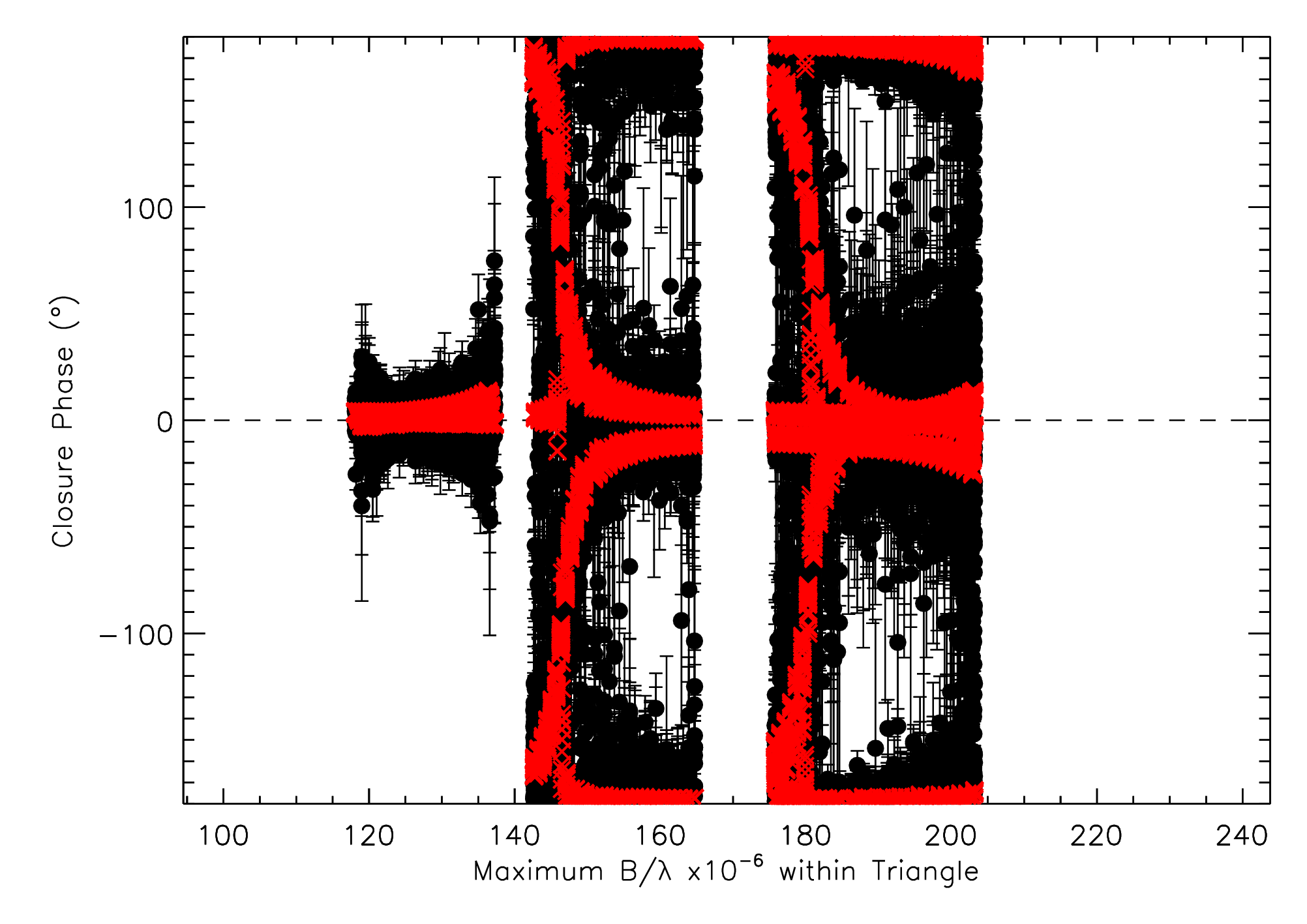}
\figsetgrpnote{Binary fit for Polaris on UT 2021Apr04. Top Row: ($u,v$) coverage (left) and $\chi^2$ map from the binary grid search (right). In the $\chi^2$ map, the red, orange, yellow, green, blue, purple, large black, and small black symbols correspond to solutions within $\Delta \chi^2$ = 1, 4, 9, 16, 25, 36, 49, and $>$50 from the minimum $\chi^2$. Bottom row: The filled black circles show the squared visibilities (left) and closure phases (right) measured with MIRC-X using the 30 second integration time. The red crosses show the best-fit binary model.}
\figsetgrpend

\figsetend

\begin{figure}
\figurenum{13.2}
  \begin{center}
    \scalebox{0.13}{\includegraphics{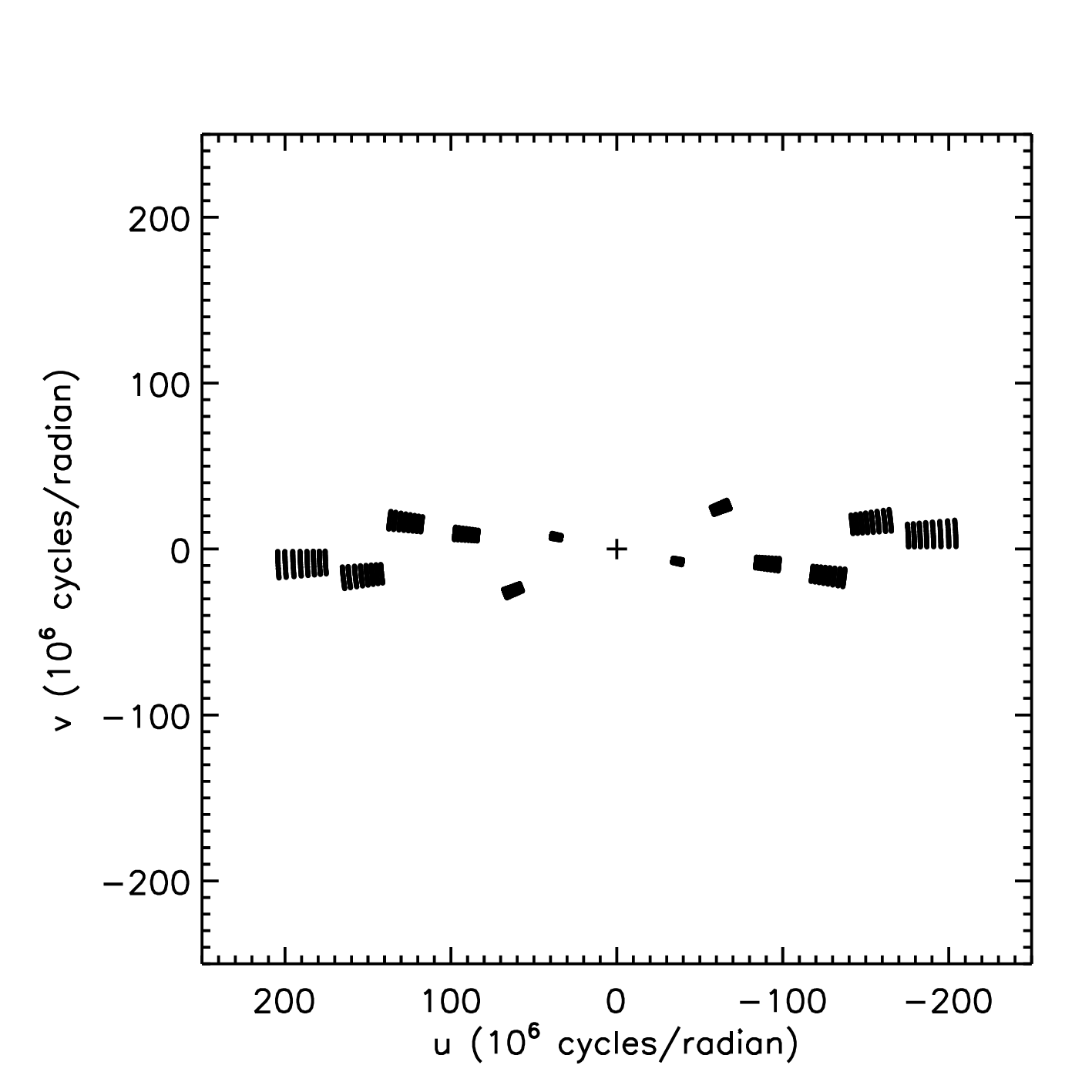}}
    \scalebox{0.095}{\includegraphics{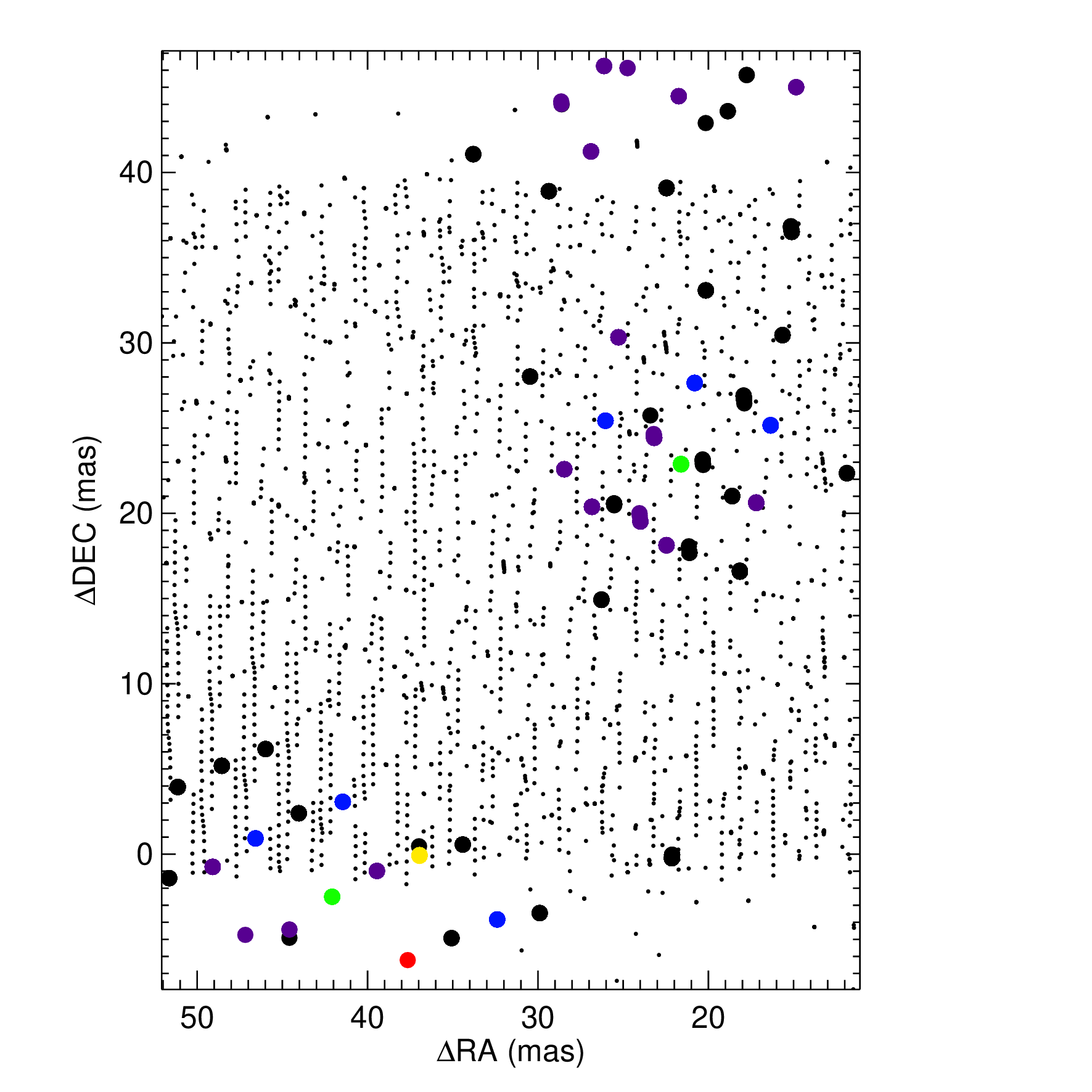}} \\
     \scalebox{0.1}{\includegraphics{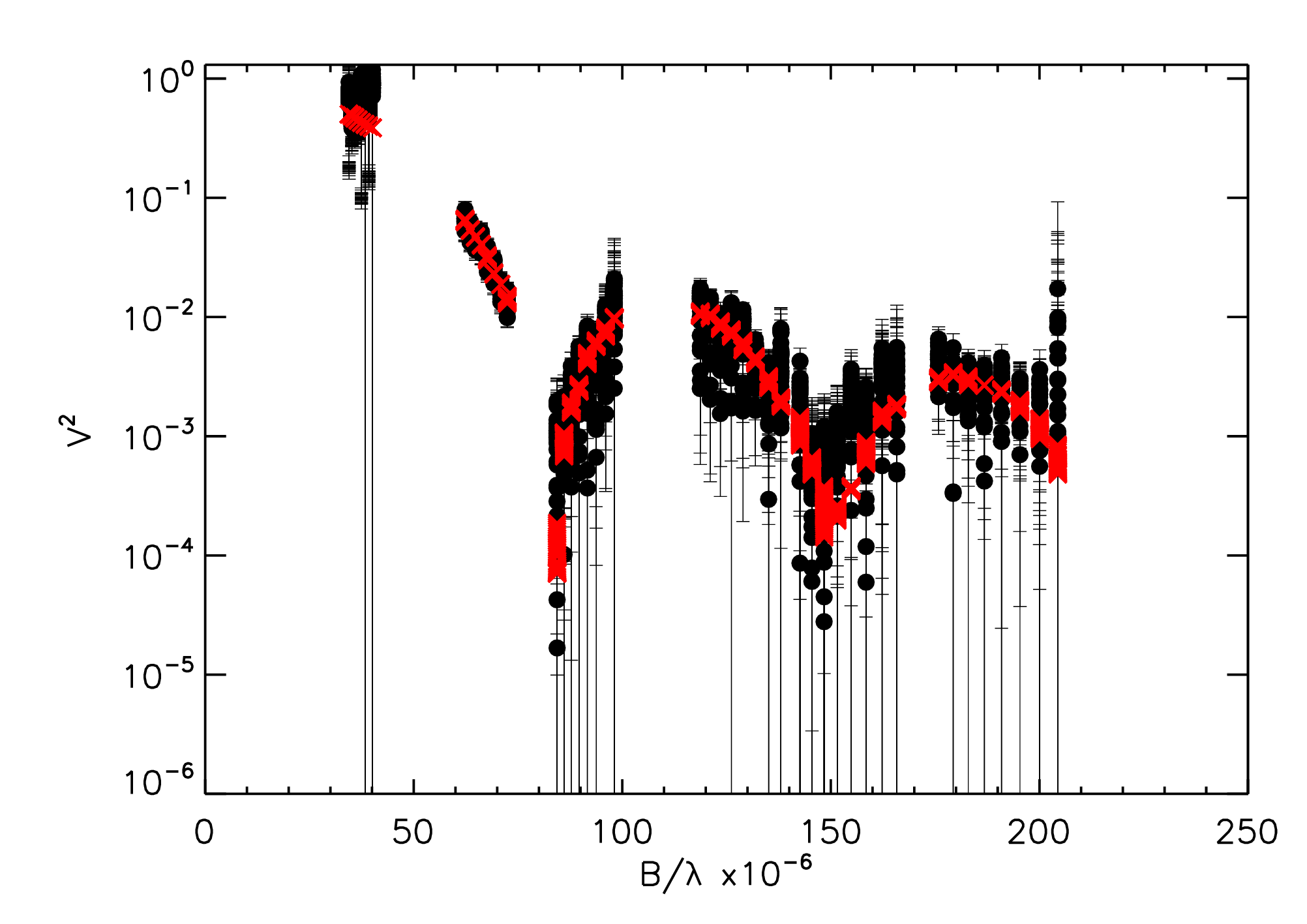}}
     \scalebox{0.1}{\includegraphics{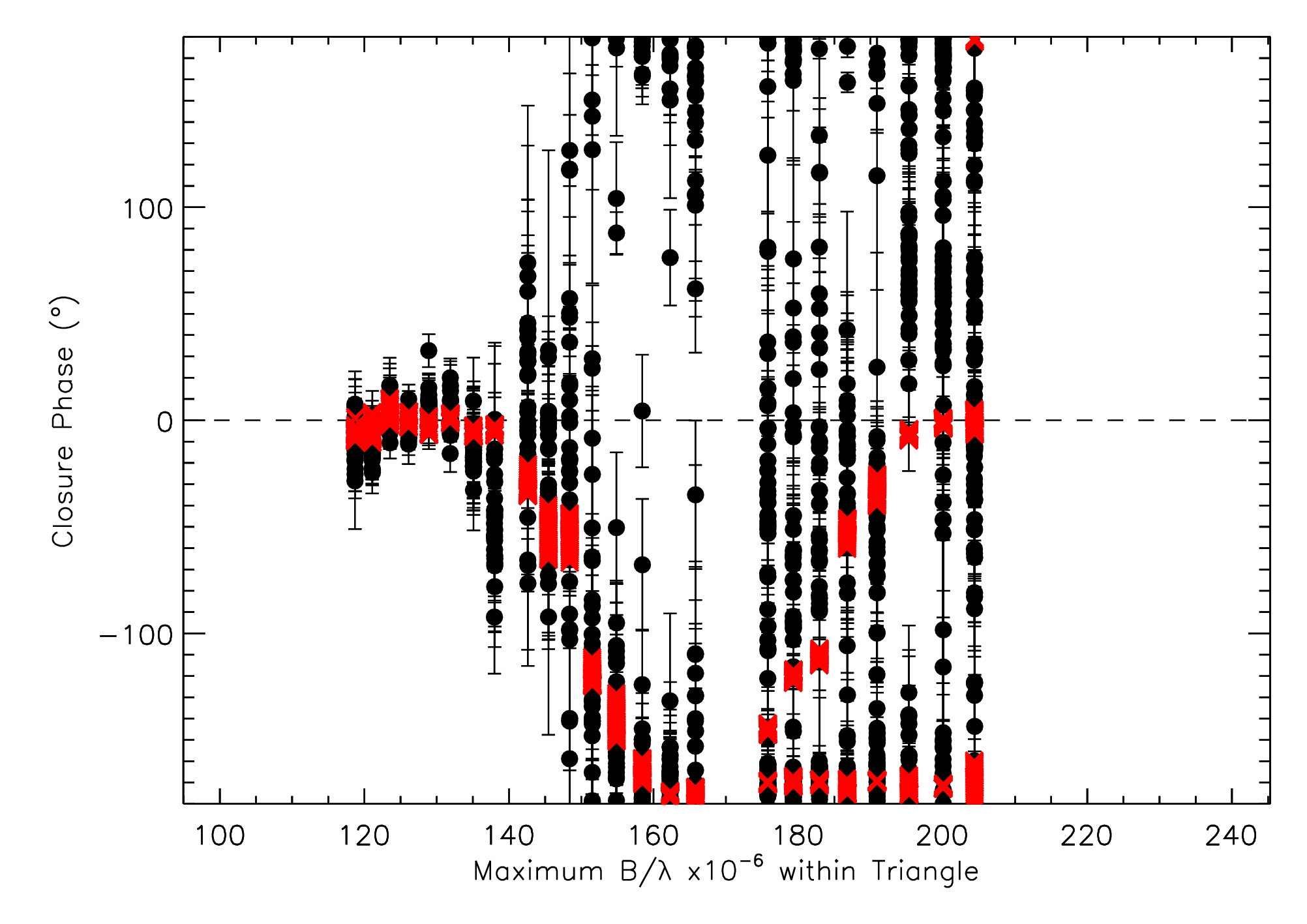}}
     \caption{Binary fit for Polaris on UT 2016Nov18. Top Row: ($u,v$) coverage (left) and $\chi^2$ map from the binary grid search (right). In the $\chi^2$ map, the red, orange, yellow, green, blue, purple, large black, and small black symbols correspond to solutions within $\Delta \chi^2$ = 1, 4, 9, 16, 25, 36, 49, and $>$50 from the minimum $\chi^2$. Bottom row: The filled black circles show the squared visibilities (left) and closure phases (right) measured with MIRC-X using the 30 second integration time. The red crosses show the best-fit binary model.}
  \end{center}
\end{figure}

\begin{figure}
\figurenum{13.3}
  \begin{center}
    \scalebox{0.13}{\includegraphics{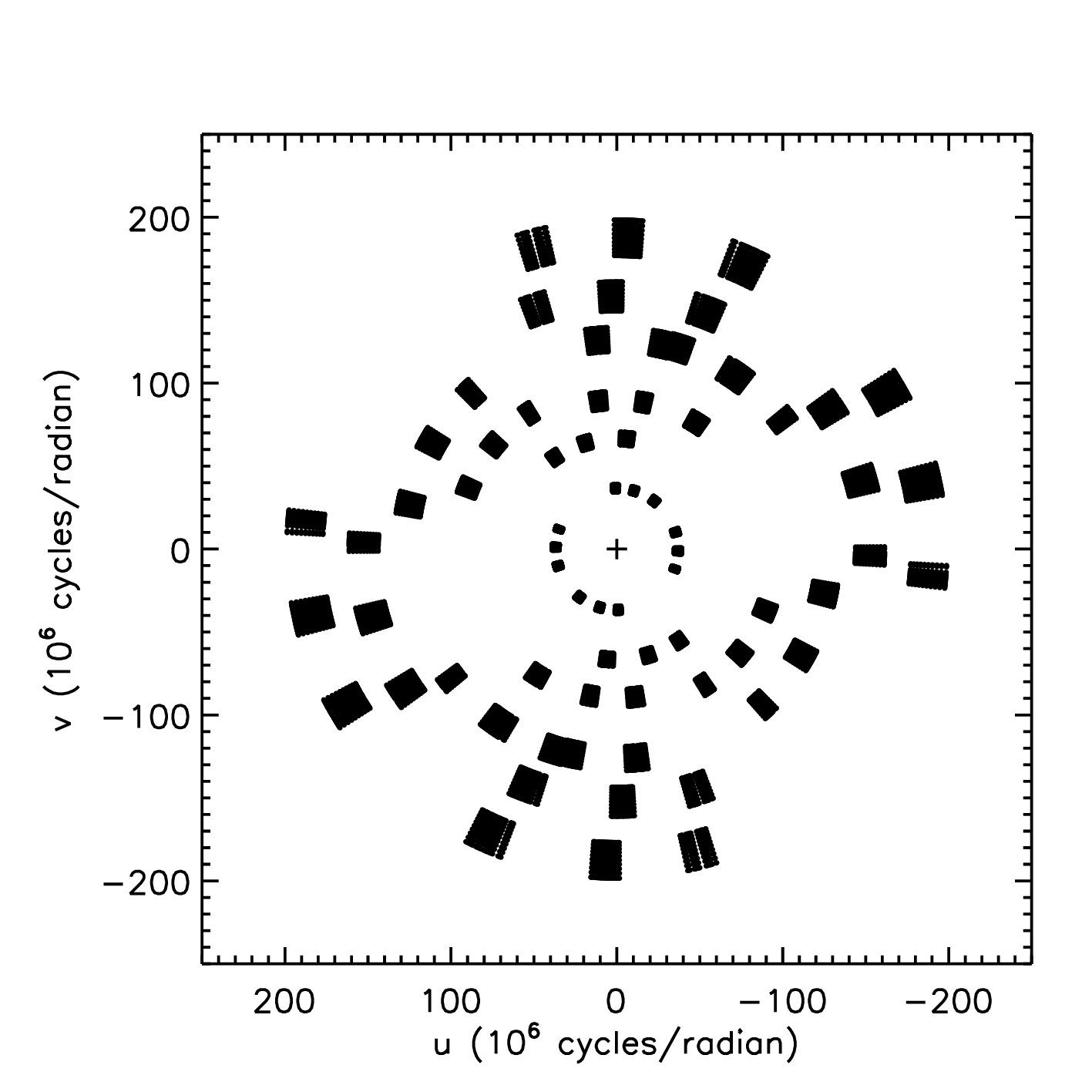}}
    \scalebox{0.10}{\includegraphics{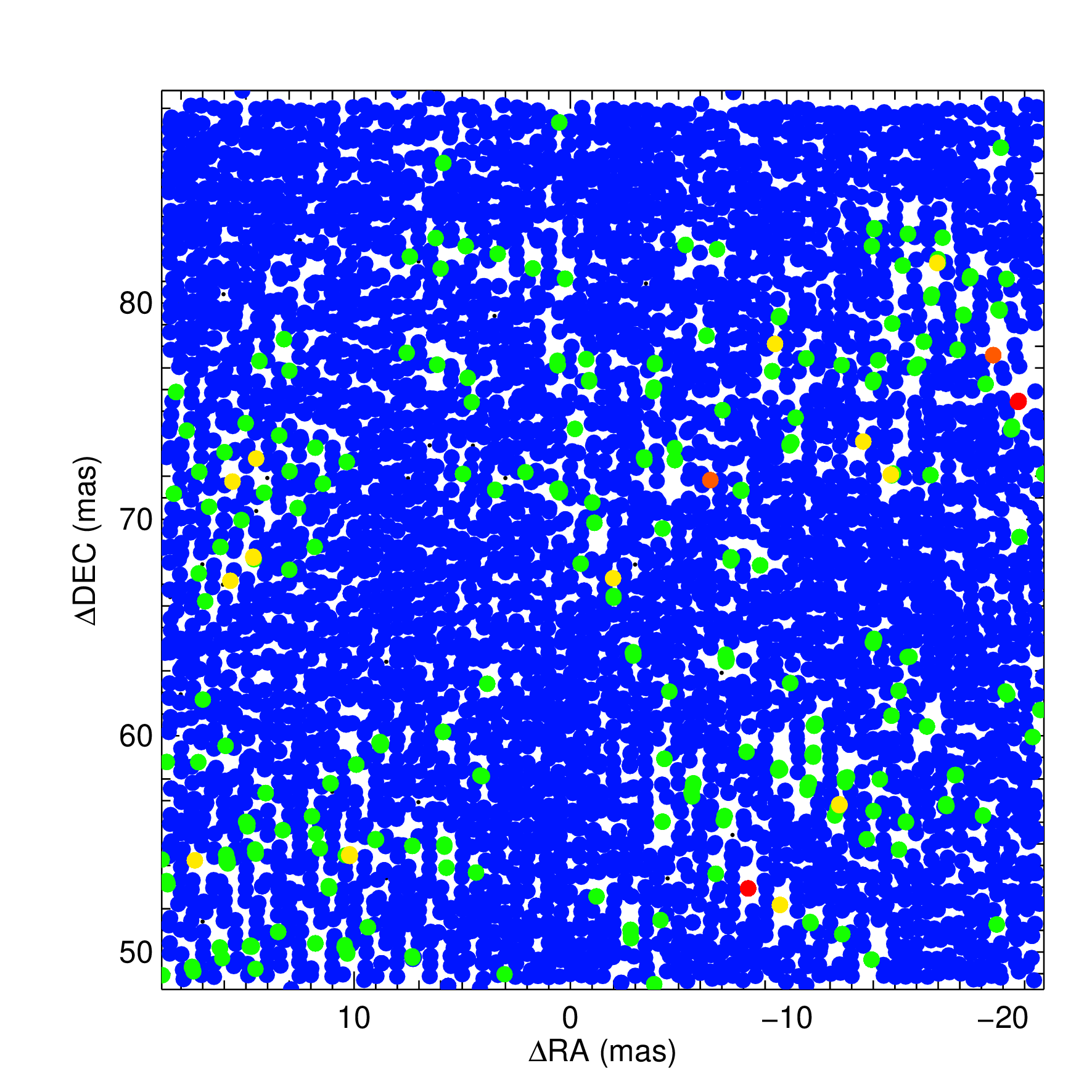}} \\
     \scalebox{0.1}{\includegraphics{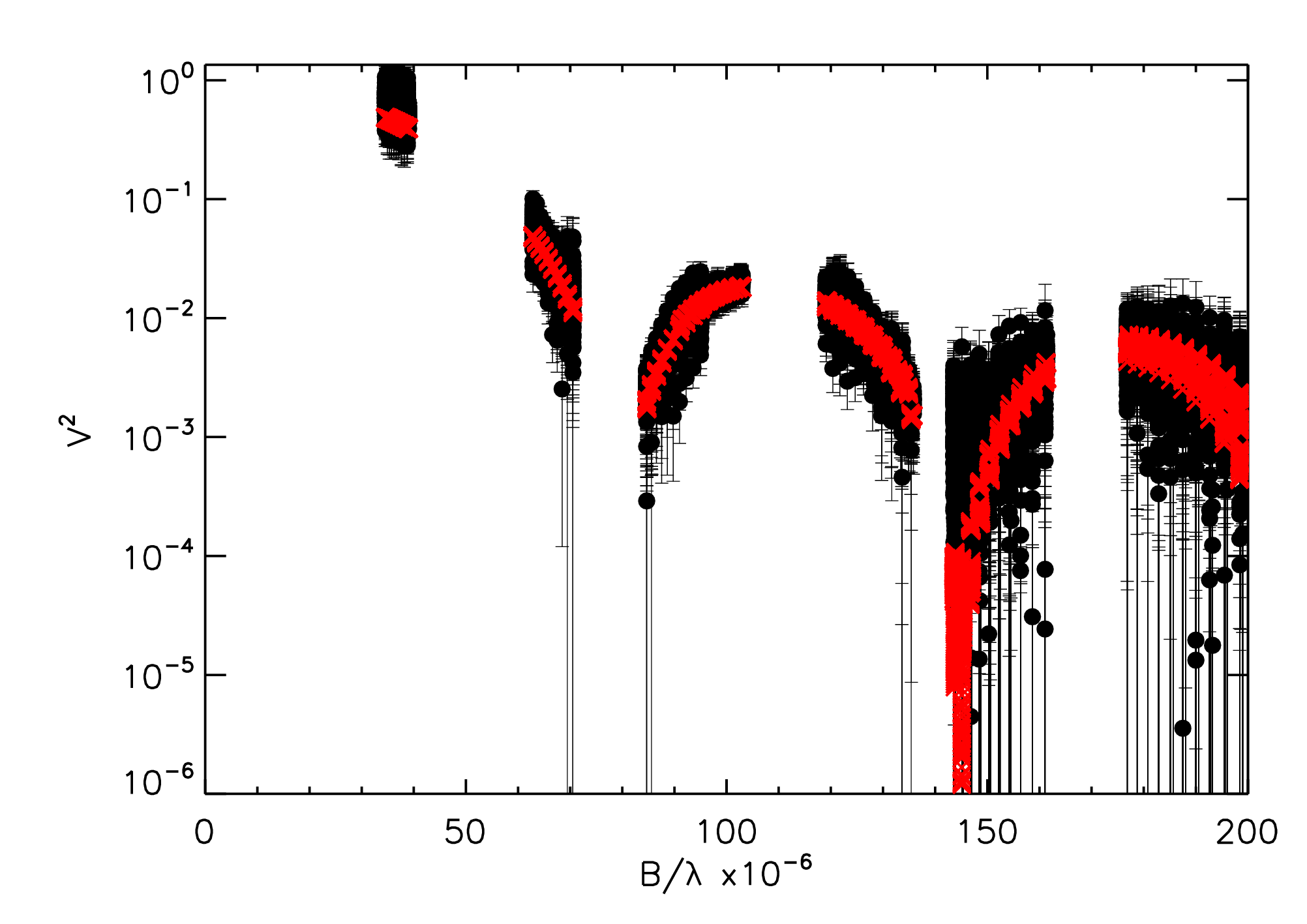}}
     \scalebox{0.1}{\includegraphics{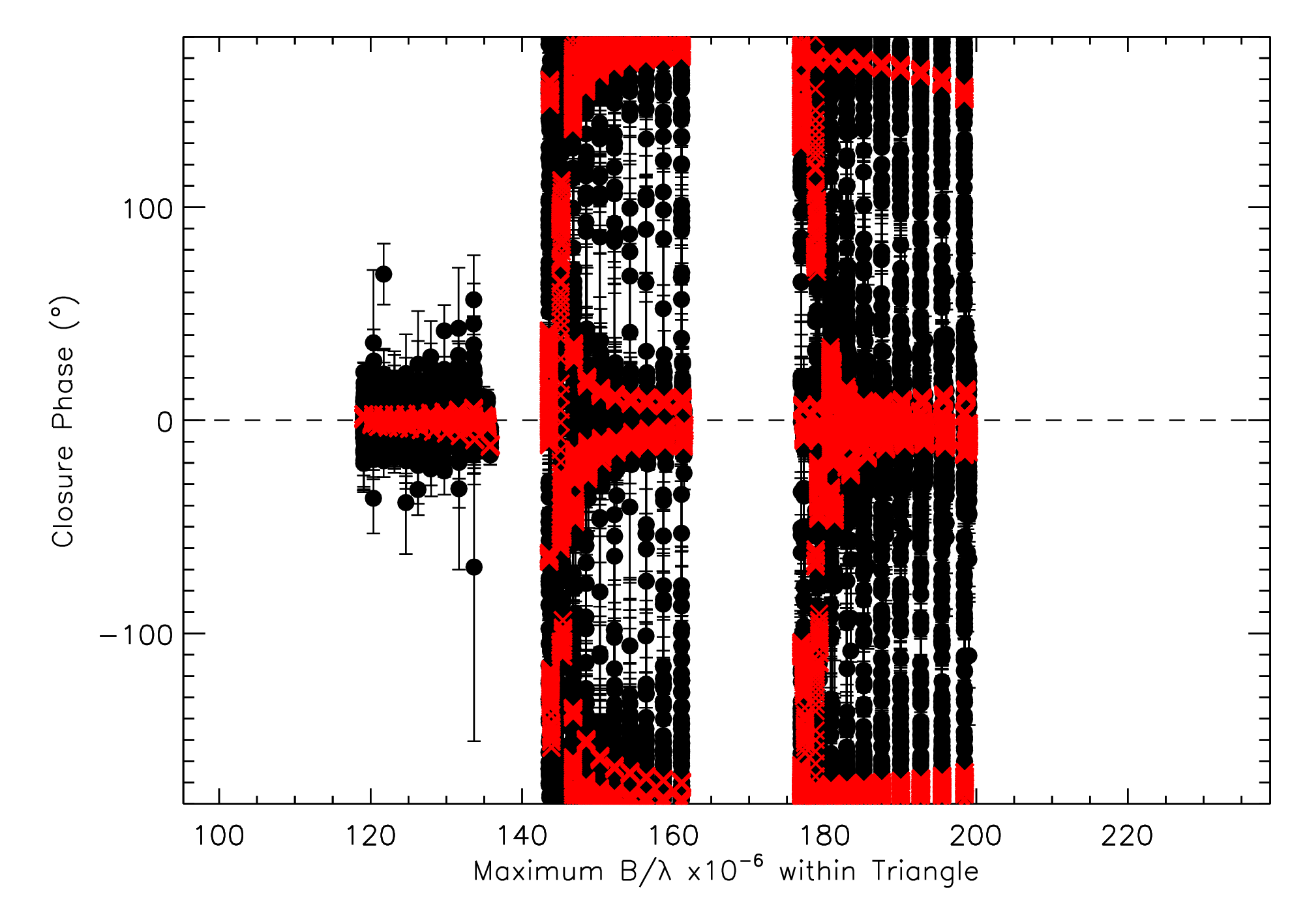}}
     \caption{Binary fit for Polaris on UT 2018Aug27. Top Row: ($u,v$) coverage (left) and $\chi^2$ map from the binary grid search (right). In the $\chi^2$ map, the red, orange, yellow, green, blue, purple, large black, and small black symbols correspond to solutions within $\Delta \chi^2$ = 1, 4, 9, 16, 25, 36, 49, and $>$50 from the minimum $\chi^2$. Bottom row: The filled black circles show the squared visibilities (left) and closure phases (right) measured with MIRC-X using the 30 second integration time. The red crosses show the best-fit binary model.}
  \end{center}
\end{figure}

\begin{figure}
\figurenum{13.4}
  \begin{center}
    \scalebox{0.13}{\includegraphics{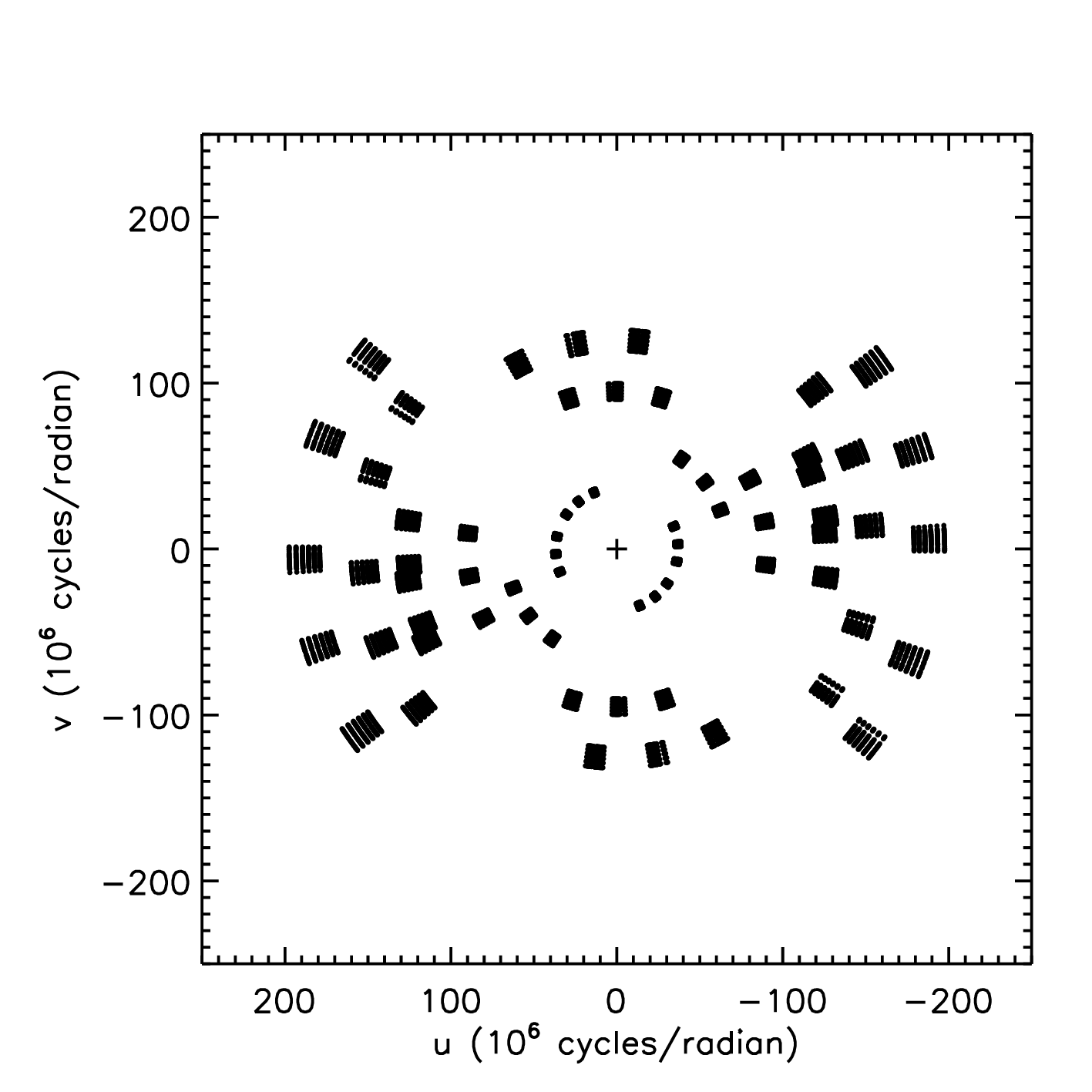}}
    \scalebox{0.10}{\includegraphics{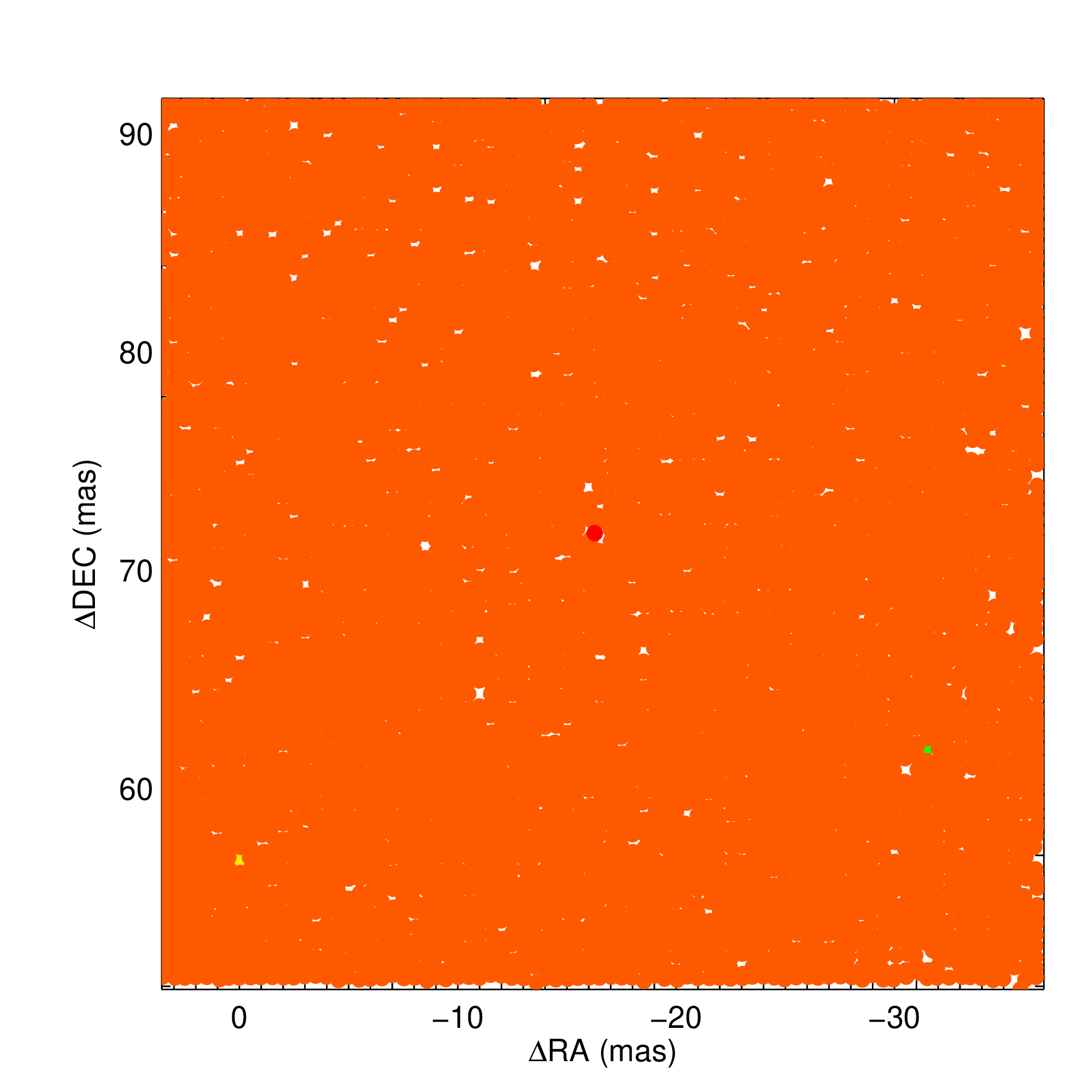}} \\
     \scalebox{0.1}{\includegraphics{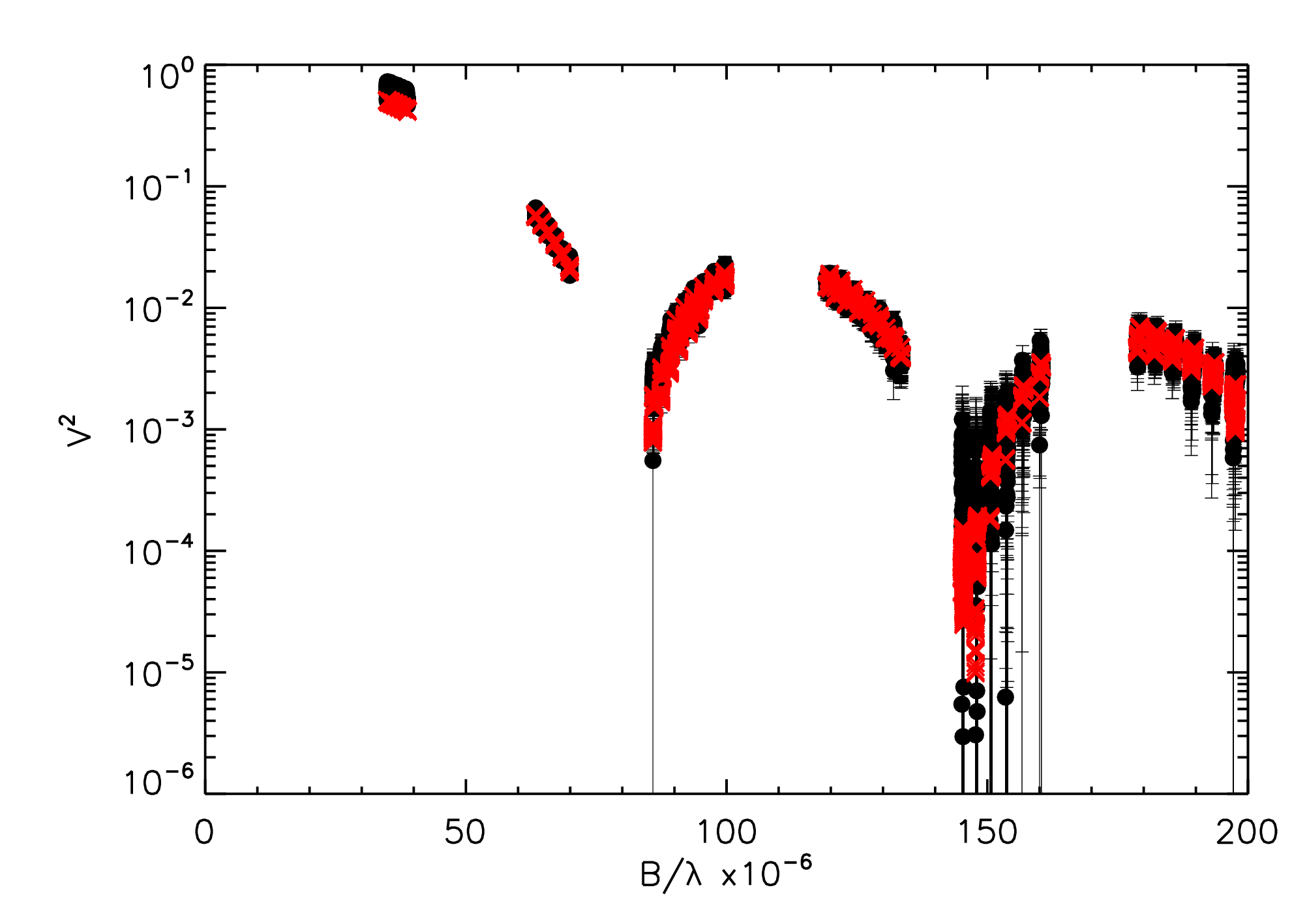}}
     \scalebox{0.1}{\includegraphics{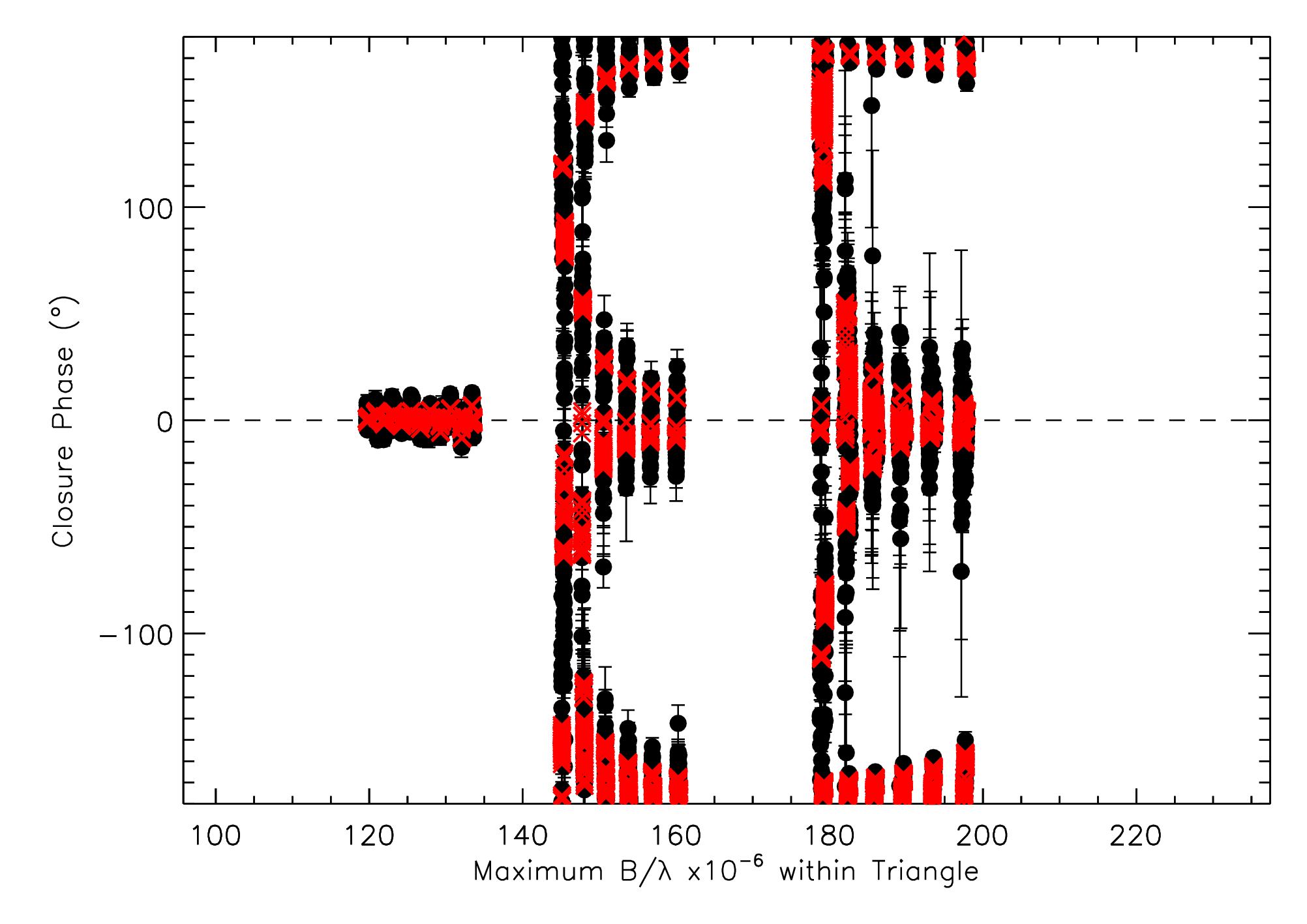}}
     \caption{Binary fit for Polaris on UT 2019Apr09. Top Row: ($u,v$) coverage (left) and $\chi^2$ map from the binary grid search (right). In the $\chi^2$ map, the red, orange, yellow, green, blue, purple, large black, and small black symbols correspond to solutions within $\Delta \chi^2$ = 1, 4, 9, 16, 25, 36, 49, and $>$50 from the minimum $\chi^2$. Bottom row: The filled black circles show the squared visibilities (left) and closure phases (right) measured with MIRC-X using the 30 second integration time. The red crosses show the best-fit binary model.}
  \end{center}
\end{figure}

\begin{figure}
\figurenum{13.5}
  \begin{center}
    \scalebox{0.13}{\includegraphics{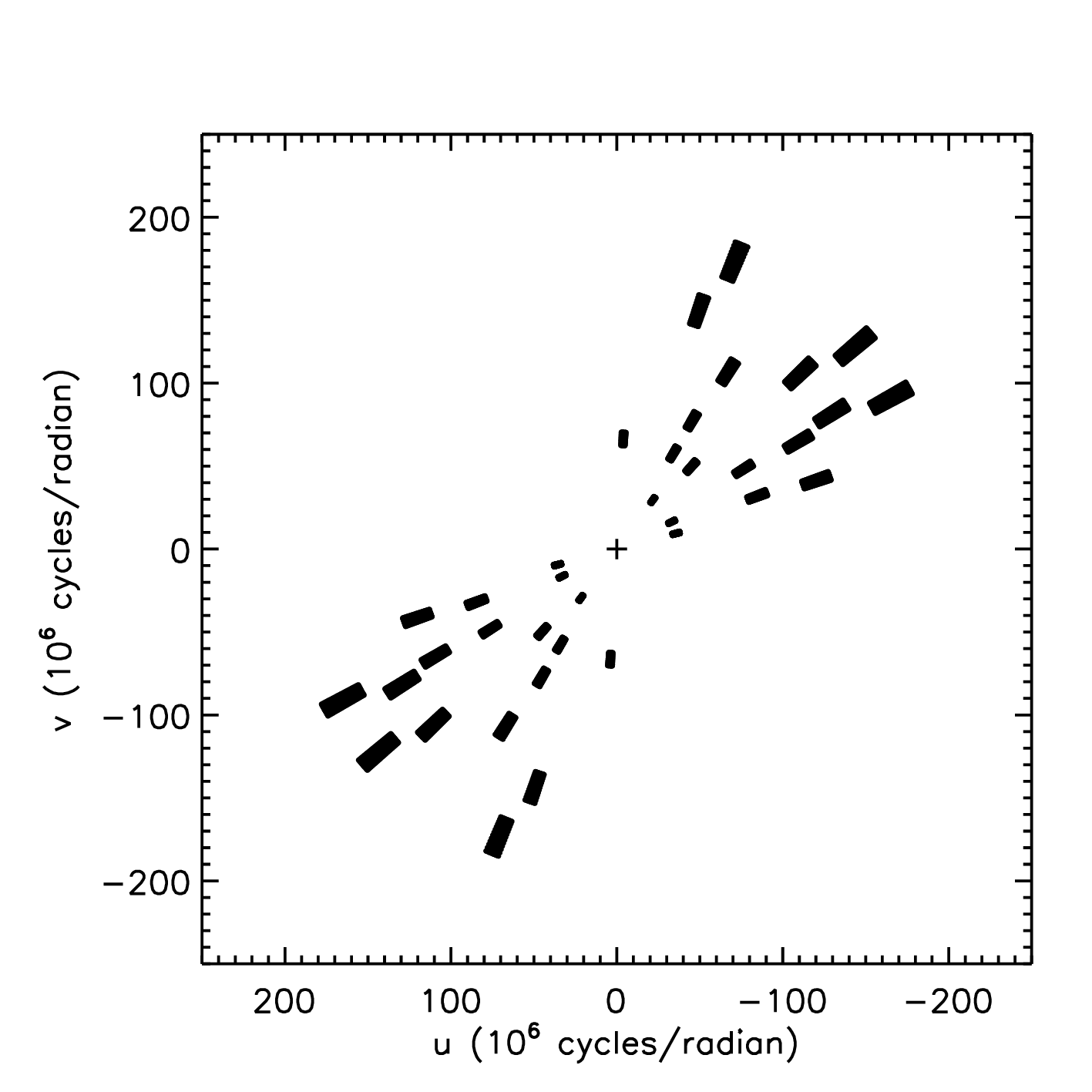}}
    \scalebox{0.10}{\includegraphics{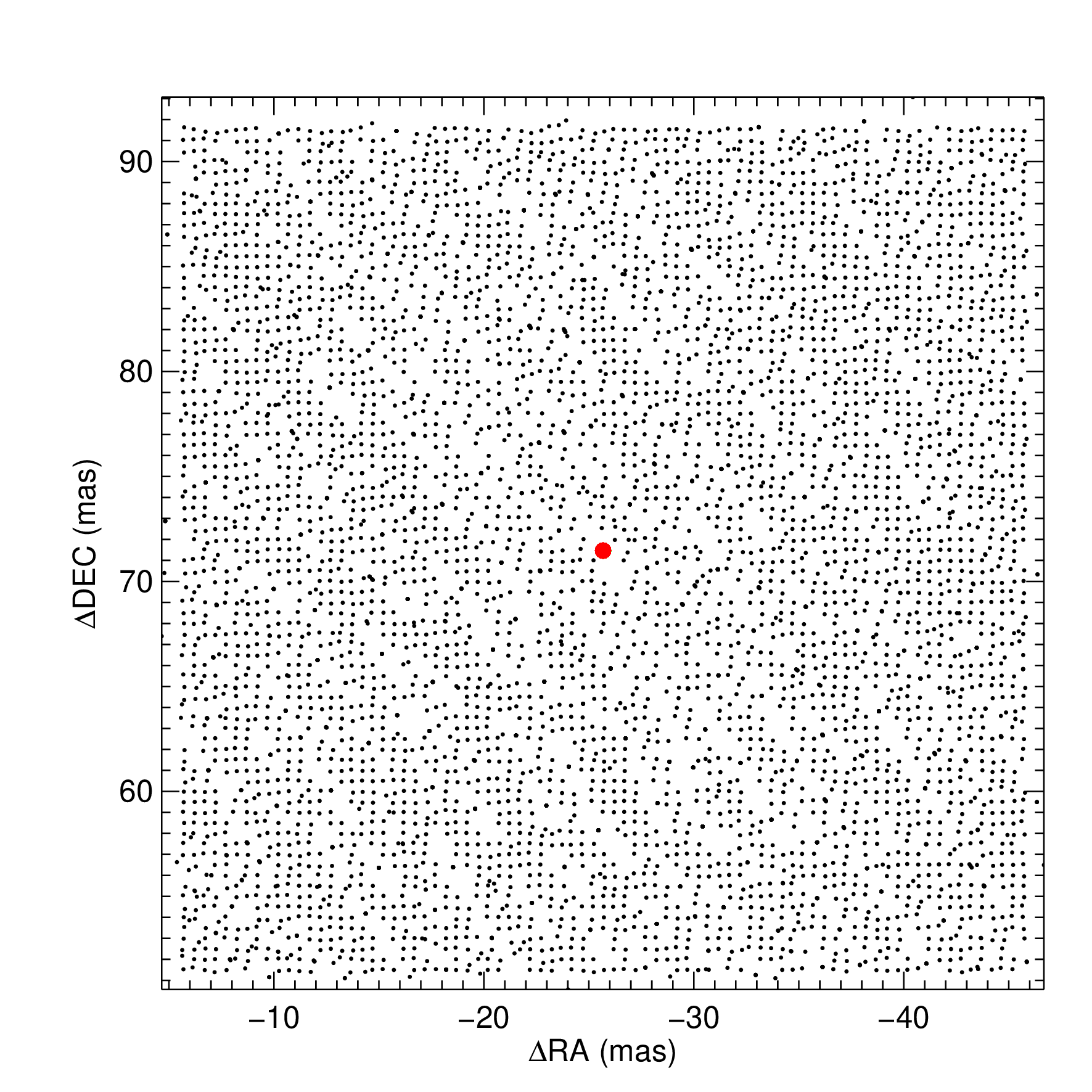}} \\
     \scalebox{0.1}{\includegraphics{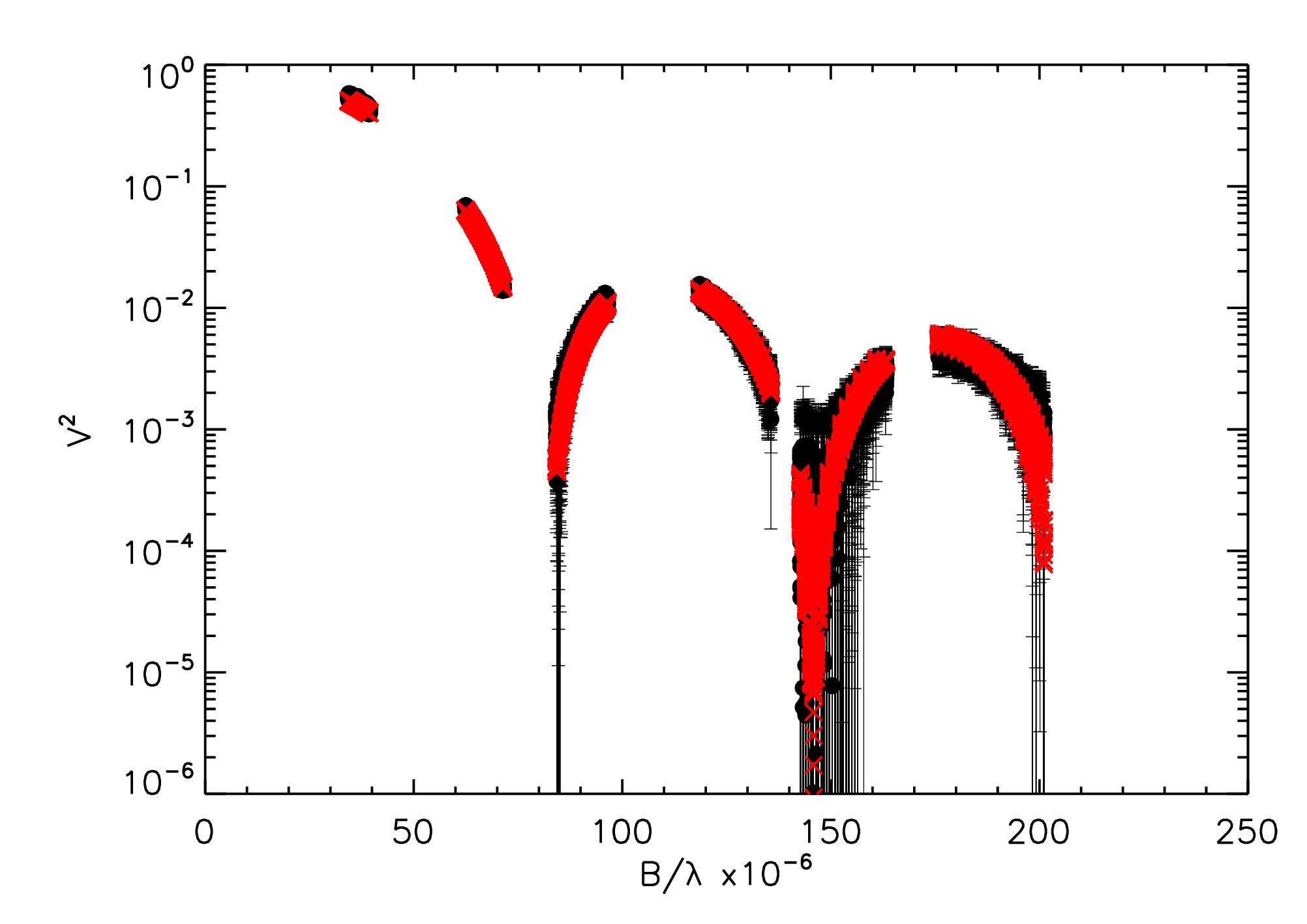}}
     \scalebox{0.1}{\includegraphics{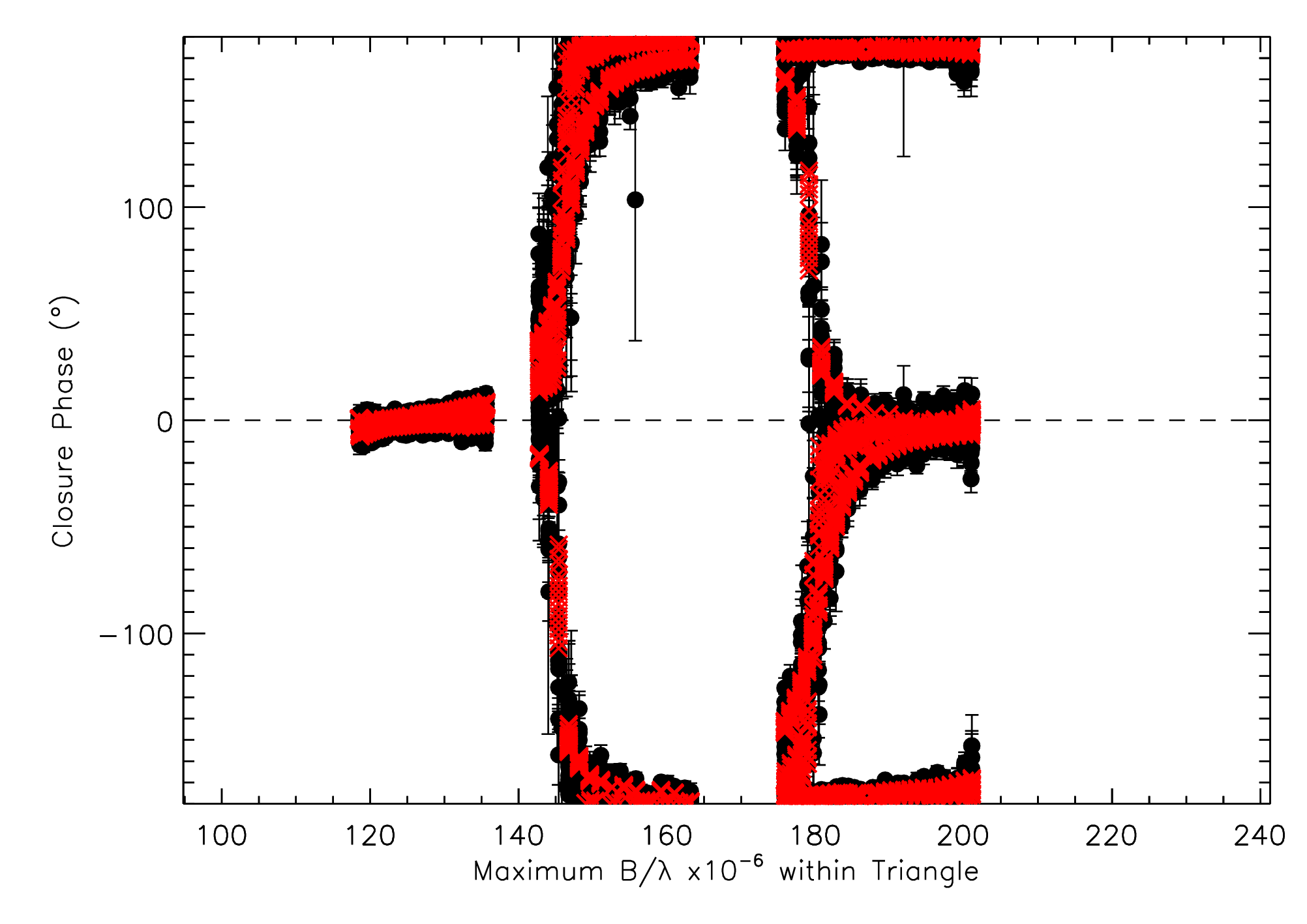}}
     \caption{Binary fit for Polaris on UT 2019Sep02. Top Row: ($u,v$) coverage (left) and $\chi^2$ map from the binary grid search (right). In the $\chi^2$ map, the red, orange, yellow, green, blue, purple, large black, and small black symbols correspond to solutions within $\Delta \chi^2$ = 1, 4, 9, 16, 25, 36, 49, and $>$50 from the minimum $\chi^2$. Bottom row: The filled black circles show the squared visibilities (left) and closure phases (right) measured with MIRC-X using the 30 second integration time. The red crosses show the best-fit binary model.}
  \end{center}
\end{figure}

\begin{figure}
\figurenum{13.6}
  \begin{center}
    \scalebox{0.13}{\includegraphics{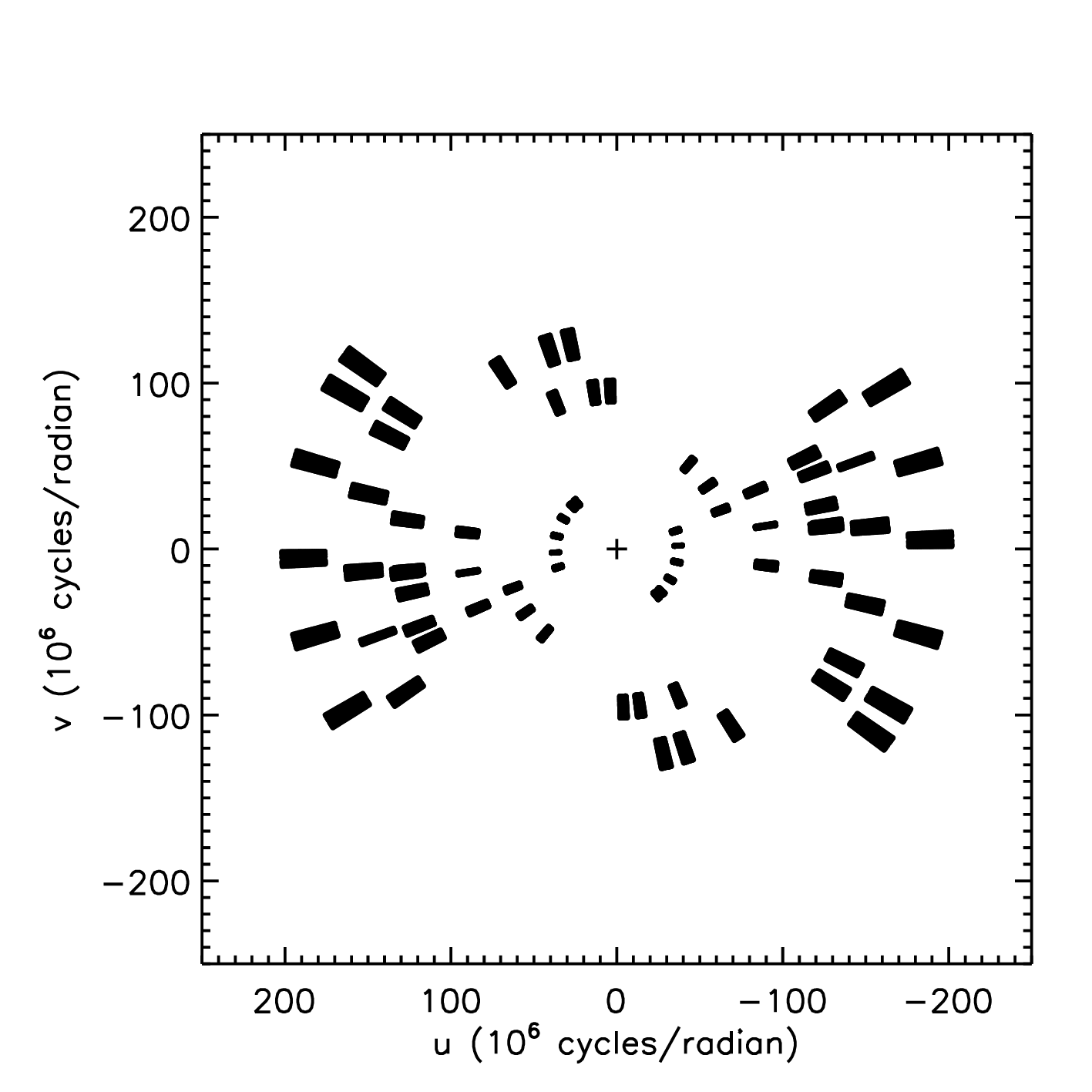}}
    \scalebox{0.10}{\includegraphics{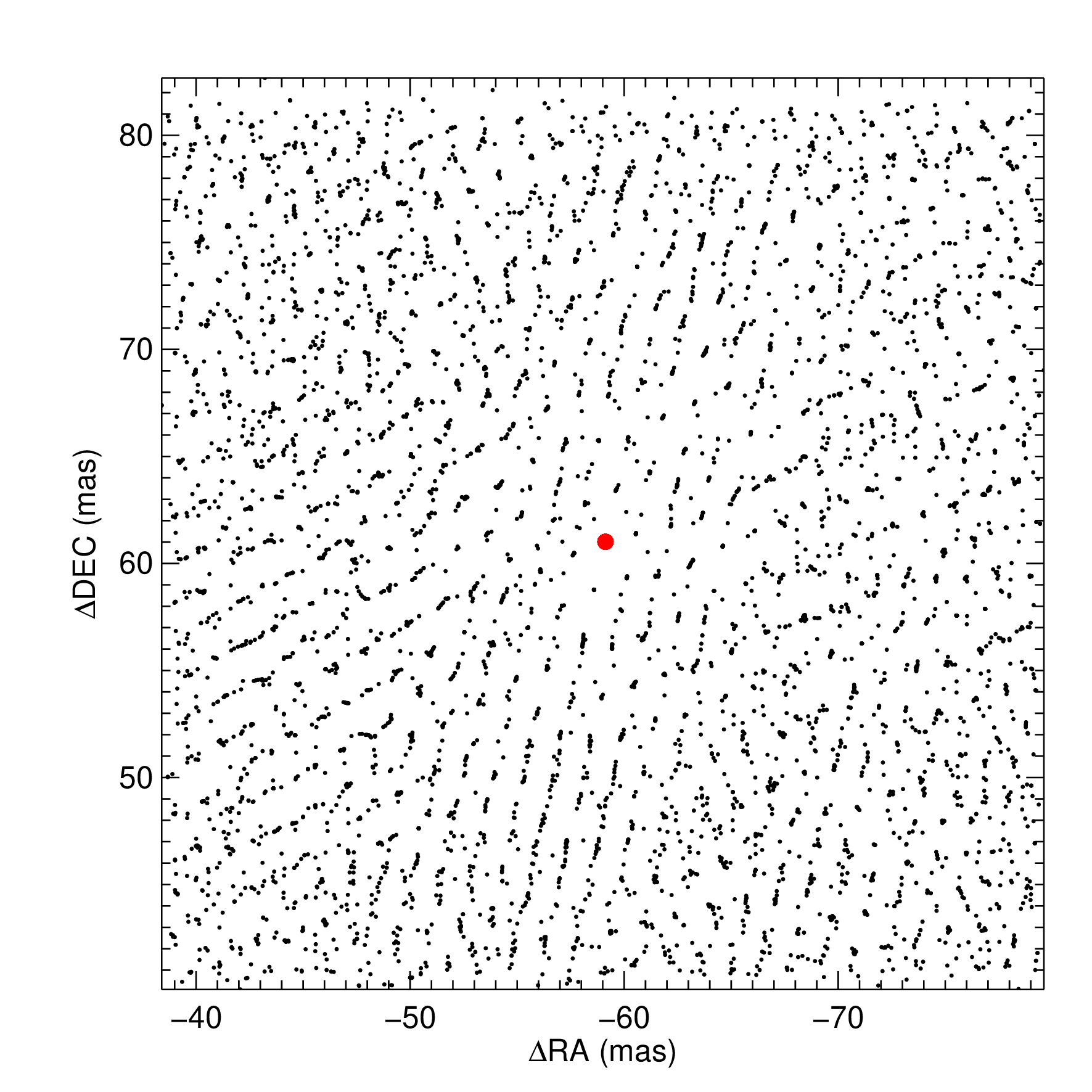}} \\
     \scalebox{0.1}{\includegraphics{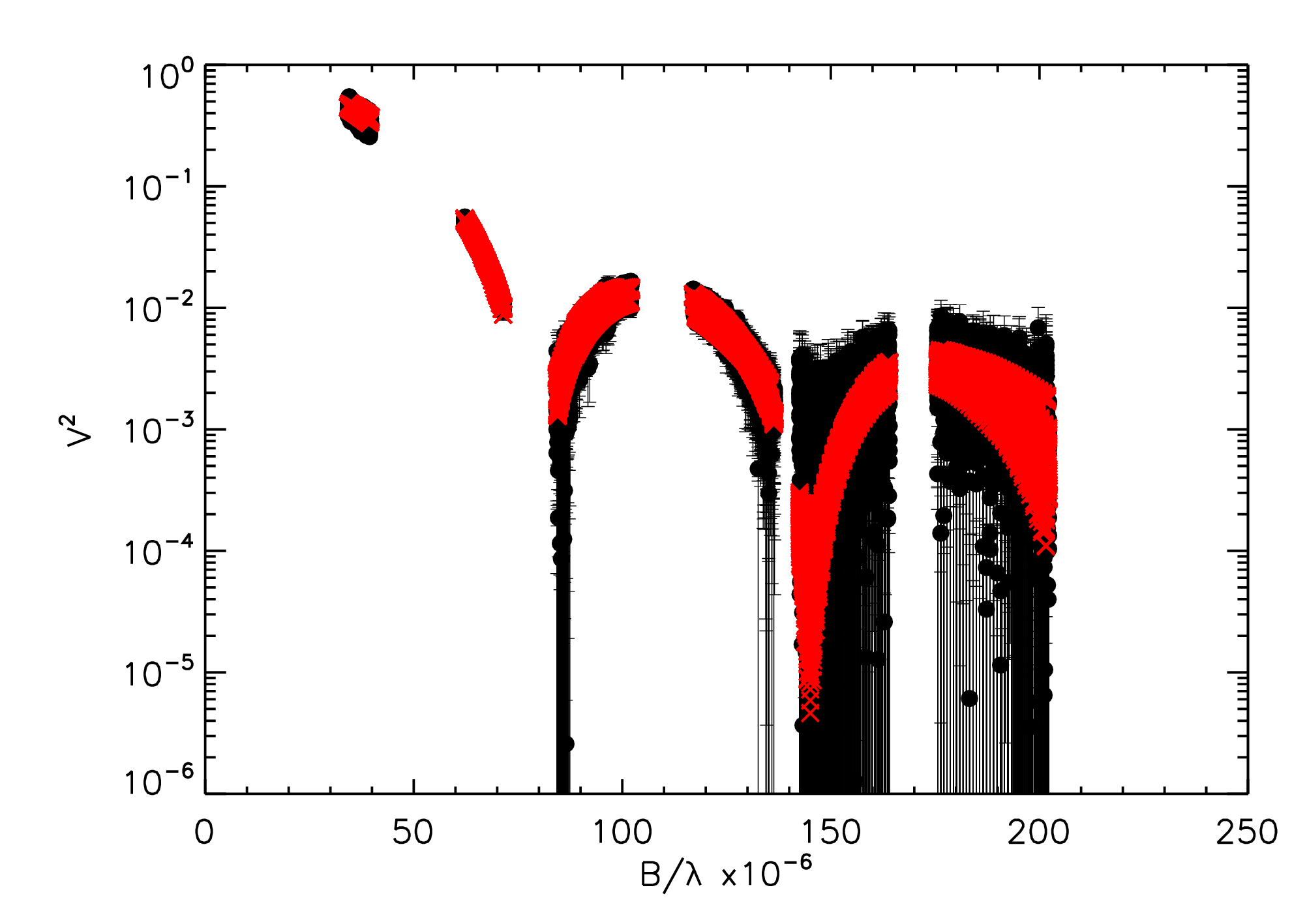}}
     \scalebox{0.1}{\includegraphics{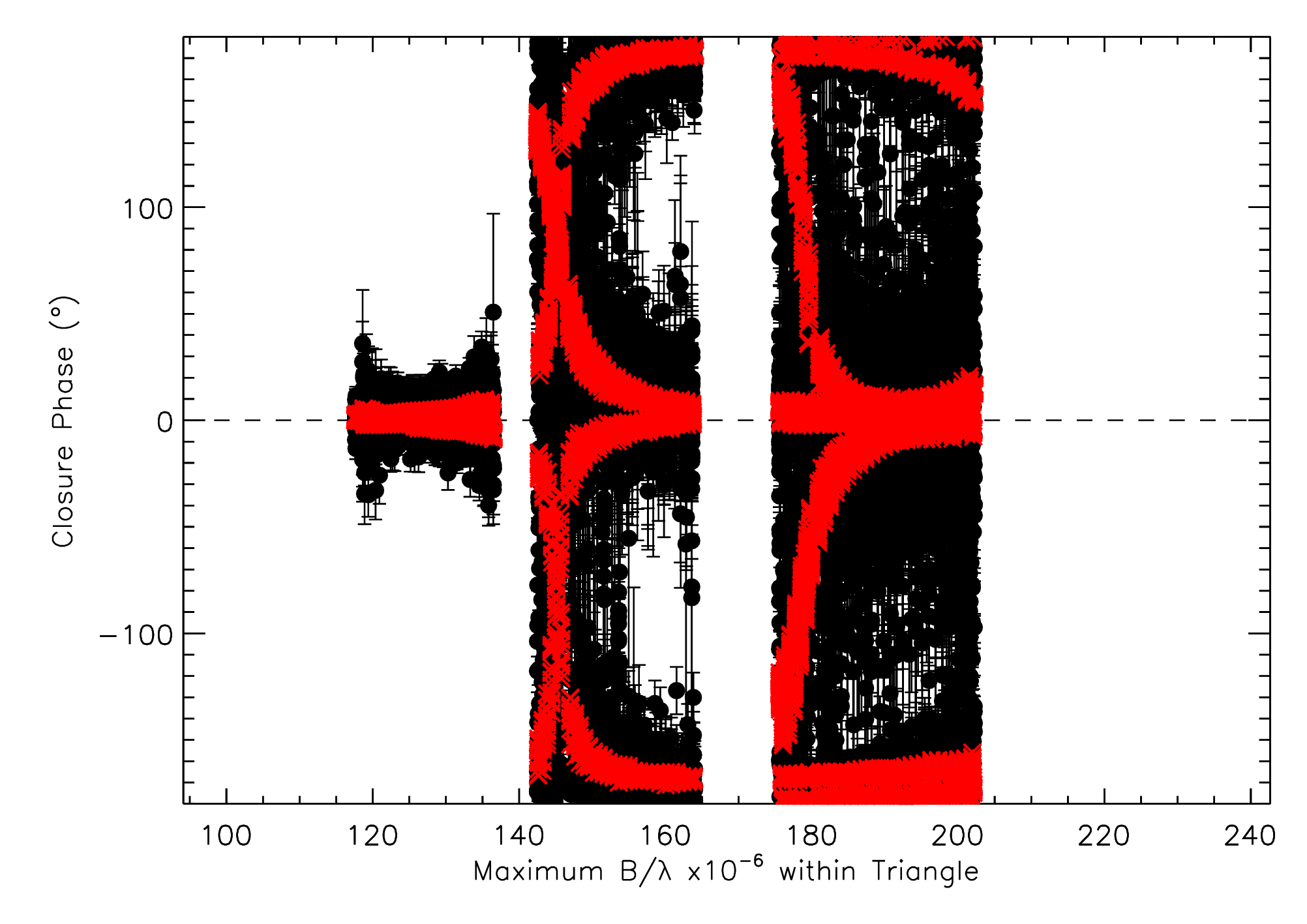}}
     \caption{Binary fit for Polaris on UT 2021Apr02. Top Row: ($u,v$) coverage (left) and $\chi^2$ map from the binary grid search (right). In the $\chi^2$ map, the red, orange, yellow, green, blue, purple, large black, and small black symbols correspond to solutions within $\Delta \chi^2$ = 1, 4, 9, 16, 25, 36, 49, and $>$50 from the minimum $\chi^2$. Bottom row: The filled black circles show the squared visibilities (left) and closure phases (right) measured with MIRC-X using the 30 second integration time. The red crosses show the best-fit binary model.}
  \end{center}
\end{figure}

\begin{figure}
\figurenum{13.7}
  \begin{center}
    \scalebox{0.13}{\includegraphics{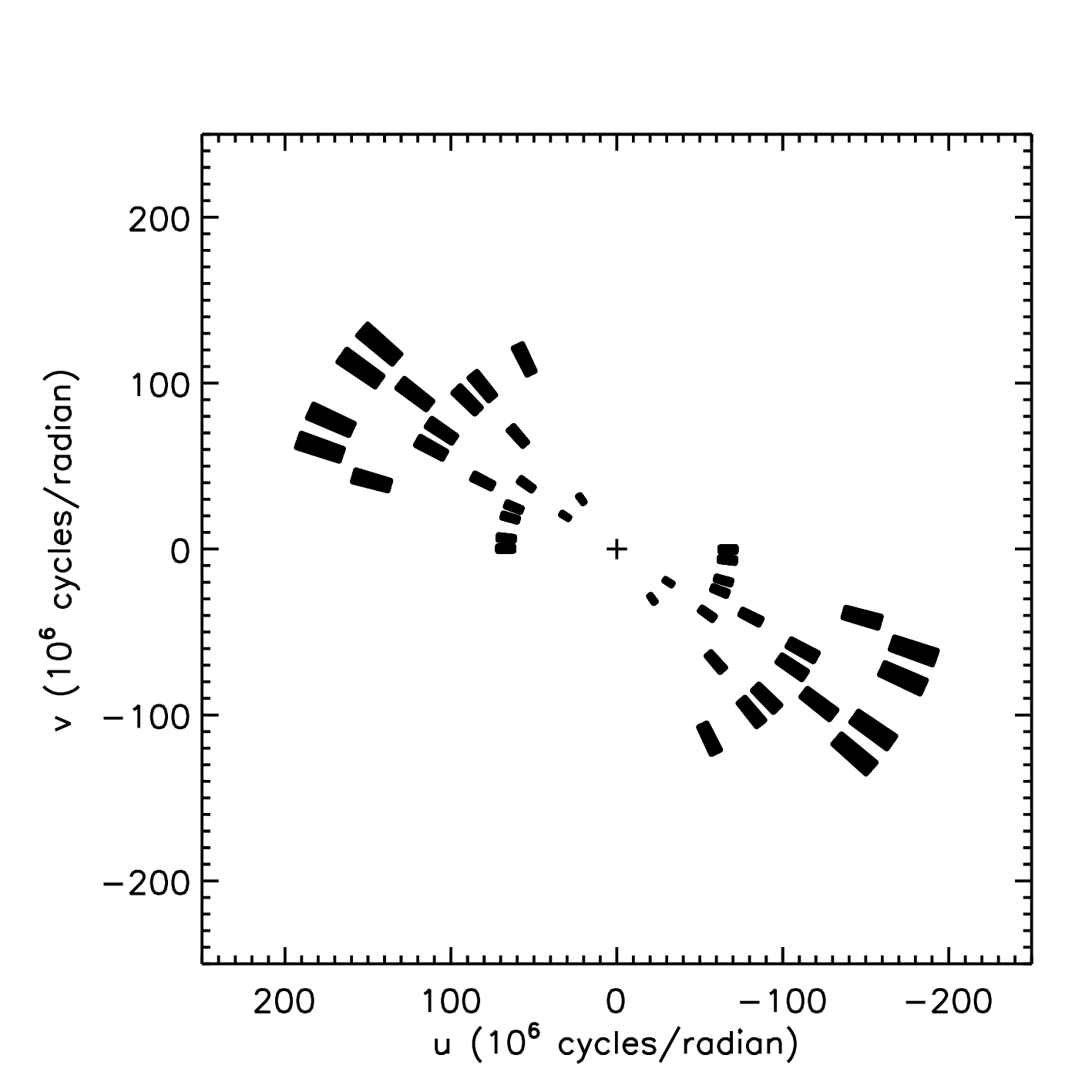}}
    \scalebox{0.10}{\includegraphics{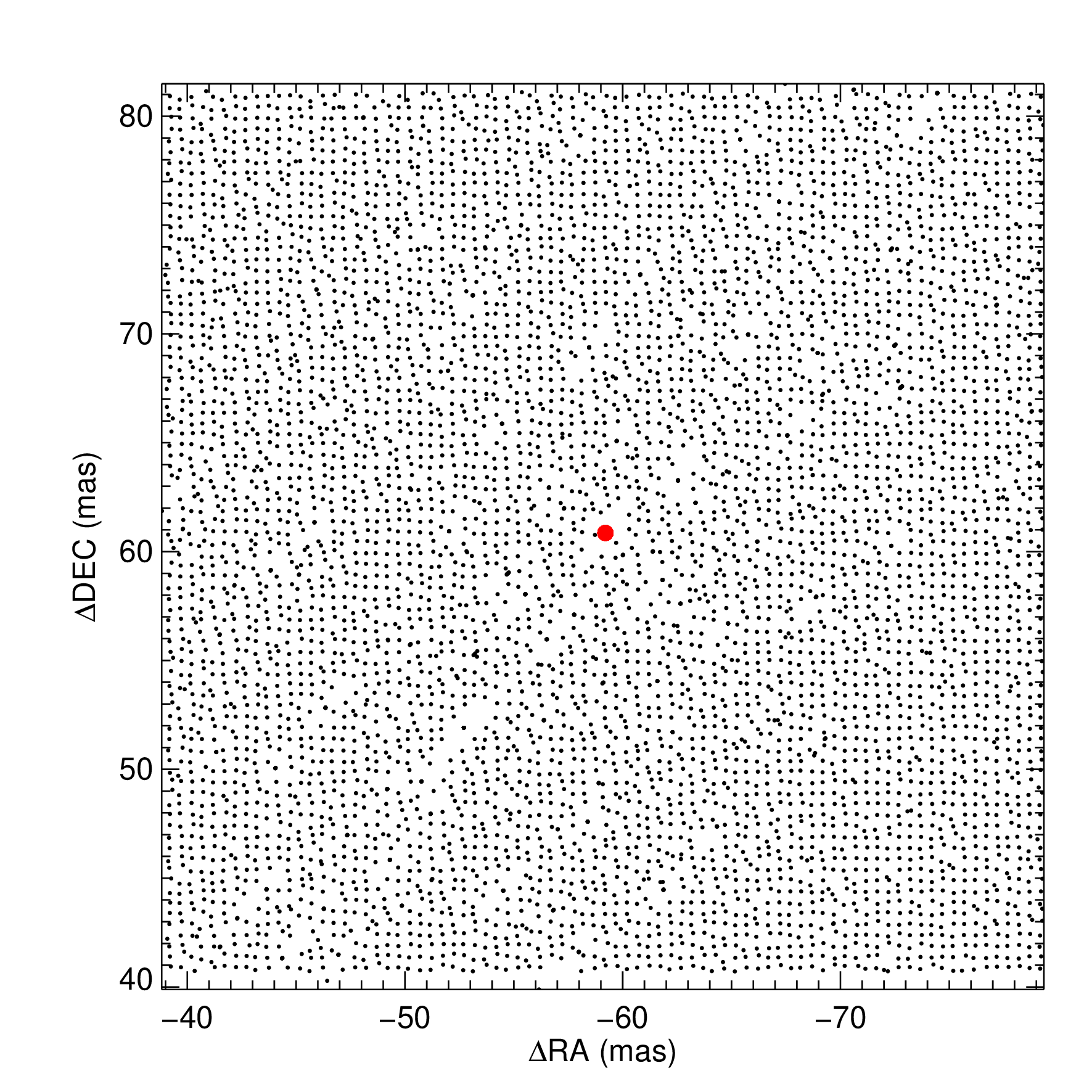}} \\
     \scalebox{0.1}{\includegraphics{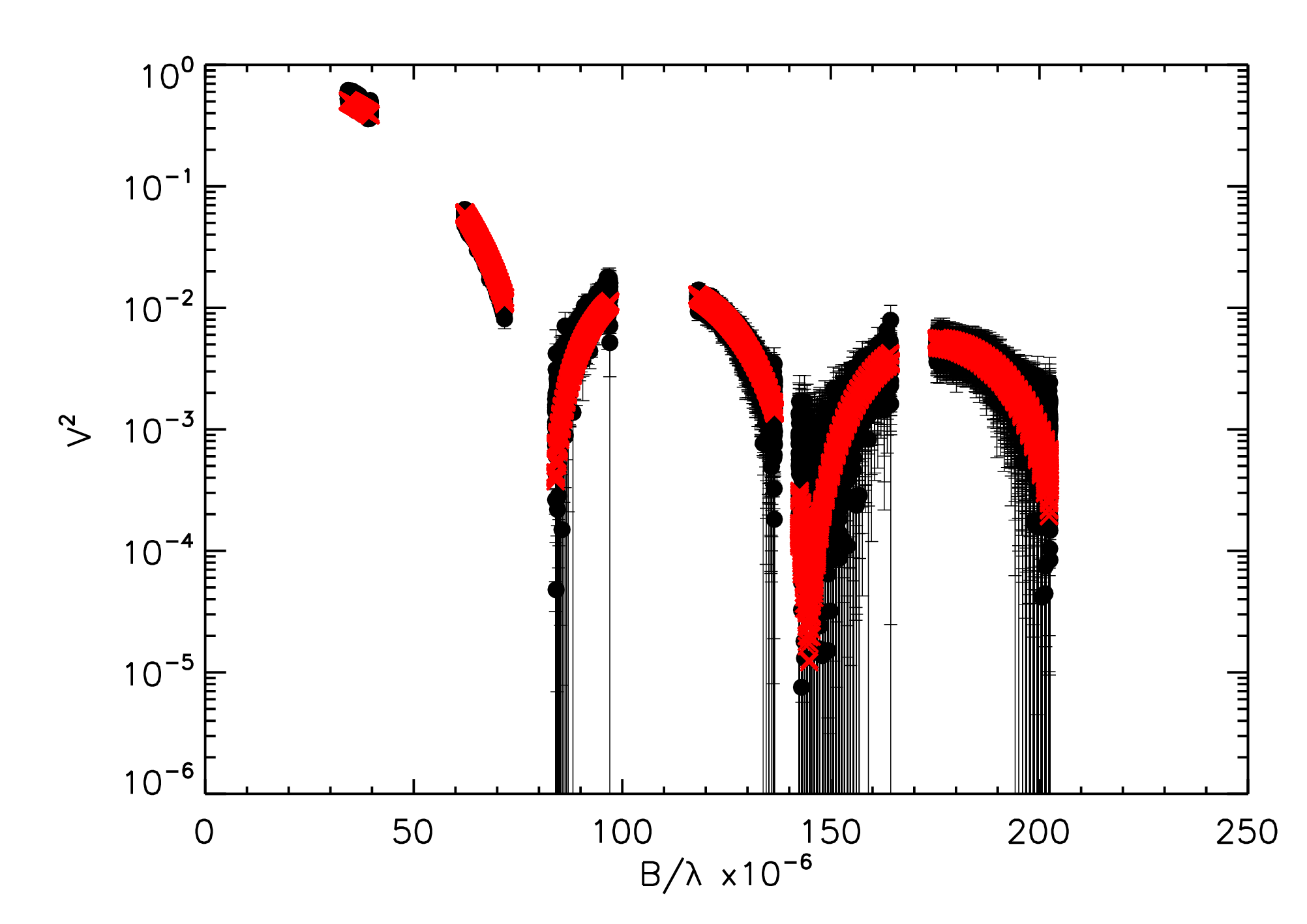}}
     \scalebox{0.1}{\includegraphics{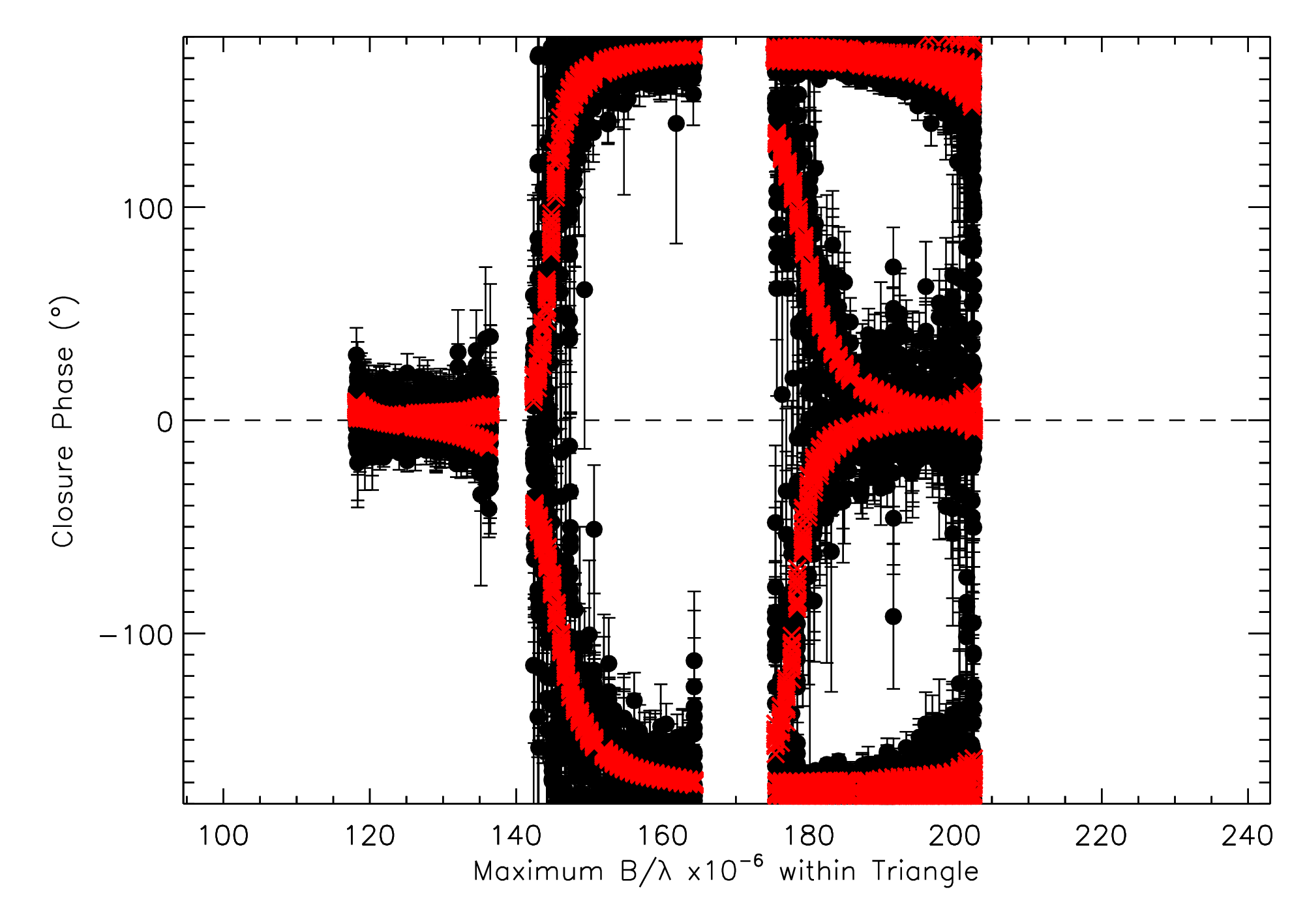}}
     \caption{Binary fit for Polaris on UT 2021Apr03. Top Row: ($u,v$) coverage (left) and $\chi^2$ map from the binary grid search (right). In the $\chi^2$ map, the red, orange, yellow, green, blue, purple, large black, and small black symbols correspond to solutions within $\Delta \chi^2$ = 1, 4, 9, 16, 25, 36, 49, and $>$50 from the minimum $\chi^2$. Bottom row: The filled black circles show the squared visibilities (left) and closure phases (right) measured with MIRC-X using the 30 second integration time. The red crosses show the best-fit binary model.}
  \end{center}
\end{figure}

\begin{figure}
\figurenum{13.8}
  \begin{center}
    \scalebox{0.13}{\includegraphics{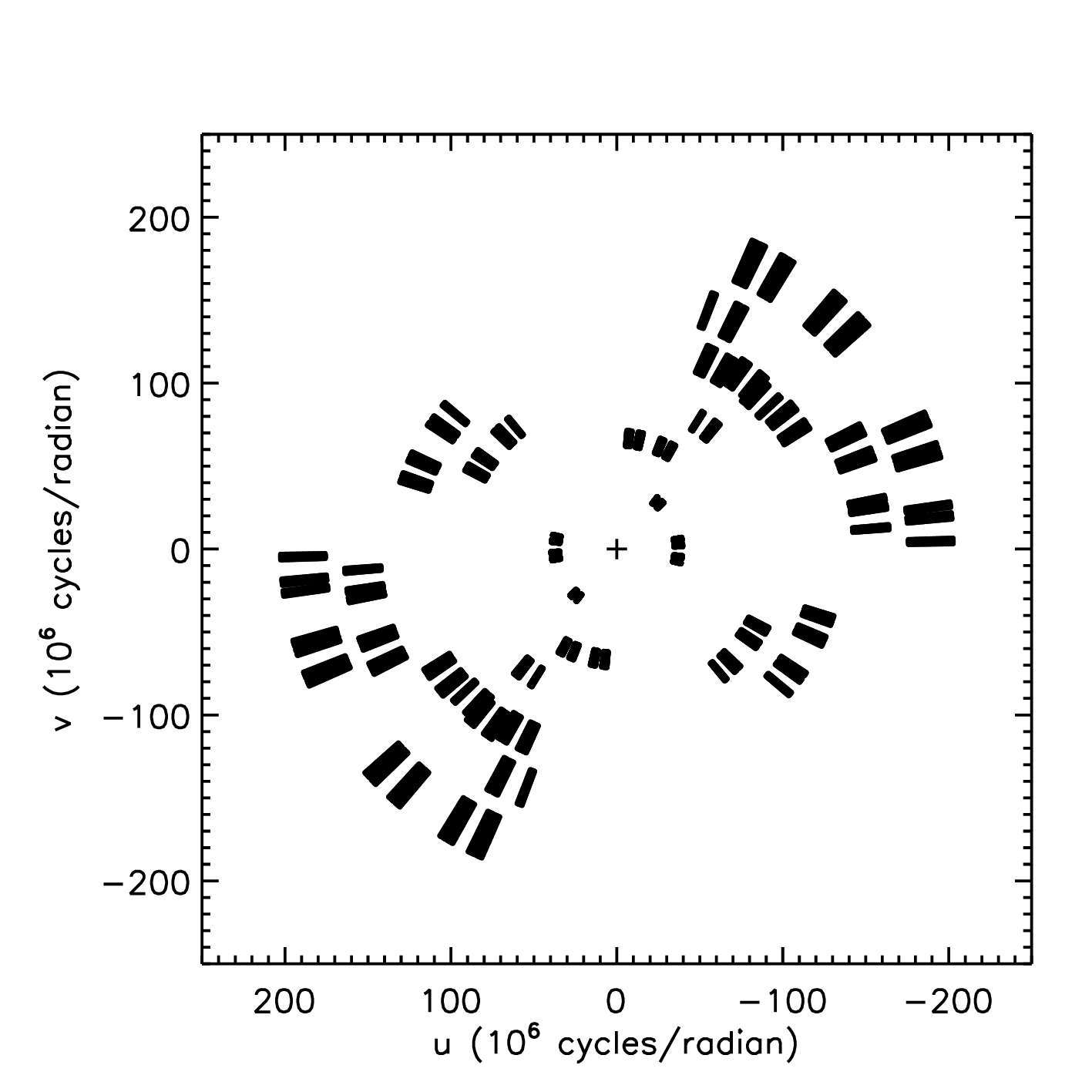}}
    \scalebox{0.10}{\includegraphics{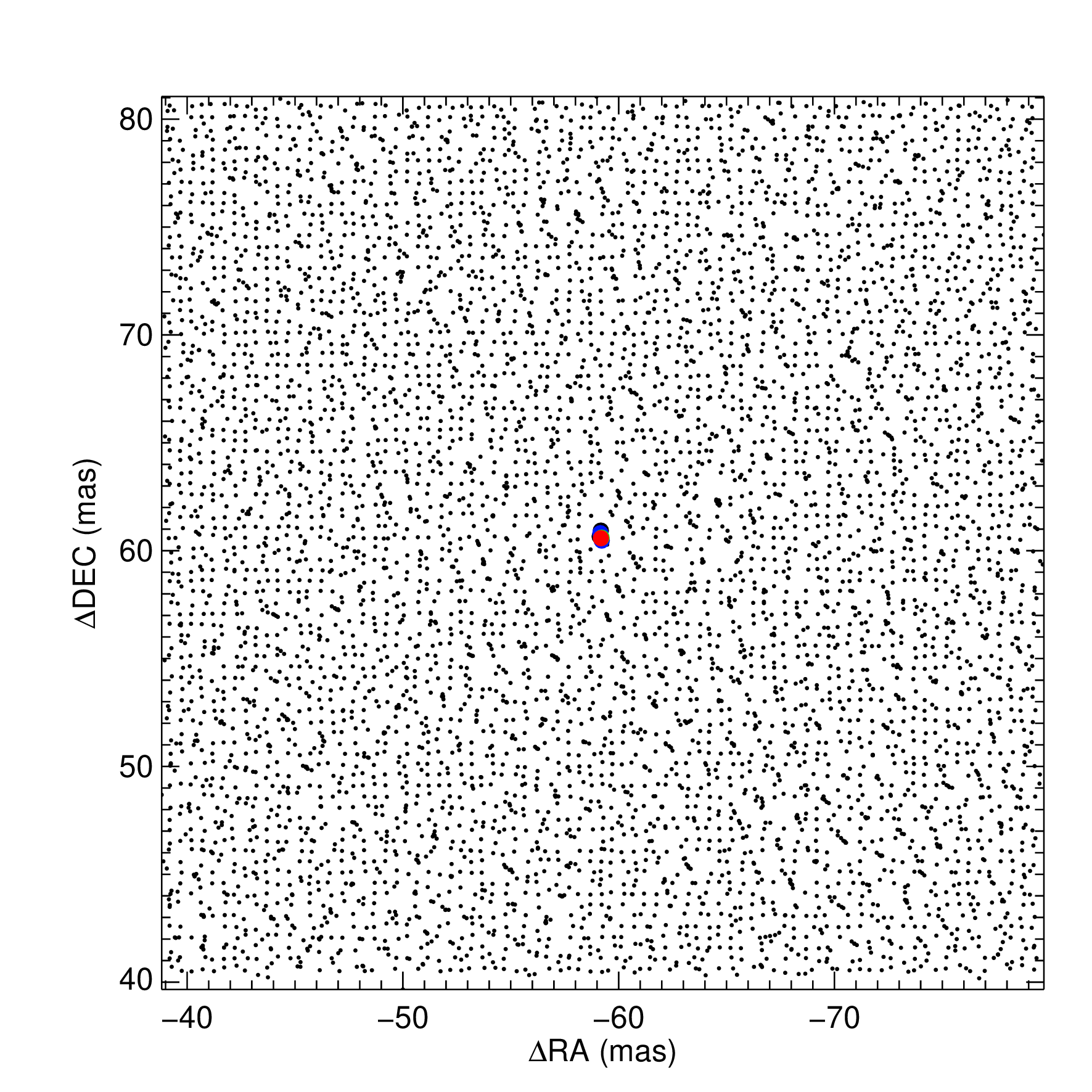}} \\
     \scalebox{0.1}{\includegraphics{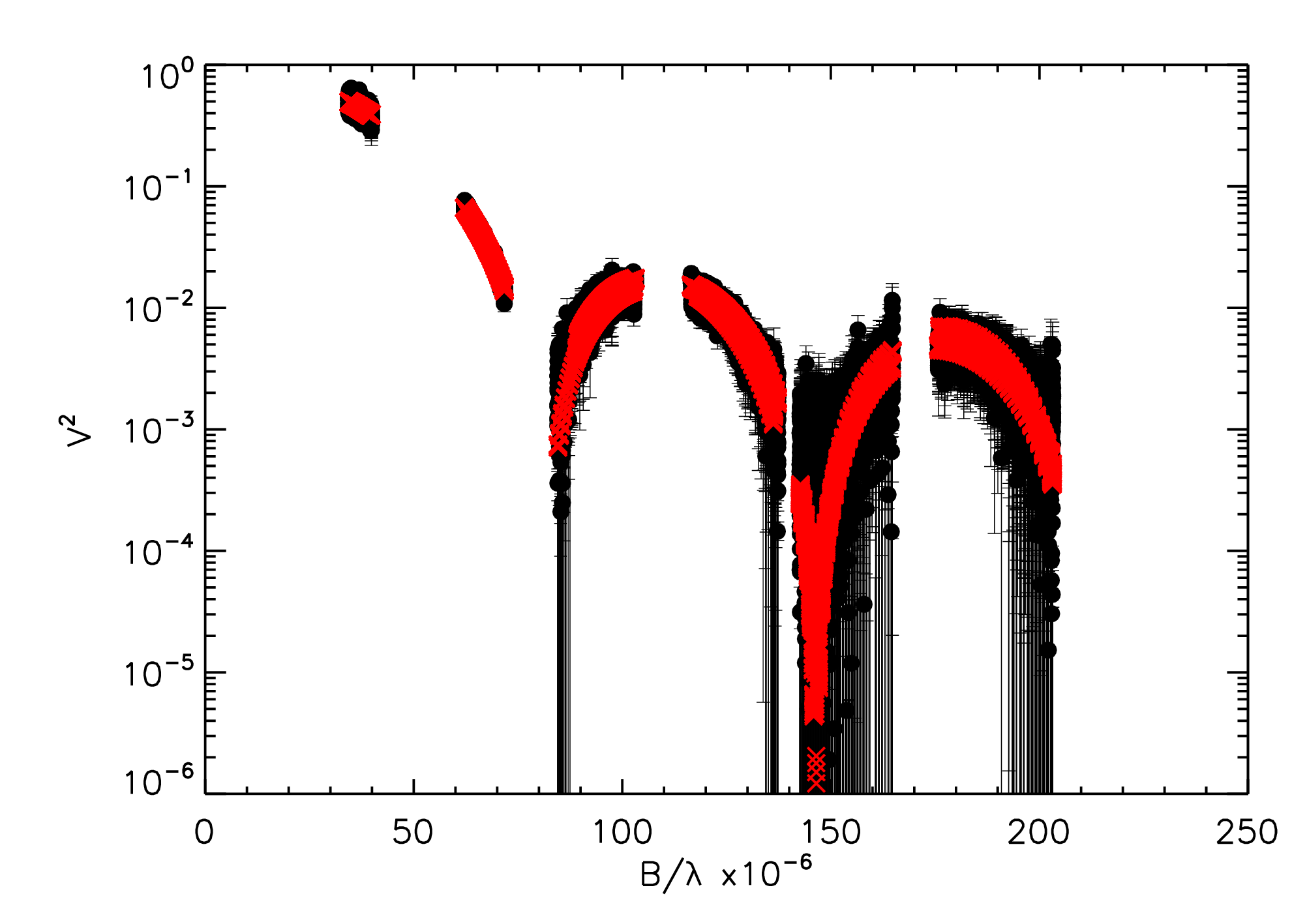}}
     \scalebox{0.1}{\includegraphics{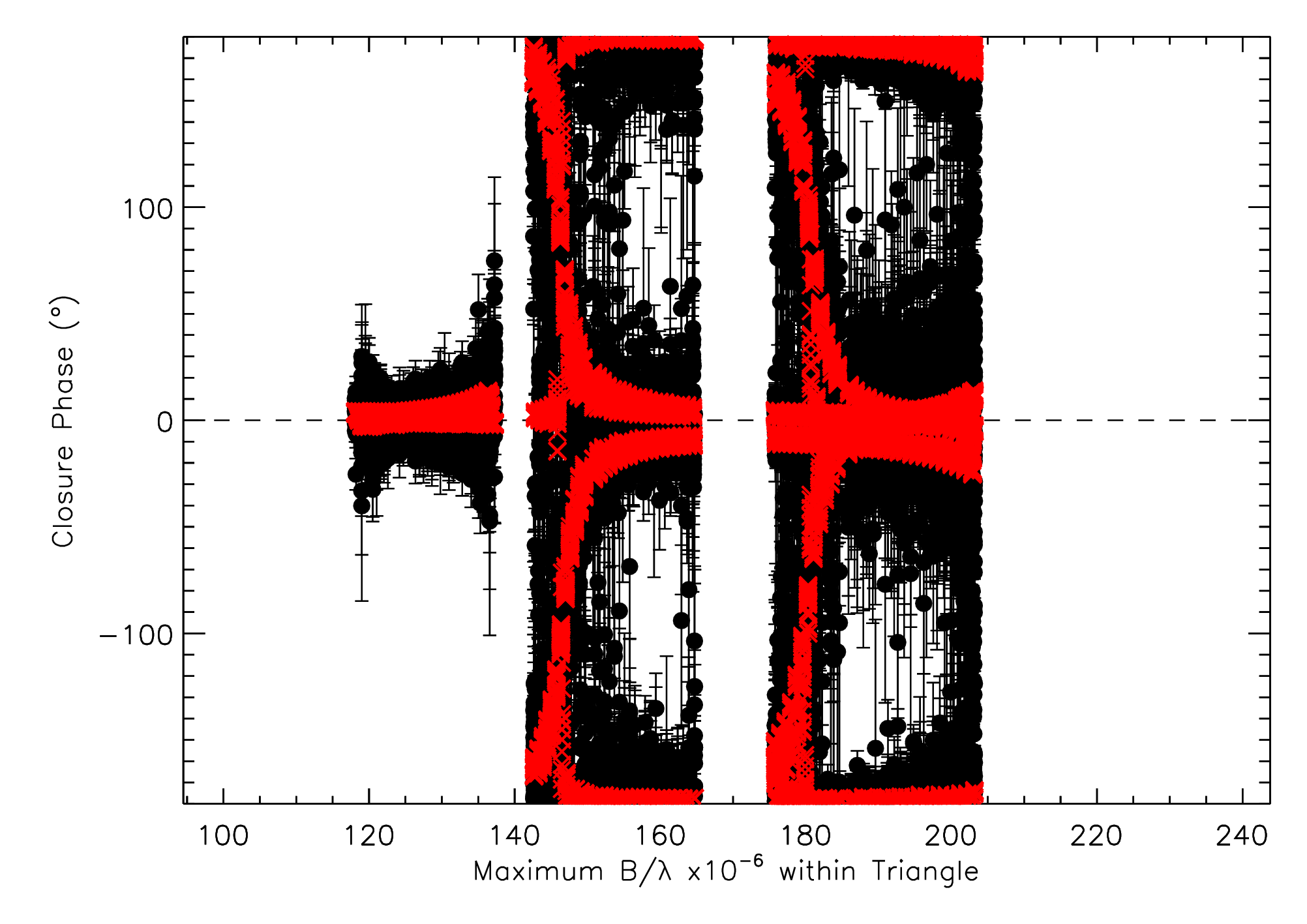}}
     \caption{Binary fit for Polaris on UT 2021Apr04. Top Row: ($u,v$) coverage (left) and $\chi^2$ map from the binary grid search (right). In the $\chi^2$ map, the red, orange, yellow, green, blue, purple, large black, and small black symbols correspond to solutions within $\Delta \chi^2$ = 1, 4, 9, 16, 25, 36, 49, and $>$50 from the minimum $\chi^2$. Bottom row: The filled black circles show the squared visibilities (left) and closure phases (right) measured with MIRC-X using the 30 second integration time. The red crosses show the best-fit binary model.}
  \end{center}
\end{figure}




\end{document}